\documentclass[a4paper,12pt]{article}
\pdfoutput=1

\usepackage{jheppub}

\usepackage{a4wide,epsfig,psfrag,amsmath,amssymb,scalefnt}
\usepackage[dvipsnames]{xcolor}
\usepackage{amsmath,comment,braket}
\usepackage{enumitem}
\usepackage{placeins}
\usepackage{breqn}
\usepackage{slashed}
\usepackage{comment}
\usepackage{subcaption}
\usepackage{hyperref}
\usepackage{url}
\usepackage{physics}

\usepackage{listings}
\allowdisplaybreaks

\begin{document}

\preprint{CERN-TH-2026-297, P3H-26-093, TTP26-042}

\title{Analytic next-to-leading order electroweak corrections to Higgs boson pair production in the forward limit}

\author[a]{Joshua Davies,}
\author[b]{Dominik Grau,}
\author[c]{Kay Sch\"onwald,}
\author[b]{Matthias Steinhauser,}
\author[b]{Daniel Stremmer}

\emailAdd{
J.O.Davies@liverpool.ac.uk,
dominik.grau@student.kit.edu,
kay.schonwald@cern.ch,
matthias.steinhauser@kit.edu,
daniel.stremmer@kit.edu}

\affiliation[a]{Department of Mathematical Sciences, University of
    Liverpool, Liverpool, L69 3BX, UK}
    
\affiliation[b]{Institut f{\"u}r Theoretische Teilchenphysik, Karlsruhe Institute of Technology (KIT),  Wolfgang-Gaede Stra\ss{}e 1, 76131 Karlsruhe, Germany}

\affiliation[c]{Theoretical Physics Department, CERN, 1211 Geneva 23, Switzerland}

\abstract{

We calculate next-to-leading order electroweak corrections for
Higgs boson pair production in the gluon fusion channel, in the
forward limit. We perform an analytic calculation and obtain a final
result for the partonic cross section that depends explicitly on all
kinematic parameters and particle masses. We describe our method,
which can be readily applied to similar processes. We observe
rapid convergence within the validity range of the forward
expansion. Furthermore, we demonstrate that the entire phase space can
be covered when this approach is combined with the high-energy
expansion.
}

\maketitle
\flushbottom

\newpage

\vspace*{2em}


\section{Introduction}

One of the main goals of the Large Hadron Collider (LHC) and its future operation
as the ``High Luminosity (HL)-LHC'' is the detailed study of the scalar sector of the
Standard Model (SM). In particular, while we have a good understanding of the
Higgs boson's mass and its couplings to gauge bosons and third generation Yukawa couplings, we do not yet have a measurement of
its self-coupling, $\lambda$, with any precision. Current limits on the observed
value of $\lambda$ are provided by CMS and ATLAS as $-0.71<\lambda_{obs}/\lambda<6.1$ \cite{CMS:2026nuu}. Projections 
for the performance of the HL-LHC are very encouraging, with an anticipated
measurement of $\lambda$ with a precision at the 30\%-level~\cite{ATLAS:2025eii}. However,
these projections assume a reduction by a factor of two in the uncertainties
of the associated theoretical predictions, motivating a large variety of theoretical
efforts.

The most important process to study in the context of measurements of $\lambda$
is Higgs boson pair production. Due to the large gluon luminosity at Hadron
colliders, the (loop-induced) production of Higgs pairs through gluon fusion is
the most dominant production mechanism by about a factor of 20. Next-to-leading
order (NLO) QCD corrections to $gg\to HH$ prove to be large \cite{Borowka:2016ehy,Baglio:2018lrj,Wang:2020nnr} compared to the leading order (LO) \cite{Glover:1987nx}, so there is
great interest in computing further higher-order contributions. The dominant
uncertainty contributing to the NLO cross section and to approximate NNLO
predictions is due to the renormalization of the top quark mass \cite{Baglio:2020wgt,Bagnaschi:2023rbx}, motivating
efforts to compute the missing top quark mass dependence at NNLO \cite{Davies:2025ghl,Davies:2024znp,Davies:2023obx,Jaskiewicz:2024xkd}, however
the NLO electro-weak (EW) corrections are anticipated to be of a similar size,
and are the topic of this paper.
We would like to point out that, in the limiting case of large $m_t$ values, results are even available at the next-to-next-to-next-to-leading order (see, e.g., Refs.~\cite{Chen:2019lzz,AH:2022elh,Chen:2026zmi}).

In the literature, a number of calculations are already available. Using fully numerical
approaches, the complete EW corrections have been computed in Ref.~\cite{Bi:2023bnq}.
Sub-sets of the contributing diagrams have been computed in Refs.~\cite{Heinrich:2024dnz,Bhattacharya:2025wnr}; work
is on-going to extend the methods used there to the full amplitude, thus providing
a valuable cross-check of Ref.~\cite{Bi:2023bnq}.
Contributions from factorizable EW diagrams are available \cite{Zhang:2024rix}, as well as the
light-quark \cite{Bonetti:2025vfd} and quark-induced \cite{Bonetti:2026cih} contributions.

On the analytic side, a full calculation which
retains dependence on all kinematic parameters is not feasible. However, an effective
approach is to expand the amplitude in complementary kinematic regions and to cover
the full phase space by combining them appropriately. The complete set of EW diagrams
involving at least one top quark has already been computed in the high-energy limit \cite{Davies:2025wke,Davies:2026wbx}
and in the large top quark mass limit~\cite{Davies:2023npk} (see also Ref.~\cite{Muhlleitner:2022ijf} for partial results). In the context of NLO QCD corrections
to the processes $gg\to HH$ \cite{Davies:2018qvx,Davies:2019dfy,Bonciani:2018omm,Bellafronte:2022jmo,Davies:2023vmj,Davies:2025qjr}, $gg\to Z^\star Z^\star$~\cite{Davies:2020lpf,Davies:2025out} (see Ref.~\cite{Degrassi:2024fye} for on-shell $ZZ$ production), $gg\to ZH$ \cite{Davies:2020drs,Degrassi:2022mro,Chen:2022rua,Davies:2026uxl} and
$gg\to W^\star W^\star$ \cite{Czaja:2026hha},
the combination of such expansions has been very successful, resulting in amplitudes
which are fast to evaluate and flexible in terms of input parameter values. These
amplitudes are available, or will soon be implemented, in the package \texttt{ggxy} \cite{Davies:2025qjr}.

The goal of this
paper is to compute the NLO EW corrections in the forward-scattering limit,
providing the missing ingredient required to produce a comparable description for
the EW amplitudes. Here the situation is more complex, as a second non-zero mass
appears in the propagators of loop integrals. This necessitates either computing
significantly more complicated integrals, or performing further expansions; here
we take the latter approach.
We consider only Feynman diagrams that contain at least one top-quark propagator. The form factors for the massless contribution were calculated analytically in Ref.~\cite{Bonetti:2025vfd}.

The remainder of the article is structured as follows: in the next
section, we present some technical details of our calculation, and in
Section~\ref{sec::scalar} we show that our approximation method works
at the level of scalar Feynman integrals.  Results for the
form factors are given in Section~\ref{sec::results}.  We conclude in
Section~\ref{sec::concl} and provide a brief outlook.
In Appendix~\ref{app::photon} we show results for the photon contribution, which is not discussed in the main part of the paper since it can be derived from existing QCD results. In Appendix~\ref{app::data} we briefly describe the format of the supplementary file which contains results for the form factors for sample phase-space points.


\section{\label{sec::framework}Framework}

\subsection{Form factors}

We consider the process $g(q_1)g(q_2)\rightarrow H(q_3)H(q_4)$ and follow the convention of Section 2 of Ref.~\cite{Davies:2026wbx}. We decompose the amplitude in terms of a linear combination of two Lorentz structures with scalar coefficients $\mathcal{M}_1$ and $\mathcal{M}_2$. These coefficients are then further split into ``triangle'' and ``box'' form factors as follows 
\begin{align}
    \mathcal{M}_1 &= X_0\,s\,(F_{\text{tri}1}+F_{\text{box}1}) \nonumber \\
    \mathcal{M}_2 &= X_0\,s\,F_{\text{box}2}. \label{eq:amplitude}
\end{align}
The prefactor in Eq.~(\ref{eq:amplitude}) is defined as 
\begin{align}
    X_0=\frac{G_F}{\sqrt{2}}\frac{\alpha_s(\mu)}{4\pi}
\end{align}
where $G_F$ is the Fermi constant, $\alpha_s$ the strong coupling constant and $\mu$ is the renormalization scale. We define the perturbative expansion of the form factors as follows
\begin{align}
    F=F^{(0)}+\frac{\alpha_s(\mu)}{\pi}F^{(1,0)}+\frac{\alpha}{\pi}F^{(0,1)}+...,
\end{align}
where $\alpha$ is the fine structure constant. The contribution $F^{(1,0)}$ refers to the two-loop QCD corrections. In this paper we compute the top quark contribution to $F^{(0,1)}$. Analytic results for the 
light-quark contribution to $F^{(0,1)}$ can be found in Ref.~\cite{Bonetti:2025vfd}.

We define all external momenta as incoming, such that the Mandelstam variables are given by
\begin{align}
    s=(q_1+q_2)^2, \qquad t=(q_1+q_3)^2, \qquad u=(q_2+q_3)^2,
\end{align}
with the additional relations 
\begin{align}
    q_1^2=q_2^2=0, \qquad q_3^3=q_4^2=m_H^2, \qquad s+t+u=2m_H^2.
\end{align}
Later, we will often use $\sqrt{s}$, the transverse momentum ($p_T$) of the Higgs bosons, and the opening angle $\theta$ with respect to the beam axis in the centre-of-mass system, which fulfil the following relations:
\begin{align}
    p_T^2 = \frac{tu-m_H^4}{s} = \frac{s}{4}\beta^2\sin^2\theta\,,\quad {\rm with}\quad\beta=\sqrt{1-\frac{4m_H^2}{s}}\,,
\end{align}
to parametrize the phase space.

\begin{figure}[t]
    \centering
    \includegraphics[width=1\textwidth]{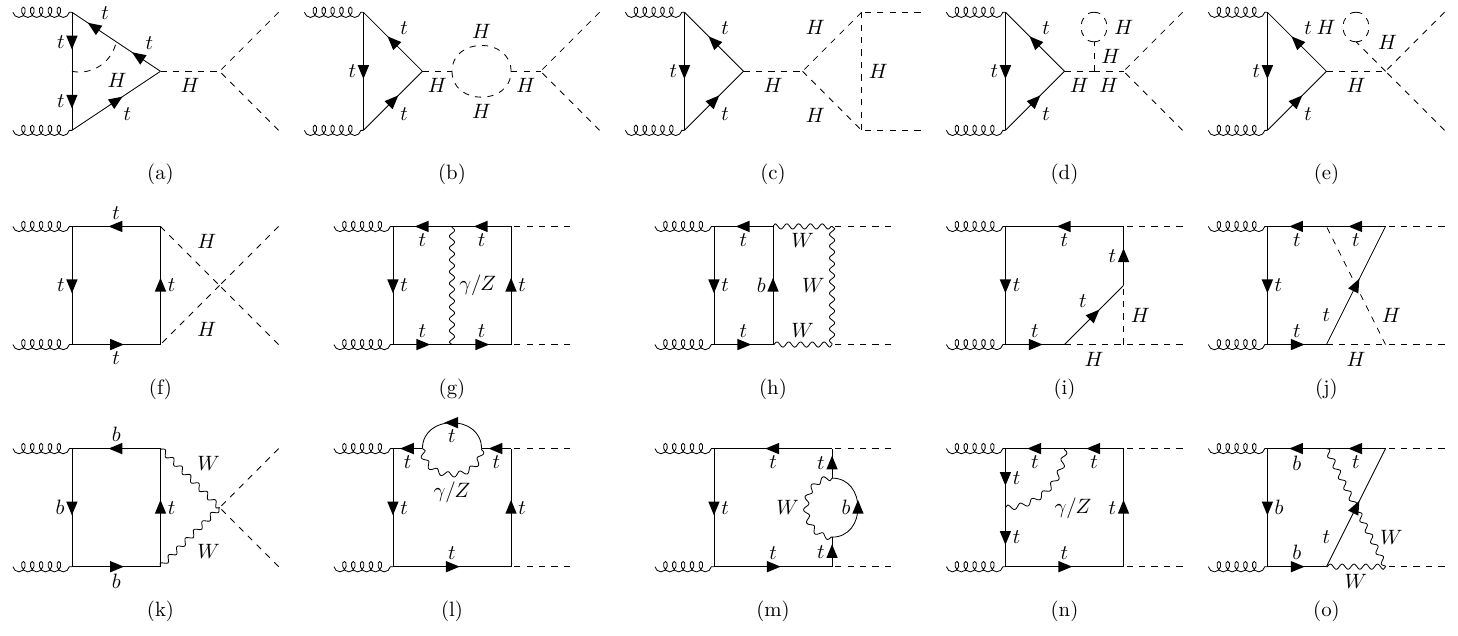} 
    \caption{Sample Feynman diagrams contributing to the NLO EW corrections to $gg\rightarrow HH$. Dashed, solid, wavy and curly lines correspond to scalar particles, fermions, electroweak gauge bosons and gluons, respectively. The diagrams (a) to (e) contribute to $F_{\text{tri1}}$ and the diagrams (f) to (o) to $F_{\text{box1}}$ and $F_{\text{box2}}$.}
    \label{fig:feynmandiagrams}
\end{figure}

In Fig.~\ref{fig:feynmandiagrams} sample 
two-loop Feynman diagrams which contribute to $F^{(0,1)}$ are shown.
For our calculation we use the Feynman gauge. In this case all integrals which appear depend on at most two different internal masses: the mass of the top quark $m_t$ and the mass of one of the EW gauge bosons $m_B$ (with $m_B\in\{m_W,m_Z,m_H\}$). 
Since we keep all scales explicit we can treat all integrals with a general second mass $m_B$ and only for the numerical evaluation specify the corresponding particle masses. By doing so we avoid redundancy in the definition of integral families and subsequent computational steps. In total we find 127 one-particle irreducible  integral families for this process.

\subsection{Amplitude generation}

For our calculation, we employ a well-tested setup that has already been
used in a number of similar calculations. We use
\texttt{qgraf}~\cite{Nogueira:1991ex} to generate the amplitudes. These are
subsequently processed using \texttt{tapir}~\cite{Gerlach:2022qnc} to produce the
input for \texttt{exp}~\cite{Seidensticker:1999bb,Harlander:1998cmq,exp_git}, which in turn generates
\texttt{FORM}~\cite{Davies:2026cci} code for each individual diagram. At this
stage, we utilize the in-house ``\texttt{calc}'' setup,
which is largely written in 
\texttt{FORM}, to manipulate the
amplitude; specifically, we apply Feynman rules and projectors,
compute traces, and map onto the integral families identified by \texttt{tapir}.

This results in expressions in which the form factors are represented
as a linear combination of scalar integrals, with each of these
integrals belonging to a specific integral family. In
Ref.~\cite{Davies:2026wbx}, a method was presented for performing an
expansion of the form factors in the high-energy limit; that is, in a
regime where the Mandelstam variables $s$ and $t$ are 
larger than the masses of the top quark, the Higgs boson, and the
gauge bosons. In the following, we discuss two approaches for
obtaining expansions in the forward limit.

\subsection{Forward expansion}

\begin{figure}[t]
    \centering
    \includegraphics[width=0.85\linewidth]{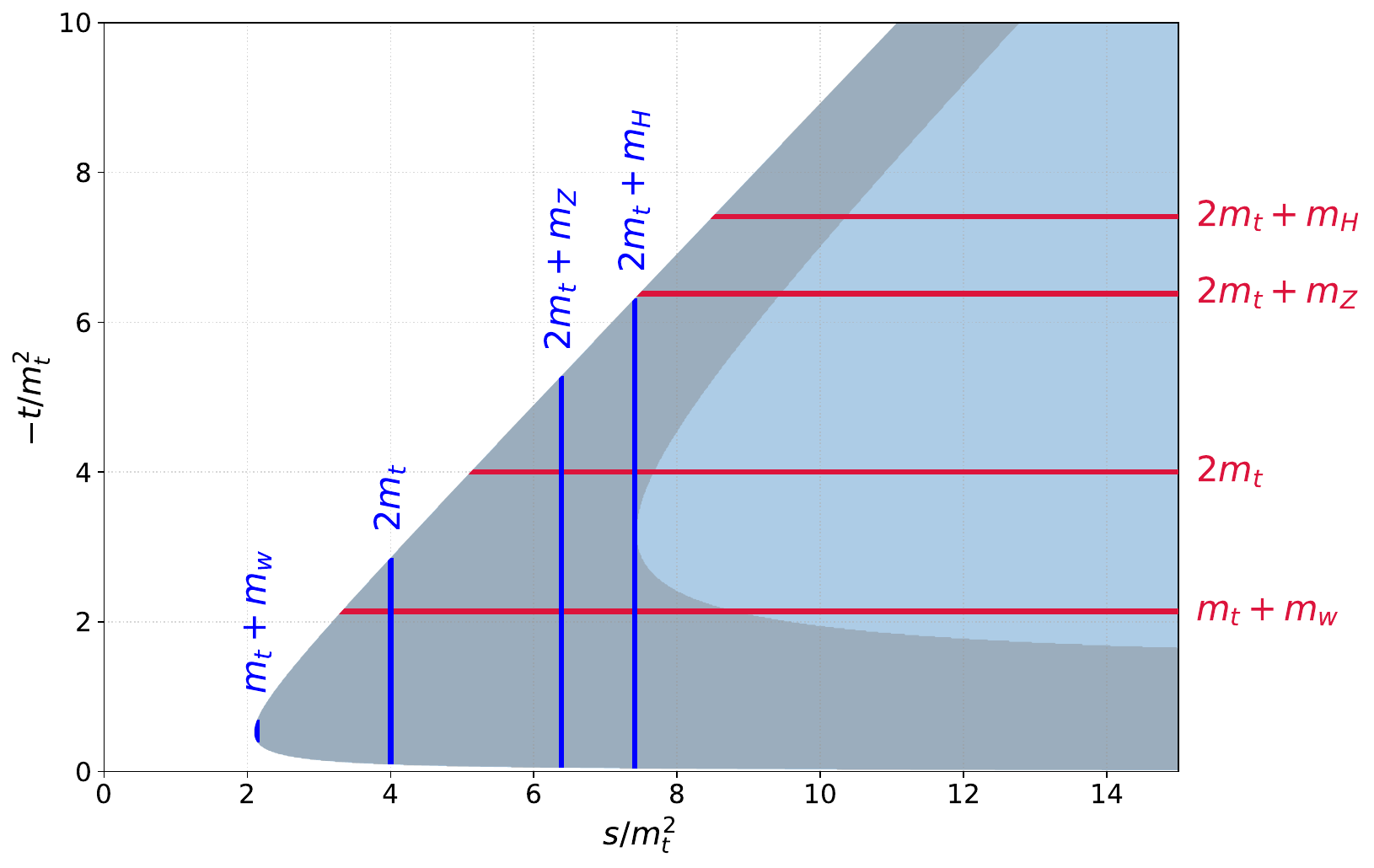}
    \caption{Physical phase space in the $s/m_t^2$---$(-t/m_t^2)$ plane shown as gray area. The darker region corresponds to $p_T<200$~GeV. The red and blue lines indicate the possible cuts in $s$ and $(-t)$.}
    \label{fig:PScuts}
\end{figure}

In Fig.~\ref{fig:PScuts}, we show the physical phase space,
parametrized in terms of the (normalized) Mandelstam variables $s$
and $t$.  We also indicate the various two- and three-particle cuts
that can appear in the Feynman diagrams.  These determine the radii of
convergence for the different expansions. For example, we expect the
high-energy expansion, where one expands in the limit $m_t\rightarrow0$, to be limited by the highest cut
which is for $\sqrt{-t}=(2m_t+m_H)$,
$(2m_t+m_Z)$, or $2m_t$ for diagrams involving virtual Higgs, $Z$, or $W$
bosons.\footnote{In this discussion, we do not consider diagrams
involving photons, as they represent a special case of the QCD
corrections discussed in Ref.~\cite{Davies:2023vmj}.}  Similarly, if we consider the scaling $|t|,m_H^2\ll s,m_t^2,m_B^2$ which we assume for the forward expansion, then one is limited by the lowest cuts in $\sqrt{-t}$ for each particle type: $2m_t$
for Higgs and $Z$, and $m_t+m_W$ for $W$. It follows, that the regions of the phase space where the respective expansions are expected to converge do not intersect, but are separated by the gap between the lowest and highest cuts. Later we will discuss how the application of various Pad\'e approximants allows both expansions to converge through the thresholds and agree with each other in the gap.

In Section~\ref{sub::naive}, we discuss an approach for the forward expansion that is closely related
to the high-energy expansion addressed in Ref.~\cite{Davies:2026wbx}. This method
offers the advantage of enabling high expansion orders in $t$.
A drawback, however, is that the expansion does not converge near particle cuts in $s$, even with the application of Pad\'e approximations. To circumvent this issue, we discuss an alternative
approach in Section~\ref{sub::fwd} that incorporates information
regarding these thresholds. Since the corresponding expansion is more
complex, it is only feasible to carry it out to a lower order in
$t$. In Section~\ref{sub::combine} we show that combining both forward-expansion approaches covers the
phase-space region up to $p_T \approx 500 - 700$~GeV for all
values of $\sqrt{s}$.

\subsubsection{\label{sub::naive}Naive-$t$ expansion}

In the first approach, labelled the ``naive-$t$ expansion,'' we follow the methods applied for the high-energy expansion~\cite{Davies:2026wbx}.
We expand the integrand in the external
Higgs boson mass using \texttt{LiteRed}~\cite{Lee:2013mka}
and perform a Taylor expansion of
the internal masses ($m_H, m_W, m_Z$) around the top-quark mass, i.e.~we expand the integrand in
\begin{eqnarray}
    \delta_B = 1 - \frac{m_B}{m_t}
\end{eqnarray}
up to and including $\delta_B^3$.
Note that no mixed terms with products of different $\delta_B$ appear, since at most one particle with a mass different from $m_t$ can be present in a single diagram at two loops.

Afterwards we perform an integration-by-parts reduction of the integrals,
which still depend on $s$, $t$, and $m_t^2$,
using \texttt{Kira}~\cite{Lange:2025fba}.
The master integrals can then be expanded, using their differential equation with respect to $t$, by inserting a simple Taylor series ansatz into the differential equation.
The boundary conditions for this expansion are master integrals at $t=0$ which depend only on the dimensionless ratio $s/m_t^2$ and can be solved using the ``expand-and-match''
approach~\cite{Fael:2021kyg,Fael:2021xdp,Fael:2022miw}. The boundaries for this step can be fixed in the large-$m_t$ limit in the same way as has been done in the case of QCD corrections~\cite{Davies:2023vmj}.
Using this approach, we compute expansion terms
up to $m_H^4$ and $t^{10}$.

\subsubsection{\label{sub::fwd}$\delta q$ expansion}

It turns out that the naive-$t$ expansion looses precision for values of $\sqrt{s}$
close to particle thresholds as, e.g., $\sqrt{s}=2m_t$ or $\sqrt{s}=2m_t+m_B$. To address these phase space regions, we adopt a second strategy, where we do not expand in $\delta_B$.
The core of this strategy is the so-called ``$\delta
q$ expansion''~\cite{grauMasterThesis,Davies:2025out}, which performs a simultaneous
expansion in $t$ and $m_H^2$. While formally equivalent to
the so-called ``$p_T$ expansion''~\cite{Bonciani:2018omm}, it is technically simpler to
implement, as it relies essentially on a $d$-dimensional Taylor
expansion of the amplitude in terms of the momentum
$\delta q=q_1+q_3$.

The $\delta
q$ expansion is applied at the integrand level to each Feynman integral $I(q_1,q_2,q_3,m_t,m_B,\epsilon)$, where $d=4-2\epsilon$ is the space-time dimension, and to their coefficients. For the coefficients, which are rational functions in $s$, $t$, $m_H$, $m_t$, $m_B$ and $\epsilon$, a simultaneous Taylor expansion is performed in the Mandelstam variable $t$ and the mass of external Higgs bosons $m_H^2$.\footnote{In our calculation we must distinguish between the Higgs boson in the final state and the 
Higgs boson within the loop diagrams, as they are treated 
differently. For the former we use $m_H$ and the
latter is always referred to with $m_B$; thus, there is 
no risk of confusion. }   In the integrands we substitute $q_3=-q_1+\delta q$ and perform a Taylor expansion in the four-momentum vector $\delta q$ which 
has the following form:
\begin{eqnarray}
    I(q_1,q_2,q_3,m_t,m_B,\epsilon) & = &
    I(q_1,q_2,\delta q,m_t,m_B,\epsilon) |_{\delta q\rightarrow0} \nonumber \\[2mm]
    && + \left ( \frac{\partial I(q_1,q_2,\delta q,m_t,m_B,\epsilon)}{\partial\delta q_\mu} \biggl |_{\delta q\rightarrow0} \right) \delta q_\mu  \nonumber \\
    && + \left ( \frac{\partial^2 I(q_1,q_2,\delta q,m_t,m_B,\epsilon)}{\partial\delta q_\mu\: \partial\delta q_\nu} \biggl |_{\delta q\rightarrow0} \right) \delta q_\mu \delta q_\nu + \ldots \, .
\end{eqnarray}
The derivative w.r.t.~$\delta q$ leads to tensor integrals, such that the coefficient of the $n^{\rm th}$ expansion term is a linear combination of tensor integrals with rank up to $n$. 
In fact, one of the computationally demanding tasks in this approach
is the calculation of these high-rank
tensor integrals. For this purpose, we use the \texttt{FORM} package
\texttt{OPITeR}~\cite{Goode:2024cfy}. 

After applying the tensor reductions, there
are three different  scalar products that include $\delta q$, 
given by
\begin{align}
    q_1\cdot\delta q = \frac{t-m_H^2}{2}, \qquad q_2\cdot\delta q = \frac{m_H^2 - t }{2}, \qquad \delta q^2 = t \,. \label{eq:scalarprod}
\end{align}
They are proportional to the small scales $t$ and $m_H^2$.
Afterwards, the initial Feynman integrals can be expressed as follows
\begin{align}
     I(q_1,q_2,q_3,m_t,m_B,\epsilon) = \sum_{j,k=0}^{j+k\leq N_q} \sum_i c_{i,j,k}(s,m_t,m_B,\epsilon)\: I_{0,i}(q_1,q_2,m_t,m_B,\epsilon)\: \left(\frac{t}{s}\right)^j\:
     \left(\frac{m_H^2}{s}\right)^k \,. \label{eq:forwardlimitexp}
\end{align}
The sum of the powers of $t$ and $m_H^2$ is limited by the expansion depth $N_q$. Note that, due to the last scalar product in Eq.~(\ref{eq:scalarprod}), the depth $N_q$ corresponds to the expansion order in $\delta q$ divided by two. The sum over $i$ runs over the set of all scalar integrals arising from the tensor reduction. These integrals are independent of the external momentum $q_3$ and therefore only depend on $s,m_t$ and $m_B$. For some integral families this leads to an overdetermined set of denominators;
we use \texttt{LIMIT}~\cite{Herren:2020ccq} to perform a partial fraction decomposition of the scalar integral families, and to map them onto a new minimal set.

The representation in Eq.~(\ref{eq:forwardlimitexp}) is not unique,
and the number of terms depends strongly on how the loop and external
momenta are assigned within the Feynman integrals. Instead of
optimizing the denominators of each individual integral, it proves
useful to apply the following rules to the global definition of the
integral families before performing the $\delta q$ expansion:
\begin{enumerate}
    \item Align $\delta q$ along the shortest path through the graph.
    \item Align all other momenta along the shortest paths through the graph for which the intersection with the $\delta q$ path becomes minimal.
\end{enumerate}
Using an in-house \texttt{Mathematica} script, we
automatically modified the so-called topology files generated by
\texttt{tapir}. Table~\ref{tab:dqterms_amp} shows the number of terms
at intermediate stages of the two-loop amplitude calculation, expanded
up to orders $N_q\in\{0,1,2\}$.  The table lists
these counts for both the original \texttt{tapir} topology families
and the improved versions. 
After the $\delta q$ expansion the number of produced terms is smaller by a factor of approximately  two for the improved version  once an expansion up to order $N_q=2$ is considered. After tensor reduction,
the reduction factor is four, while after partial fraction decomposition,
it is three. The advantage of the improved families increases with the
depth of the expansion; furthermore, it is intuitive that a smaller
number of terms or scalar integrals reduces the computational effort
required for subsequent steps.
This enables us to expand the
amplitude up to order $N_q=5$ with a reasonable amount of
computational resources and time.  It should be noted that the final
results, after reduction to master integrals, are of course identical.

Afterwards all scalar integrals,
which now depend only on $\{s, m_t, m_B\}$,
can be reduced to a minimal set of 200~master integrals.
We use
\texttt{Kira}~\cite{Lange:2025fba} together with \texttt{FireFly}~\cite{Klappert:2019emp} for the reduction. We format the integration-by-parts relations with \texttt{ibppp}~\cite{IBPPP_git}
to produce a \texttt{FORM} TableBase which stores only the required orders in the $\epsilon$ expansion
of the reduction relations.

\begin{table}[t]
    \centering
    \begin{tabular}{l|rrr|rrr}
        \hline
        & \multicolumn{3}{c|}{default ($10^6$)}
        & \multicolumn{3}{c}{optimized ($10^6$)} \\
        & $N_q=0$ & $N_q=1$ & $N_q=2$
        & $N_q=0$ & $N_q=1$ & $N_q=2$ \\
        \hline
        $\delta q$ expansion
        & 0.83 & 6.82 & 30.88
        & 0.84 & 5.01 & 16.80 \\

        Tensor reduction
        & 0.26 & 3.32 & 23.08
        & 0.12 & 1.07 & 5.56 \\

        Partial fraction decomposition
        & 0.13 & 1.77 & 12.41
        & 0.08 & 0.75 & 4.01 \\
        \hline
    \end{tabular}
    \caption{Comparison between the standard and optimized  momentum routing in the topology files generated by
    \texttt{tapir}. We show the number of generated 
    terms for expansion depths $N_q=0$, $N_q=$ and $N_q=2$
    following the calculation steps listed on the left-hand side.}
    \label{tab:dqterms_amp}
\end{table}
In the next step, we again employ the ``expand-and-match''
approach~\cite{Fael:2021kyg,Fael:2021xdp,Fael:2022miw} to calculate the master integrals. Since in this case we have to deal with a dependence on two dimensionless variables, we have to generalize the algorithms as described in Section~\ref{sub::eam}. After substituting the master integrals into the amplitude, we obtain semi-analytic expressions with explicit dependence on $s$, $m_t$ and $m_B$ and high-precision numerical coefficients. We perform the numerical evaluation using
\texttt{FORM}; since version 5.0, this program offers the
capability to compute expressions with arbitrary floating-point
precision.

As previously mentioned, the $\delta q$ expansion requires a more
elaborate tensor reduction and also leads to more complicated
integrals.
We succeed in calculating the expansion terms up to $N_q=5$, i.e.,
terms of the form $t^j(m_H^2)^k$ with $j+k\le5$.

\subsection{\label{sub::eam}Two-dimensional expand-and-match}

We compute the master integrals using the ``expand-and-match''
method~\cite{Fael:2021kyg,Fael:2021xdp,Fael:2022miw}. The basic idea behind this approach is to
construct high-order, generalized series expansions around suitably
chosen expansion points, with coefficients determined to high
numerical precision. Combining these expansions yields a
semi-analytic result for each master integral across the entire
parameter space, which can be evaluated numerically very quickly.

The ``expand-and-match'' method has so far been applied only to master
integrals with a single scale (see, e.g., Refs.~\cite{Fael:2026fxp,Fael:2026gqz}
for applications to three-loop four-point and four-loop two-point
processes). In the following, we discuss the extension to problems
involving two dimensionless parameters.

The expansions are constructed using differential equations, which we set up using \texttt{LiteRed}~\cite{Lee:2013mka}.
For our two-parameter problem they take the form
\begin{align}
    \partial_{\tilde{s}} \vec{M} = A_{\tilde{s}} \vec{M}, \qquad \partial_{\tilde{m}_B^2} \vec{M} = A_{\tilde{m}_B^2} \vec{M} ,
\end{align}
where for convenience we have introduced
dimensionless variables
$\tilde{s}=s/m_t^2$ and $\tilde{m}_B^2=m_B^2/m_t^2$. 
$\vec{M}$ is the vector of master integrals and $A_{\tilde{s}}$ and
$A_{\tilde{m}_B^2}$ are the coefficient
matrices.
The matrix elements are rational functions in $\tilde{s}$, $\tilde{m}_B^2$ and $\epsilon$ that exhibit numerous singularities,
which can be divided in two groups. 
The first group consists of physical singularities corresponding to the particle cuts in the $s$ channel.
As we will discuss below they 
require a particular treatment.
The second group are spurious singularities, which have no physical origin. We observed that they have no
influence on our approach. 

It is convenient to split the original $200\times 200$ system of
differential equations into three smaller systems, corresponding to
the physical sectors for the combined Higgs- and $Z$-boson
contribution (100 master integrals), the $W$ boson contribution (87
master integrals) and the photon contribution (38 master
integrals).\footnote{We take master integrals for the photon
contribution from the NLO QCD calculations~\cite{Davies:2023vmj}.}

\begin{figure}[t]
    \centering
    \includegraphics[width=0.8\linewidth]{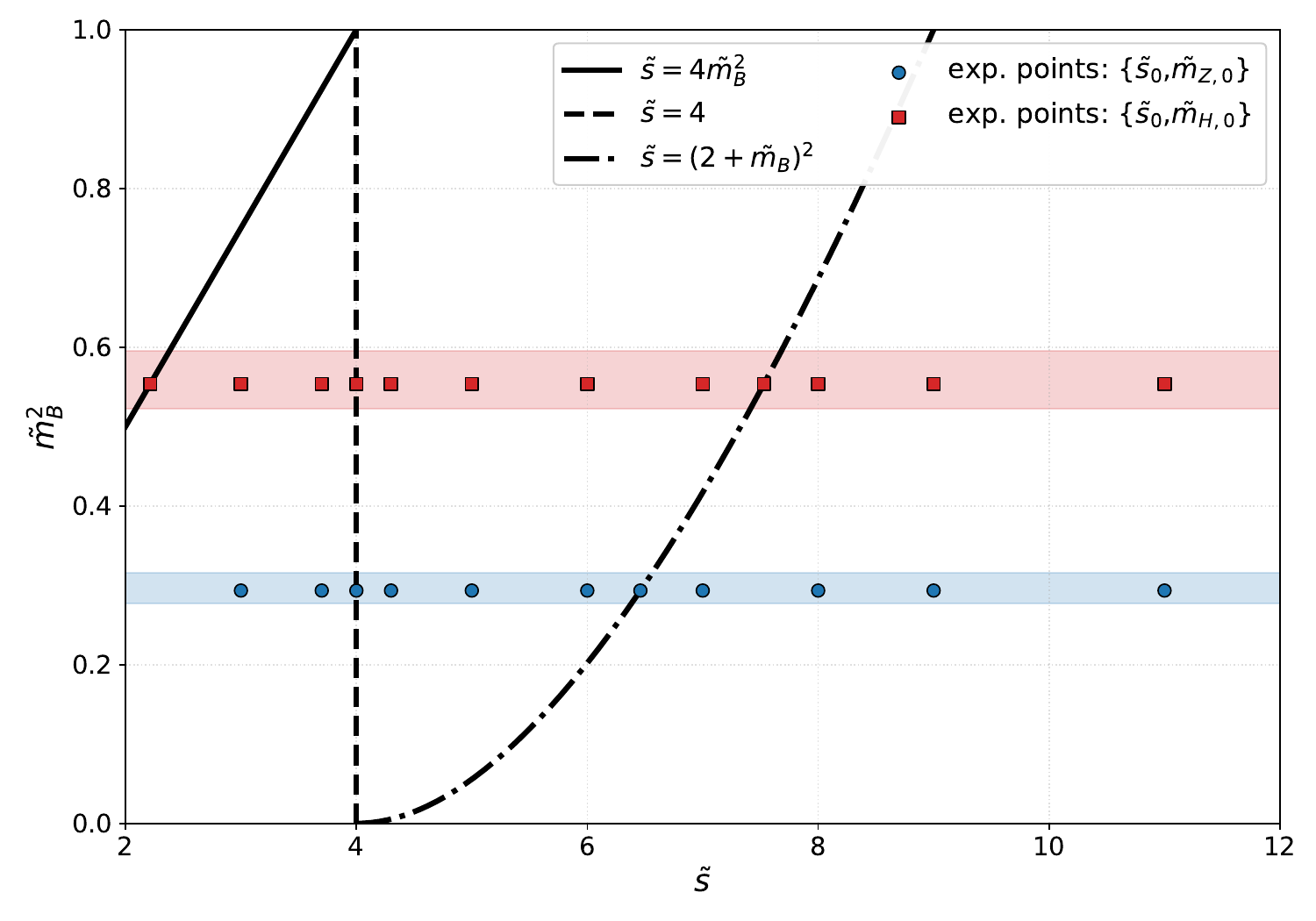}
    \caption{Section of the two-dimensional $\tilde{s}$---$\tilde{m}_B^2$ plane 
    which is covered by the ``expand-and-match'' procedure for the $Z$ and Higgs boson contribution.
    The black (solid, dashed and dash-dotted) lines
    indicate the cuts (see also Fig.~\ref{fig:PScuts}). The coloured bands indicate the region which we aim 
    to cover in the $\tilde{m}_B^2$ direction.
    In the case of the $W$-boson contribution,
the solid and dashed lines are missing, and
the equivalent of the dash-dotted
line starts at $\tilde{s}=1$. }
    \label{fig:eamphasespace}
\end{figure}

In the following we discuss the various ans\"atze for the series
expansions which depend on the respective expansion point.  In all
cases, our aim is to cover the range for $\tilde{s}$ for
$2M_H\le\sqrt{s}\le\infty$.  For $\tilde{m}_B^2=m_B^2/m_t^2$ we aim to
encompass all phenomenologically relevant values of the top-quark
mass.  In particular, it must be possible to switch from the pole to
the $\overline{\rm MS}$ scheme in the final result.
In
Fig.~\ref{fig:eamphasespace}, we show the $\tilde{s}$---$\tilde{m}_B^2$
plane for the Higgs- and $Z$-boson contributions. The required ranges for
$\tilde{m}_B^2$ are depicted as red and blue bands.

Let us begin by discussing the ansatz for the case where the
expansion point corresponds to a physical singularity. As a first
step, we perform a variable transformation from $\{\tilde{s},
\tilde{m}_B\}$ to $\{x, y\}$. The variable $x$ parametrizes the
distance to the respective physical singularity, while $y$ describes
the deviation from the expansion point $\tilde{m}_{B,0}$ in the
direction of $\tilde{m}_B$. 
In practice we choose 
\begin{eqnarray}
    \tilde{m}_{B,0}&=&\frac{m_B}{2}\biggl(\frac{1}{m_t^{\text{OS}}}+\frac{1}{\overline{m}_t}\biggl), \quad B\in\{H,Z,W\}\,,
\end{eqnarray}
where $m_t^{\text{OS}}$ and $\overline{m}_t\equiv \overline{m}_t(\overline{m}_t)$
correspond to the pole and $\overline{\rm MS}$ top quark mass, respectively.

The variable $x$ can depend on either $\tilde{m}_B$ or
$(\tilde{m}_B)^2$. In the former case, we define $y$ such that it is
linear in $\tilde{m}_B$ in order to avoid square roots in the 
expansion procedure; otherwise, $y$ depends on
$\tilde{m}_B^2$. For the four physical singularities $2m_t$, $2m_H$, $2m_t+m_{H,Z}$ and $m_t+m_W$, we have
\begin{align}
    \{ x( \tilde{s} , \tilde{m}_B ) ,   y( \tilde{s} , \tilde{m}_B )   \}  =   \left\{
    \begin{matrix}
         \{ \sqrt{\tilde{s} - 4}   ,\:  \tilde{m}_B^2 - \tilde{m}_{B,0}^2 \} \\
        \{ \sqrt{\tilde{s} - 4\tilde{m}_B^2 }   ,\:  \tilde{m}_B^2 - \tilde{m}_{B,0}^2 \} \\
        \{  \tilde{s} - ( 2 + \tilde{m}_B ) ^2  ,\: \tilde{m}_B - \tilde{m}_{B,0} \} \\
        \{  \sqrt{\tilde{s} - ( 1 + \tilde{m}_B ) ^2 } ,\: \tilde{m}_B - \tilde{m}_{B,0} \}
    \end{matrix}\right.
    \,,
\end{align}
chosen analogously to the expansion variables in Ref.~\cite{Becchetti:2025rrz}.
After transforming the differential equations to the new variables we expand the coefficient matrices in the limit $\{x,y\}\rightarrow\{0,0\}$ and insert the following ansatz for each master integral: 
\begin{align}
    M_a = \sum_{i=-2}^{N_x} \sum_{j=0}^{N_y} \sum_{k=-2}^{n_{a}^\epsilon} \sum_{l=0}^{k+2} T_{a,i,j,k,l}\: x^i\: y^j\:\epsilon^k\: (\log{x})^l\,,
    \label{eq:eamlogansatz}
\end{align}
where we typically use $N_x=25$ and $N_y=10$
for the expansion depths in $x$ and $y$.
Note that some of the master integrals
exhibit negative powers in $x$, which is why 
we start the summation at $i=-2$.
The upper limit of the $k$ summation,
$n_{a}^\epsilon$, depends on the
spurious $\epsilon$-poles in the coefficient
of the respective master integral as well as
the $\epsilon$-poles that may appear in the differential equation. For our application 
we can choose $n_{a}^\epsilon=2$ for all
of the Higgs- and $Z$-Boson master integrals and $n_{a}^\epsilon=4$ for the rest.

For the singular point at 
$\tilde{s}=\infty$ 
we use the variables 
\begin{align}
    \{ x( \tilde{s} , \tilde{m}_B ) ,   y( \tilde{s} , \tilde{m}_B )   \}  =  \{  1/\tilde{s}  ,\:  \tilde{m}_{B}^2-\tilde{m}_{B,0}^2  \}
\end{align}
and the ansatz in Eq.~(\ref{eq:eamlogansatz}) 
modified such that the upper limit of the sum over $l$ is $(k+4)$, in order to account for the quadratic logarithms at each order in $\epsilon$.

Substituting the ansatz into the 
differential equations yields a
linear system of equations for the coefficients $T_{a,i,j,k,l}$, which
we solve in terms of a minimal set of boundary conditions using \texttt{Kira}. The latter are
determined by numerically computing the master integrals with high
precision using \texttt{AMFlow}~\cite{Liu:2022chg} at an $x$-value close to the expansion point.

For the regular expansion points it is possible to use
a simple two-dimensional Taylor expansion
for the ansatz. We define the corresponding variable transformation by 
\begin{align}
    \{ x( \tilde{s} , \tilde{m}_B ) ,   y( \tilde{s} , \tilde{m}_B )   \}  =  \{  \tilde{s} - \tilde{s}_0   ,\:  \tilde{m}_{B}^2-\tilde{m}_{B,0}^2  \}
    \,,
\end{align}
and the ansatz for the master integrals
is given by
\begin{align}
    M_a = \sum_{i=0}^{N_x}\sum_{j=0}^{N_y}\sum_{k=-2}^{n_{a}^\epsilon} T_{a,i,j,k}\: x^i\:y^j\:\epsilon^k\,,
\end{align}
where we choose $N_x=50$ and $N_y=10$.
The radius of convergence of these expansion
is limited by the closest physical
threshold.

For our calculation we choose the following
regular expansion points 
\begin{eqnarray}
    \tilde{s}_0&\in&\{2.3,3,3.7,4.3,5,6,7,8,9,11,13,17,23,27,33,50\}\,.
    \label{eq::s0_reg}
\end{eqnarray}
This set of expansion points covers the whole parameter space with high precision, even the region around the physical thresholds.

In Fig.~\ref{fig:eamphasespace} we sketch a part of the parameter space considered for the Higgs- and $Z$-Boson master integrals.
The three relevant particle thresholds are depicted as lines, the expansion points as squares and circles
surrounded by the physically-relevant region (shown as coloured bands) that is covered by at least eight digits in the vicinity of the thresholds, and by twelve digits elsewhere.
This accuracy is estimated by considering two adjacent series
expansions that are matched at or near their respective expansion
points and evaluated at the midpoint for various values of $\tilde{m}_B$.

Choosing $\tilde{m}_{B}=1$
would lead to the same expansion as the naive-$t$ expansion presented before. 
Having the differential equation in $\tilde{m}_B$ at hand allows us to obtain deeper expansions in the mass difference,
though at the cost of a shallower expansion in $t$.
Which expansion is most appropriate depends on the specific region in phase space under investigation; this is discussed in Section~\ref{sub::combine}.

\subsection{\label{sec::pade}Pad\'e approximation and Aitken extrapolation}

To extend the radii of convergence of the simple expansions we apply an approximation procedure that was introduced in Refs.~\cite{Davies:2019dfy,Davies:2020lpf}, which is 
built around the construction of various Pad\'e approximants. We tag our expansion variables with $x$ and insert numerical values for all scales. We then construct a Pad\'e approximant in the variable $x$ 
\begin{align}
    [n/m](x)=\frac{a_0+a_1x+...+a_nx^n}{1+b_1x+...+b_mx^m}
\end{align} 
for all possible combinations of $n$ and $m$ that fulfil the following conditions:
\begin{align}
    N_{\text{low}} \leq n + m \leq N_{\text{high}}  \quad \text{and} \quad N_{\text{low}} \leq n + m - | n - m | .
\end{align}
After setting $x=1$ we obtain a numerical value for each of the Pad\'e approximants. We compute a weighted central value and a corresponding uncertainty, where the weight depends on the pole structure of the denominator, the number of expansion terms considered, and the difference $n-m$. 

For the $\delta q$ expansion we tag both $t$ and $m_H^2$ and choose $N_{\text{low}}=3$ and $N_{\text{high}}=5$. In the case of the naive-$t$ expansion we only tag $t$ and compute each coefficient of the $\delta_B$ expansion separately using $N_{\text{low}}=7$ and $N_{\text{high}}=10$. In the next step, we perform a single Pad\'e approximation with $n=2$ and $m=1$ w.r.t.~the $\delta_B$ expansion. This specific Pad\'e approximant is equivalent to the so-called ``\textit{Aitken extrapolation}'' (see Ref.~\cite{Press:2007ipz}), which is a well known sum acceleration technique to improve convergence.  

For the high-energy expansion, the expansion variable $m_t$ appears in even and odd powers, as well as in the arguments of logarithms. We tag combined odd and even orders using $m_t^{2k}\rightarrow m_t^{2k}x^k$ and $m_t^{2k-1}\rightarrow m_t^{2k-1}x^k$, while logarithmic dependence on $m_t$ is not tagged.
The Padé approximants are constructed with the bounds $N_{\text{low}}=46$ and $N_{\text{high}}=54$. As with the naive-$t$ expansion, this is done for each order in $\delta_B$, followed by the additional acceleration in $\delta_B$.

It has been shown for other processes that this procedure leads to precise approximations of the exact result (see Refs.~\cite{Davies:2023vmj,Davies:2025out,Czaja:2026hha}).
In Section~\ref{sec::scalar} we show that 
Pad\'e-improvement and sum acceleration yield good results for non-planar scalar integrals
which appear in the present calculation.
For the scalar integrals  a comparison with exact numerical results is possible.
For the form factors discussed in Section~\ref{sec::results}, no exact results
are available. However, we demonstrate that
after Pad\'e-improvement and summation acceleration,
results from the forward and the high-energy 
expansions converge toward each other at intermediate values of
$\sqrt{s}$ and $p_T$.

\subsection{Renormalization}

To obtain a finite result for the form factors, a one-loop
renormalization of the parameters and fields appearing in the LO form
factors must be performed. This procedure is described in detail in
the literature, see, e.g., Refs.~\cite{Denner:1991kt,Denner:2019vbn}. In the forward limit, the renormalization
proceeds in complete analogy to the treatment for large $m_t$ or high
energies, as discussed in Refs.~\cite{Davies:2023npk} and~\cite{Davies:2026wbx},
respectively.

As LO results, we use the expressions for $F_{\rm box1}$ and $F_{\rm box2}$ and
interpret the parameters $\alpha$, $m_t$, $m_Z$, $m_W$, and $m_H$ as
bare parameters. Note that the masses originate either from
propagators or from couplings. In particular, we replace the weak
mixing angle with $\cos\theta_W = m_W/m_Z$. Furthermore, the wave
functions of the external Higgs bosons must be treated as bare. The
bare quantities are then replaced by the renormalized quantities
multiplied by the respective renormalization constants in the on-shell
scheme, which we take from Refs.~\cite{Denner:1991kt,Denner:2019vbn}. 


\section{\label{sec::scalar}Scalar integrals}

\begin{figure}[t]
    \centering
    \includegraphics[width=.70\textwidth]{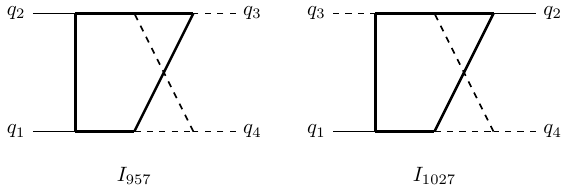} 
    \caption{
    Scalar non-planar top-level integrals with
    $q_1^2=q_2^2=0$ and $q_3^2=q_4^2=m_H^2$. 
    Thick solid internal lines have mass $m_t$ and the
    dashed lines have mass $m_H$. The integrals are related by the crossing of the external momenta $q_2 \leftrightarrow q_3$.
    }
    \label{fig:SI1dia}
\end{figure}

In this section we discuss the results of the forward and high-energy expansions\footnote{In this section we use for the Pad\'e parameters $N_{\text{low}}=20$ and $N_{\text{high}}=27$.
Larger values, which are used for the amplitude discussed in Section~\ref{sec::results}, do not lead to qualitative improvements in the given examples.} 
for the non-planar
scalar integrals shown in Fig.~\ref{fig:SI1dia}. 
In Ref.~\cite{Davies:2025wke}, 
these integrals are investigated in the context of the 
high-energy expansion (see Figs.~3 and~4 in Ref.~\cite{Davies:2025wke}), assuming massless external lines and setting $m_H=m_t$ for the dashed internal lines. For (the simplified version of) $I_{957}$, 
good results are obtained 
in the regime where the high-energy expansion is expected to converge. In the case of $I_{1027}$, however, it is not possible 
to derive a meaningful 
approximation for the imaginary part 
(for $q_3^2=q_4^2=0$ and $m_H=m_t$) 
at transverse momenta below $p_T\approx450$~GeV.

\begin{figure}[t]
    \begin{center}
    \begin{tabular}{cc}
    \includegraphics[width=.45\textwidth]{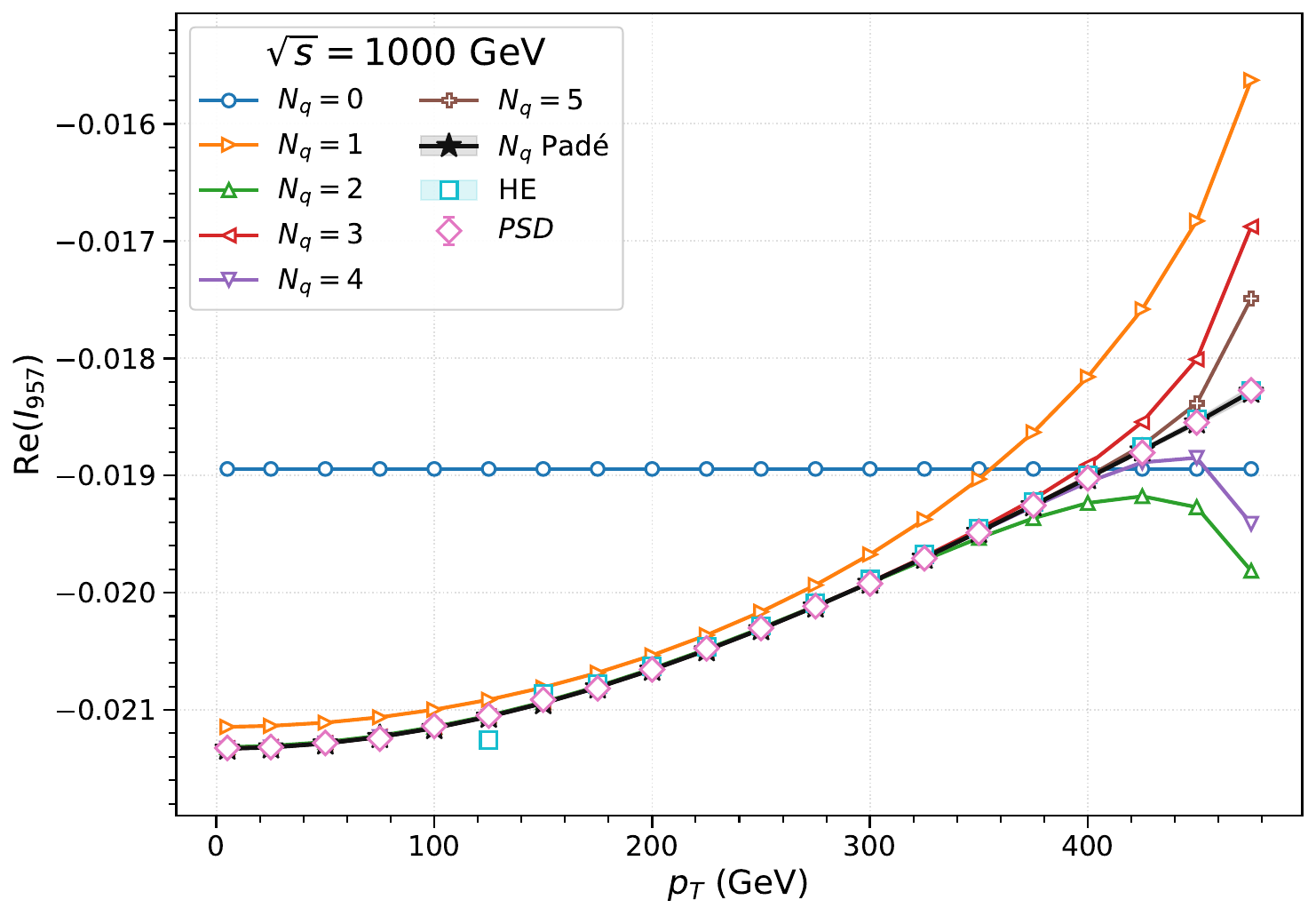} &
    \includegraphics[width=.45\textwidth]{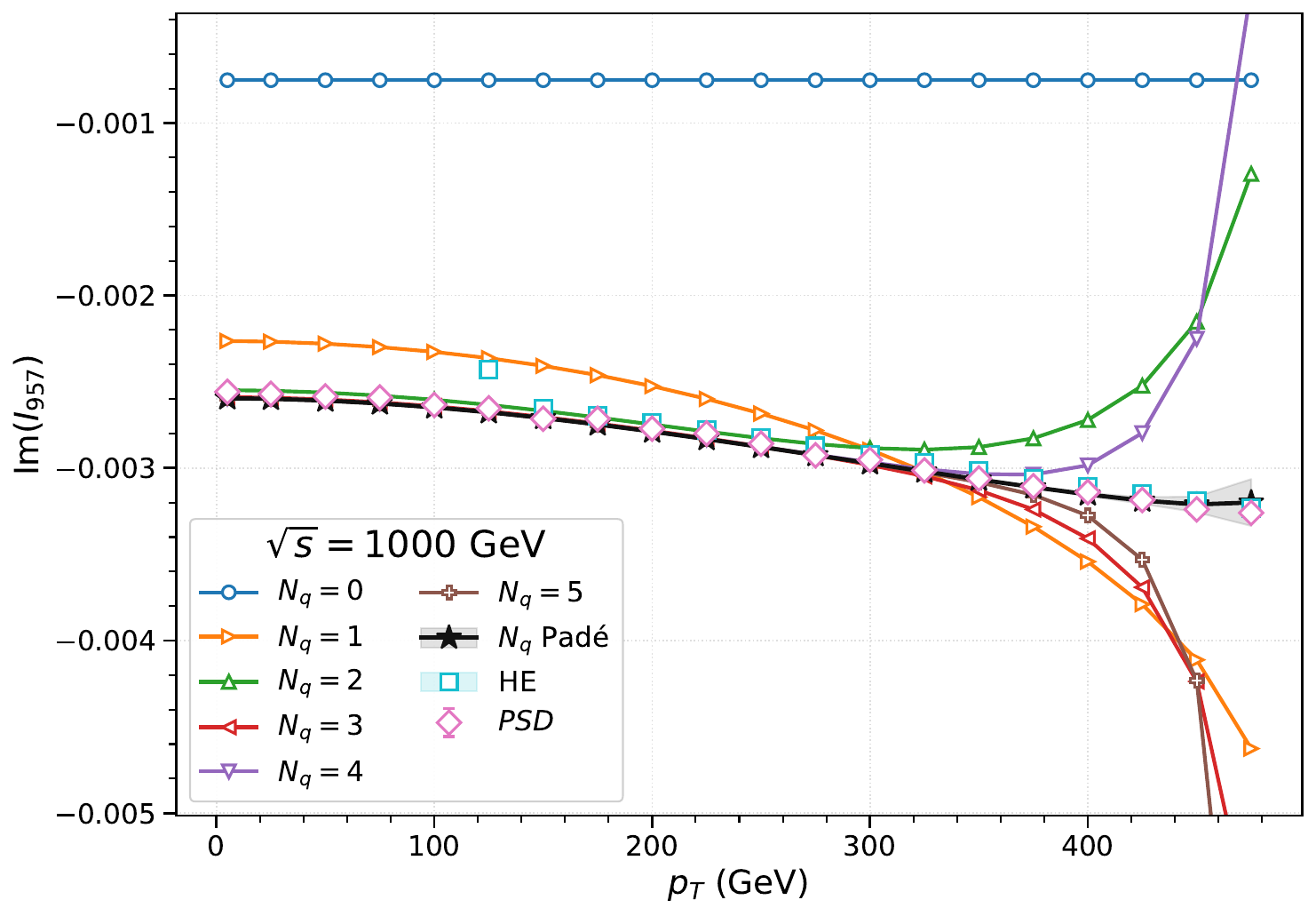} 
    \end{tabular}
    \end{center}
  \caption{
    Forward and high-energy expansions of the scalar integral $I_{957}$ shown in Fig.~\ref{fig:SI1dia}. The black dots correspond to the numeric results obtained with \texttt{pySecDec}~\cite{Heinrich:2023til}.}
    \label{fig::SI957}
\end{figure}

In the following, 
we choose $m_t=173$~GeV and $m_H=125$~GeV. 
The forward expansion is obtained following the steps outlined
in Section~\ref{sec::framework}. We
use the $\delta q$ expansion 
to obtain results up to $N_q=5$ and we take the results 
for the high-energy expansion from Ref.~\cite{Davies:2026wbx}.
Exact numerical results are
obtained with \texttt{pySecDec}~\cite{Heinrich:2023til},
requesting five significant digits.

\begin{figure}[t]
    \begin{center}
    \begin{tabular}{cc}
    \includegraphics[width=.45\textwidth]{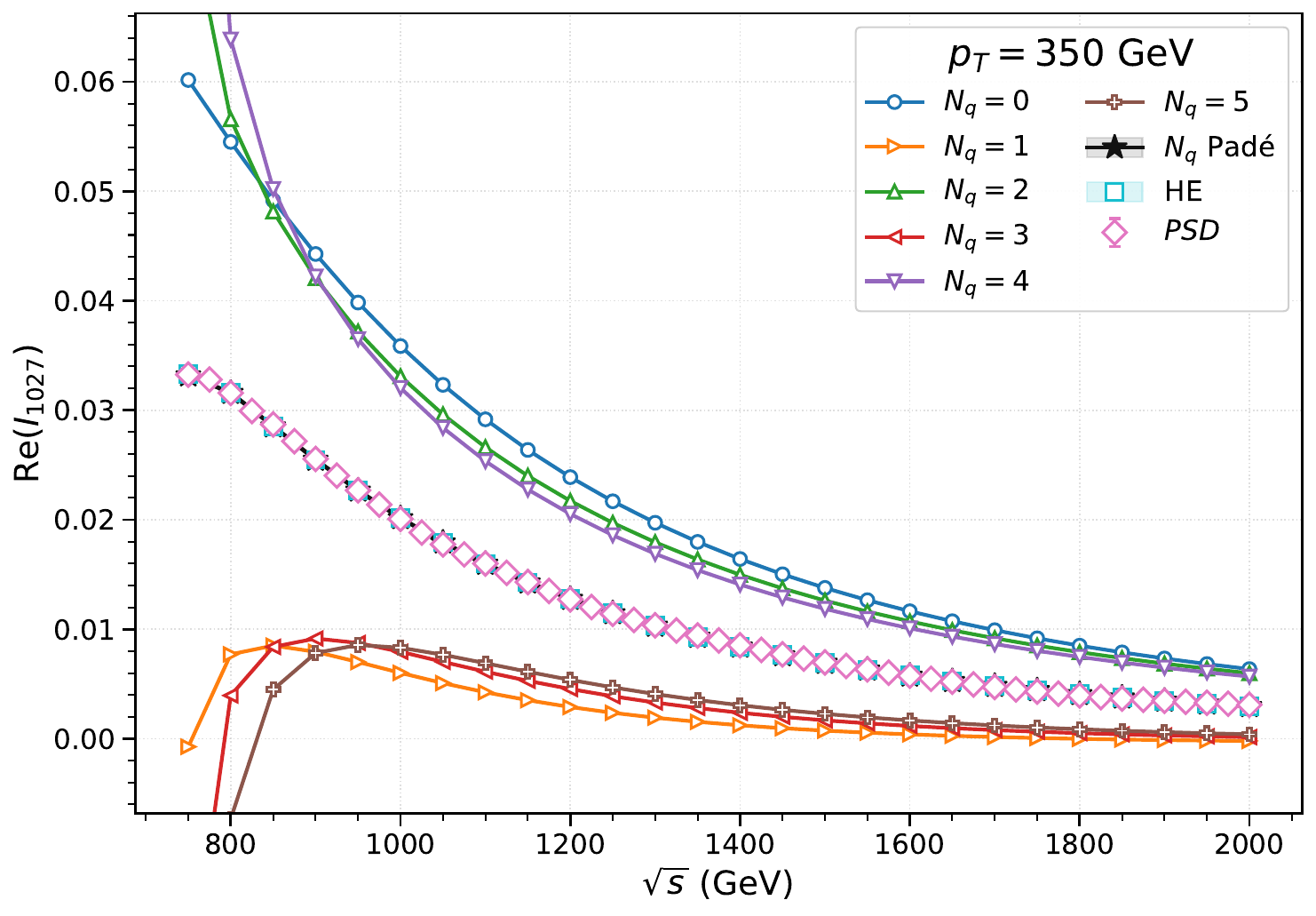} &
    \includegraphics[width=.45\textwidth]{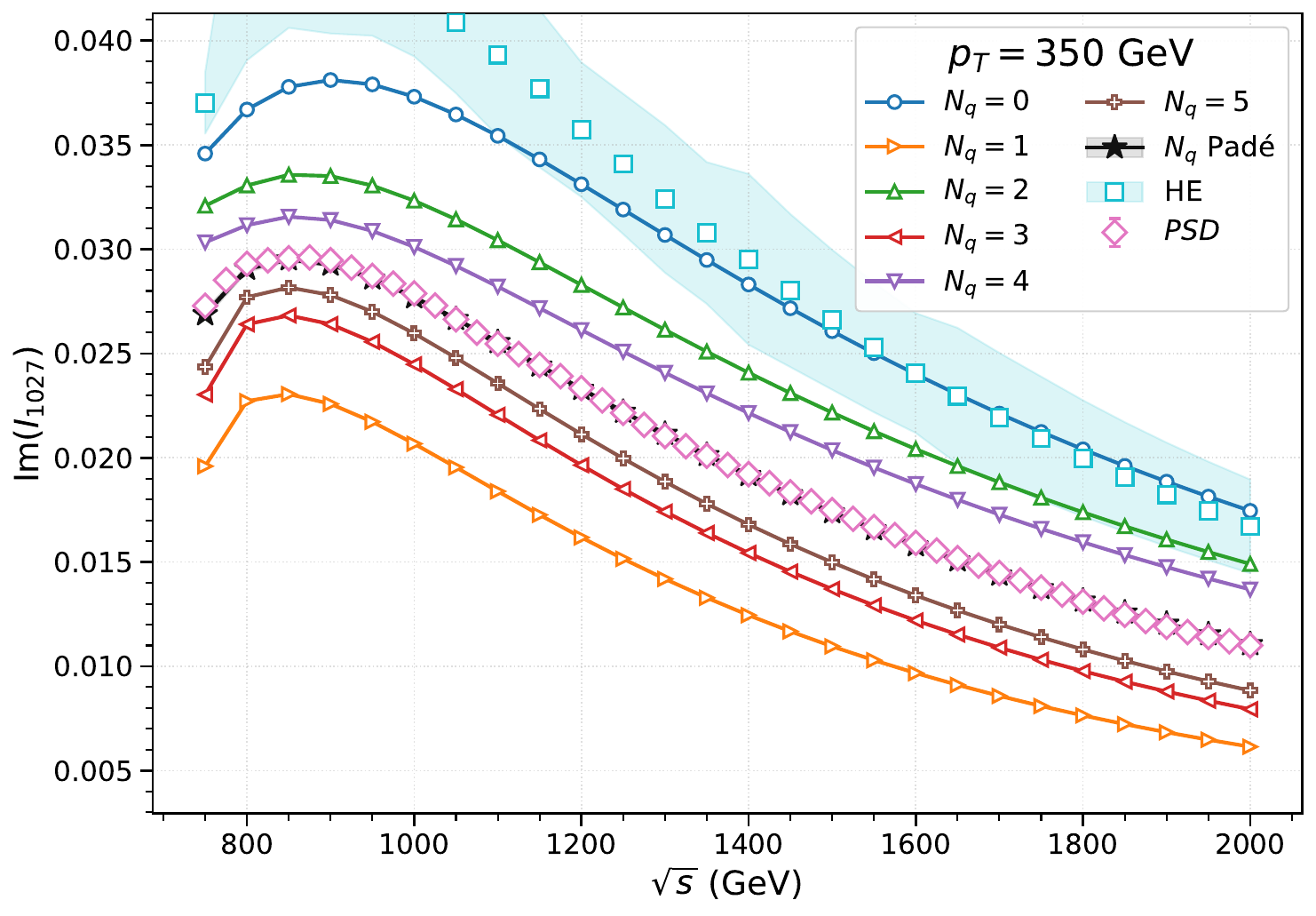} \\
    \includegraphics[width=.45\textwidth]{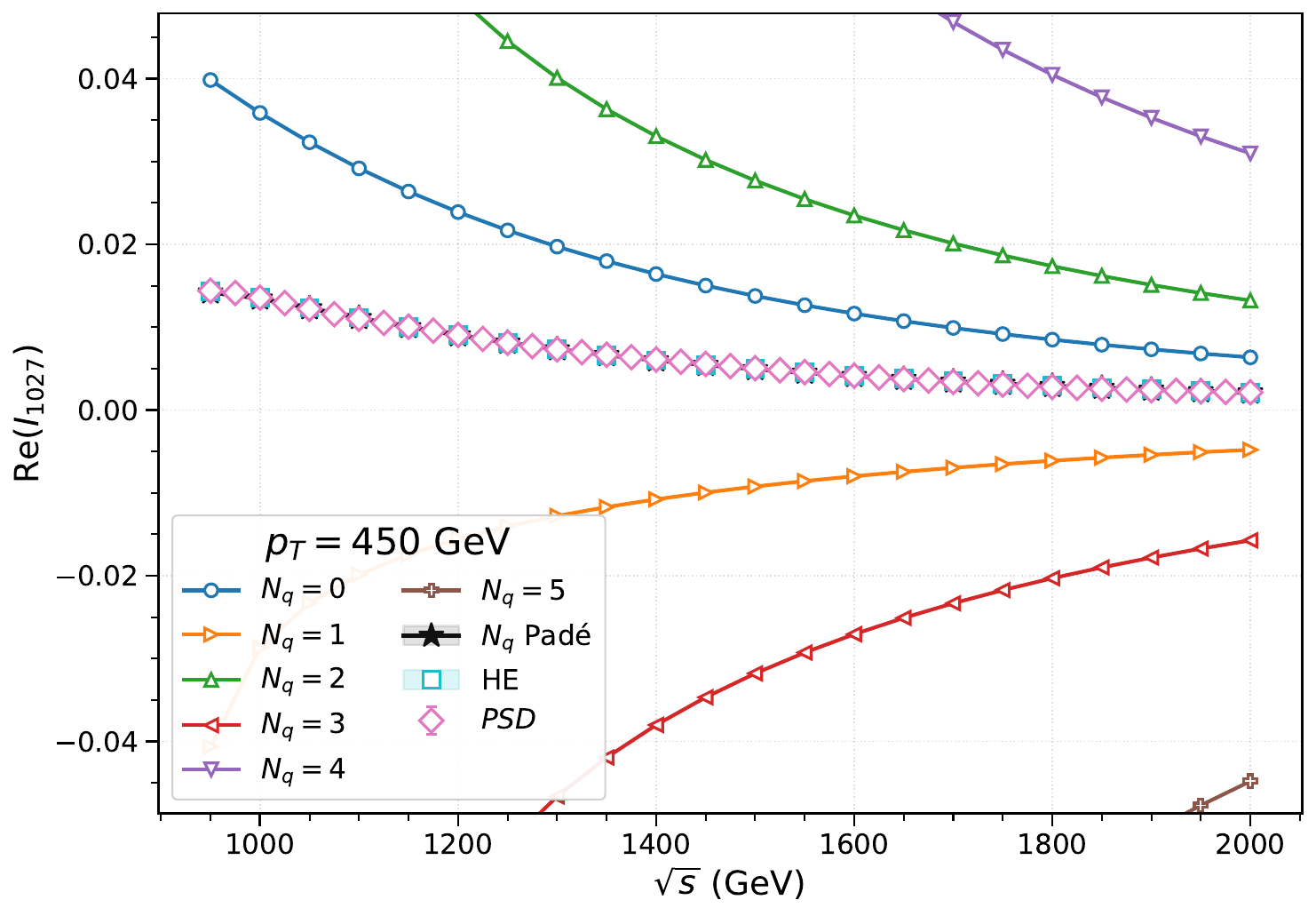} &
    \includegraphics[width=.45\textwidth]{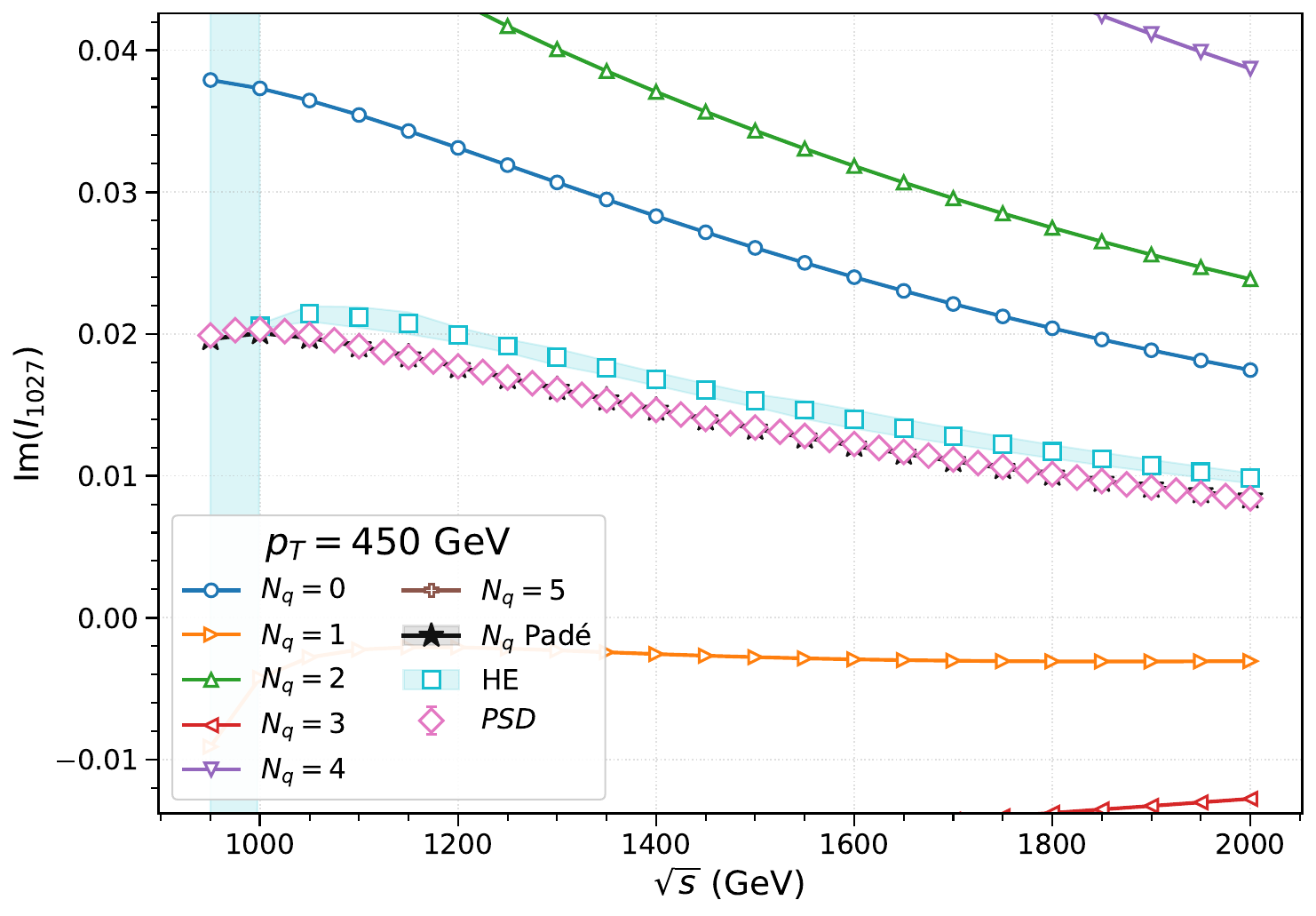} \\
    \includegraphics[width=.45\textwidth]{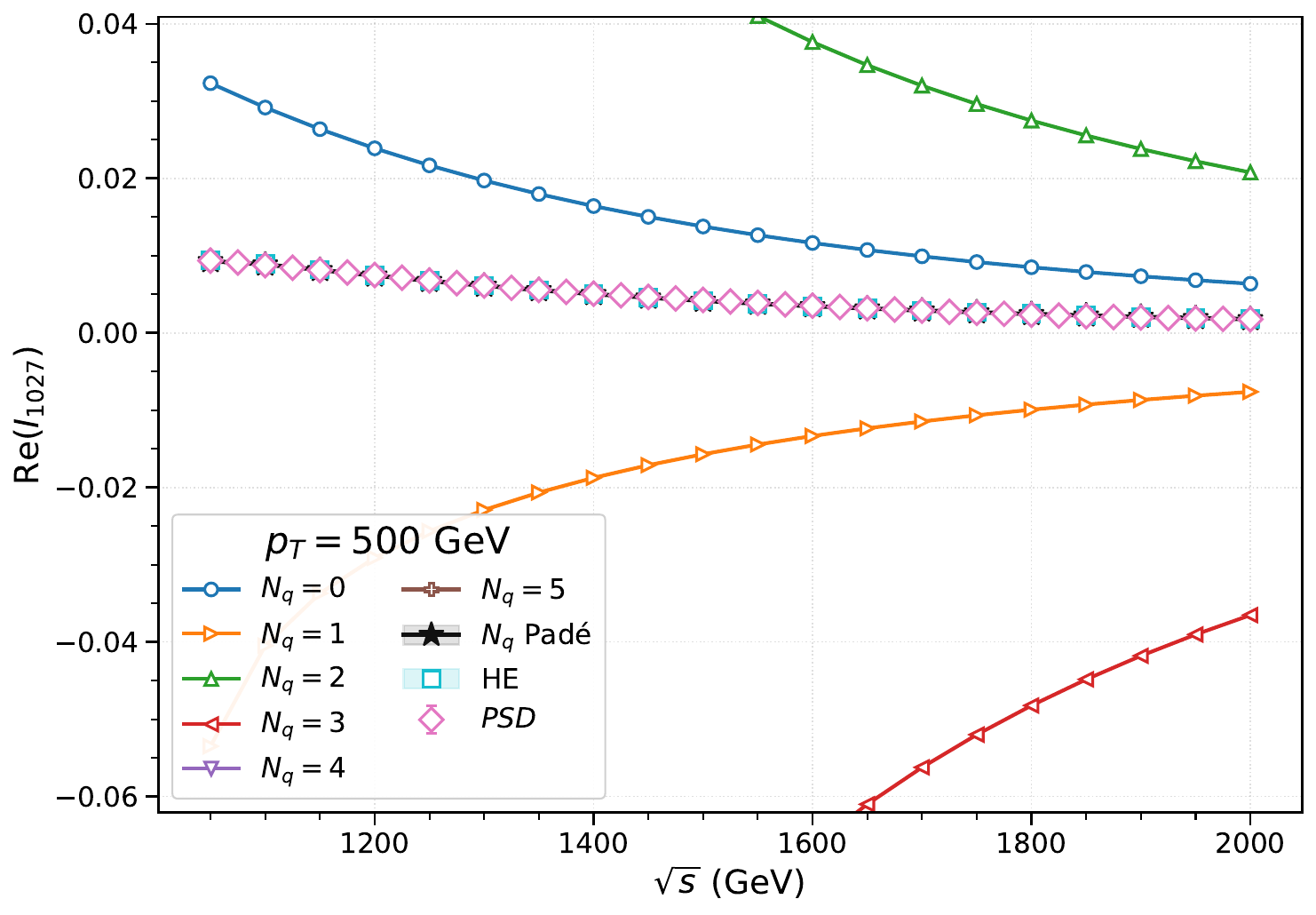} &
    \includegraphics[width=.45\textwidth]{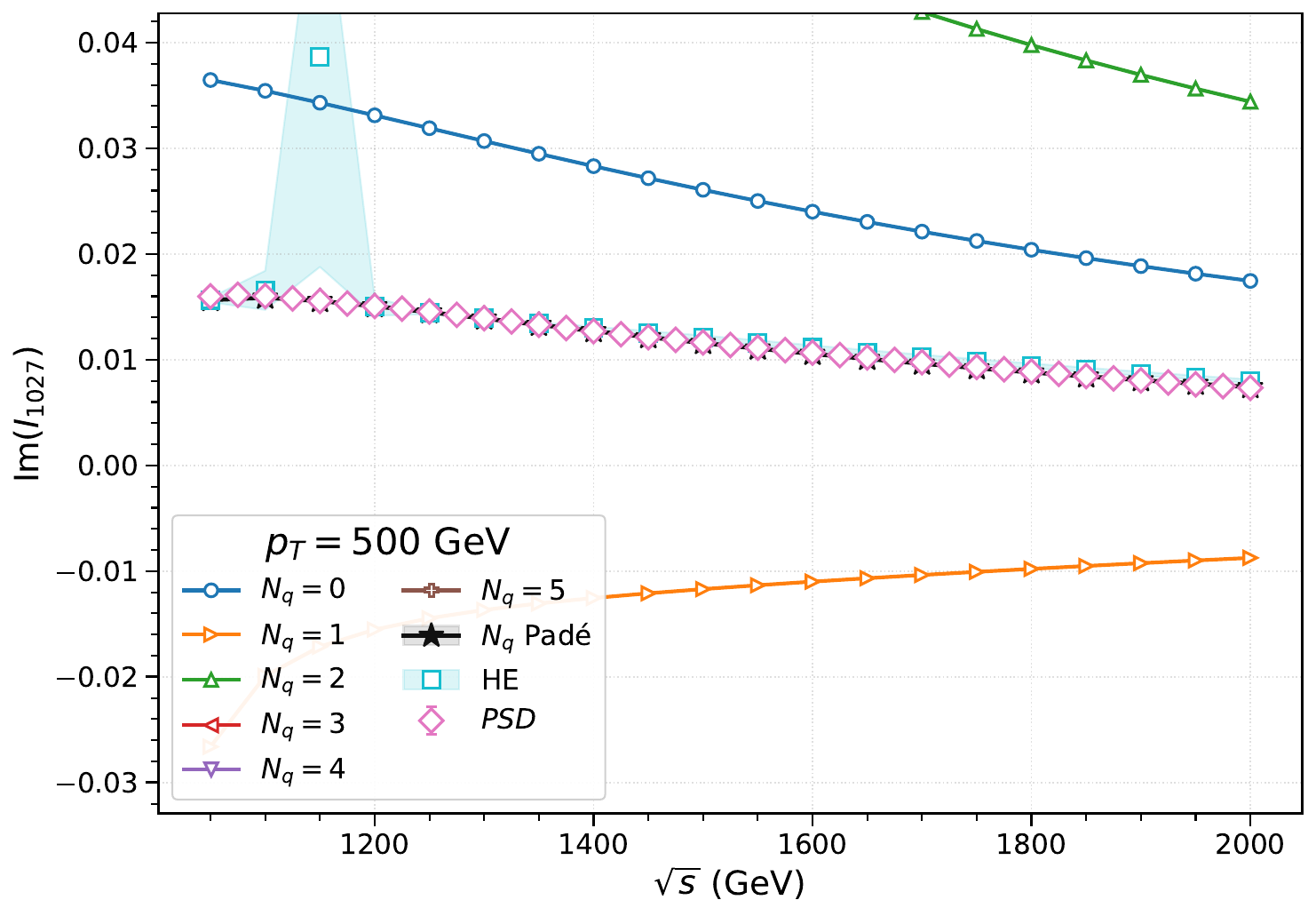} 
    \end{tabular}
    \end{center}
  \caption{
    Forward and high-energy expansions of the scalar integral $I_{1027}$ shown in Fig.~\ref{fig:SI1dia}. The black dots correspond to the numeric results obtained with the use of \texttt{pySecDec}~\cite{Heinrich:2023til}.
    Note that in those cases where the high-energy or ``$N_q$ Pad\'e'' data points are not visible, they are below the symbols used for the \texttt{pySecDec} results.
    }
    \label{fig::SI1027}
\end{figure}

In Fig.~\ref{fig::SI957}, we show results for the integral $I_{957}$.
We choose $\sqrt{s}=1000$~GeV and vary $p_T$
from zero to the highest kinematically-allowed value.
For the high-energy expansion, we show only the best
approximation, represented by light blue boxes. 
In the case of the forward expansion, we successively incorporate
higher-order terms from the $\delta q$ expansion. Furthermore, we present
the result improved via Pad\'e approximation, which we consider as our best approximation, as black dots. 

For both the real and imaginary parts, we observe that the various
terms of the forward expansion converge well. We find agreement with
the high-energy expansion as well as the exact result up to $p_T
\simeq 400$~GeV. At this point, the forward expansion is expected to
lose its validity due to the branch cut at $\sqrt{-t} = 2m_t + m_H$, which,
for $\sqrt{s} = 1000$~GeV, corresponds to a value of $p_T \simeq
405$~GeV. This is clearly visible in the coloured curves in
Fig.~\ref{fig::SI957}. In contrast, the curve improved via Pad\'e
approximation (black line) agrees with both the high-energy expansion
and the numerical results across the entire $p_T$ range. Note that for $p_T
\lesssim 150$~GeV the high-energy expansion is no longer valid
which is reflected by the large light blue uncertainty band.

In Fig.~\ref{fig::SI1027}
plots for the integral $I_{1027}$ are shown. Here we choose $p_T\in\{350,450,500\}$~GeV and vary $\sqrt{s}$
from the smallest kinematically allowed value up to 2~TeV.
With this choice, all of the displayed data points lie above the $2m_t$ $t$-channel cut of the Feynman integral $I_{1027}$ and are thus outside of the radius of convergence of the forward limit expansion. 
In fact, we observe that the expansion terms containing $t^{N_q}$
components exhibit poor convergence behaviour, or fail to converge
altogether.  In contrast, the result improved via Pad\'e approximation
shows good agreement with the numerical result.
The best approximation agrees with the exact results to within better
than 5\% for all $\sqrt{s}$ and $p_T$ values shown in Fig.~\ref{fig::SI1027}.

For the real part of the integral the high-energy expansion exhibits good convergence properties for all
chosen $p_T$ values and agrees well with the exact result to within
$1\%$. The situation is different for the imaginary part: for
$p_T=350$~GeV and $p_T=400$~GeV, the uncertainty introduced by the
Pad\'e approximation is large, and the result lies systematically above
the exact value. For $p_T=450$~GeV, the agreement is around 20\%,
significantly worse than that obtained from the
forward expansion. For $p_T=500$~GeV, the imaginary part of the
high-energy expansion shows good agreement between 3 and 10\%.

We conclude that, by using Pad\'e approximants, it is possible to
improve the expansion in the forward limit such that it agrees with
the exact results well beyond its original radius of convergence. Even
for integrals where the high-energy expansion does not yield good
results, there exists a $p_T$ interval in which both approximations
agree with each other as well as with the exact result. Outside this
range, the respective expansions perform well, and we are confident
that by combining the expansion in the forward limit with the
high-energy expansion, we can cover the entire phase space. In the
next section, we will compare the results for form factors for which
no exact results are available. The agreement between the two
expansions at the level of the form factors serves as evidence that a
good approximation of the exact amplitude has been obtained.


\section{\label{sec::results}Results for form factors}

In this section, we present numerical results for the form
factors. First, we discuss the properties of the expansion in the
region of phase space where the forward expansion is expected to
converge. This is shown in Section~\ref{sub::t_exp} using the 
$\delta q$ expansion. Subsequently, we show in Section~\ref{sub::combine} that, the combination of the $\delta q$, the naive-$t$ and the
high-energy expansions~\cite{Davies:2026wbx} covers the entire phase space.

We will show the finite,
renormalized form factors but exclude contributions involving virtual
photons; the latter represent corrections to the Abelian part of the
QCD corrections (associated with the colour factor $C_F$) that were
already addressed using expansions in Refs.~\cite{Davies:2018ood,Davies:2023vmj}. For the remaining contributions
(comprising Higgs, $W$, and $Z$ bosons as well as the corresponding
Goldstone bosons), finite results are easily obtained, since the
counterterm for the photon contribution receives contributions solely
from the renormalization of the top-quark mass. For the sake of
completeness, we list the results for the photon contribution in Appendix~\ref{app::photon}.

For the numerical evaluation we use the following input values~\cite{PhysRevD.110.030001}
\begin{eqnarray}
\label{eq::numvals}
        G_F &=& 1.166378\times
        10^{-5}\,\mbox{GeV}{{}^{-2}}\,,\nonumber\\
        m_H &=& 125.2\,\mbox{GeV}\,,\nonumber\\
        m_Z&=& 91.19\,\mbox{GeV}\,, \nonumber\\
        m_W &=& 80.37\,\mbox{GeV}\,, \nonumber\\
        m_t^{\rm OS} &=& 172.57\,\mbox{GeV}\,,\nonumber\\
        \overline{m}_t(\overline{m}_t) &=& 162.79\,\mbox{GeV}\,,
\end{eqnarray}
and introduce the coupling
\begin{eqnarray}
   \alpha_{G_\mu} &=& \frac{\sqrt{2}}{\pi} G_F m_W^2
   \left(1-\frac{m_W^2}{m_Z^2}\right)
   \,,
   \label{eq::alphaGmu}
\end{eqnarray}
which is our expansion parameter in the $G_\mu$ scheme.

\subsection{\label{sub::t_exp}Convergence properties of forward expansion}

\begin{figure}[t]
    \begin{center}
    \begin{tabular}{cc}
    \includegraphics[width=.45\textwidth]{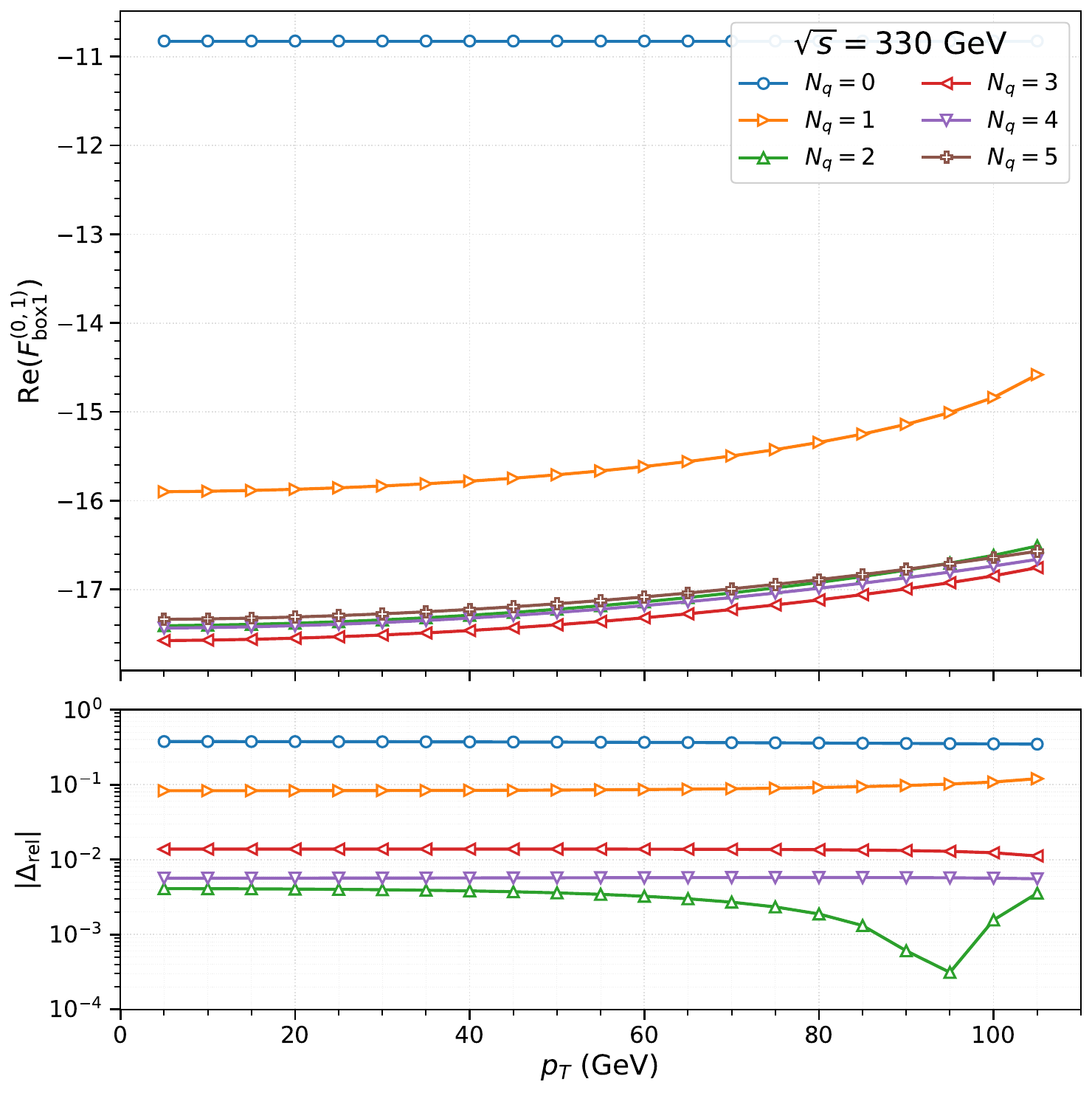} &
    \includegraphics[width=.45\textwidth]{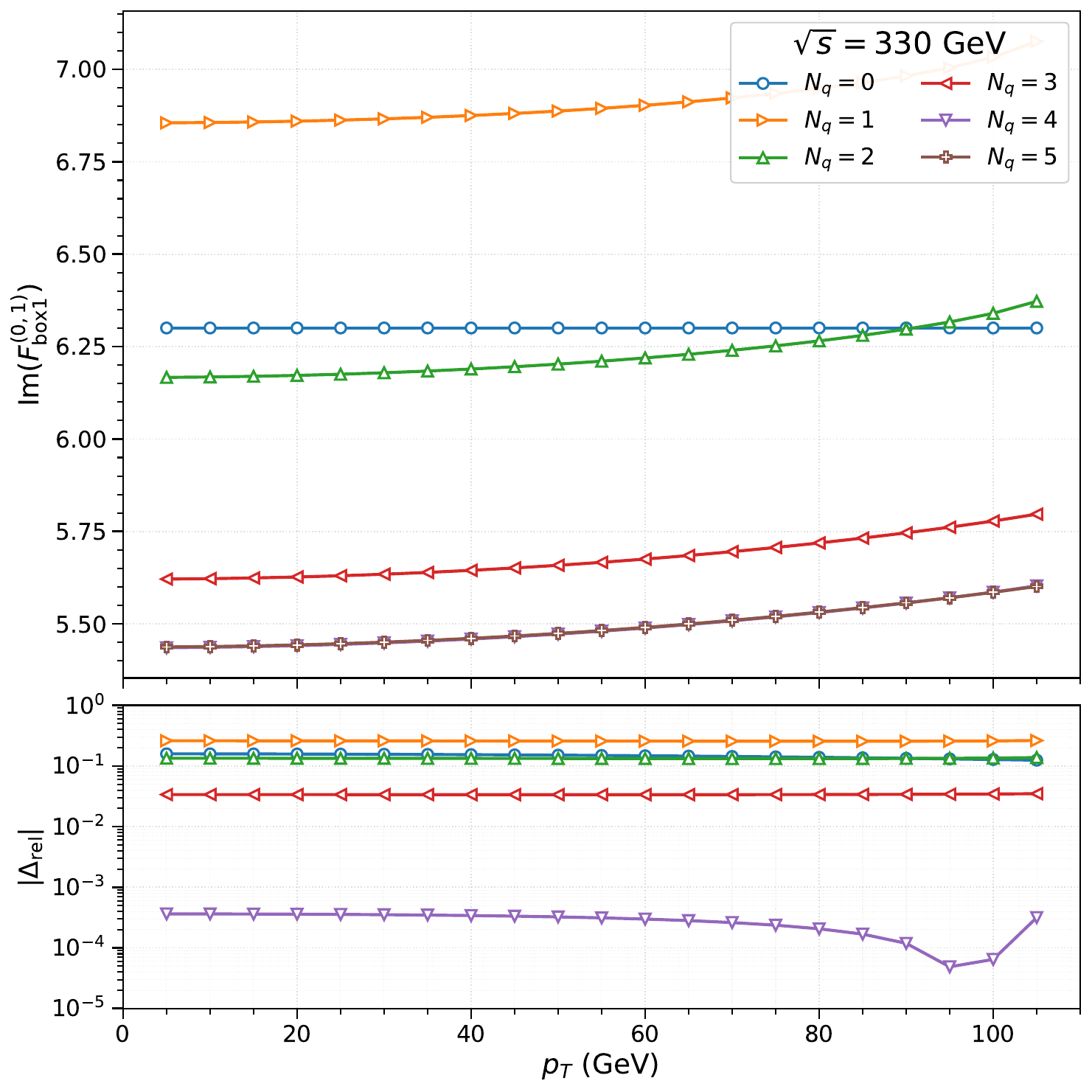} \\
    \includegraphics[width=.45\textwidth]{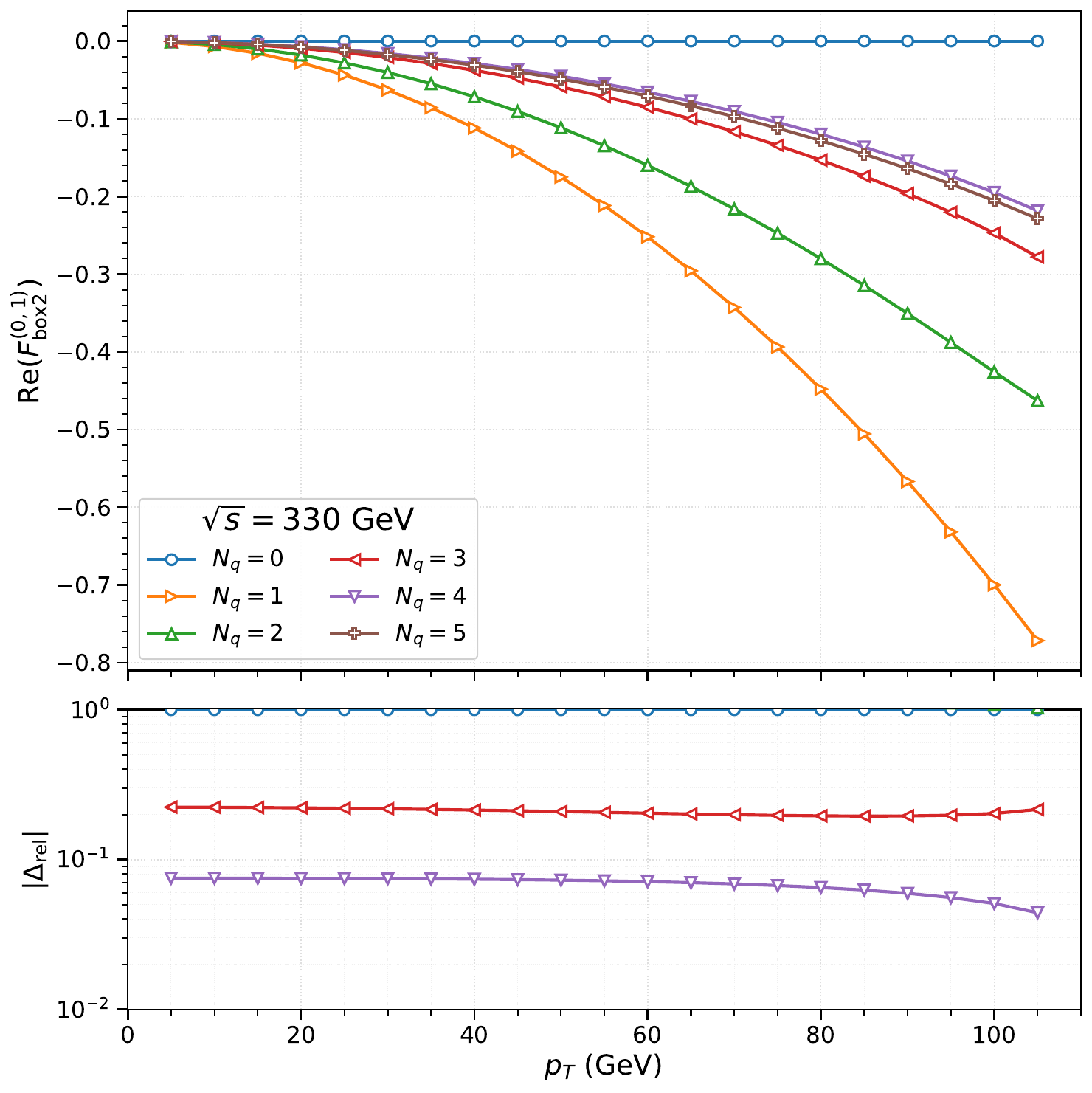} &
    \includegraphics[width=.45\textwidth]{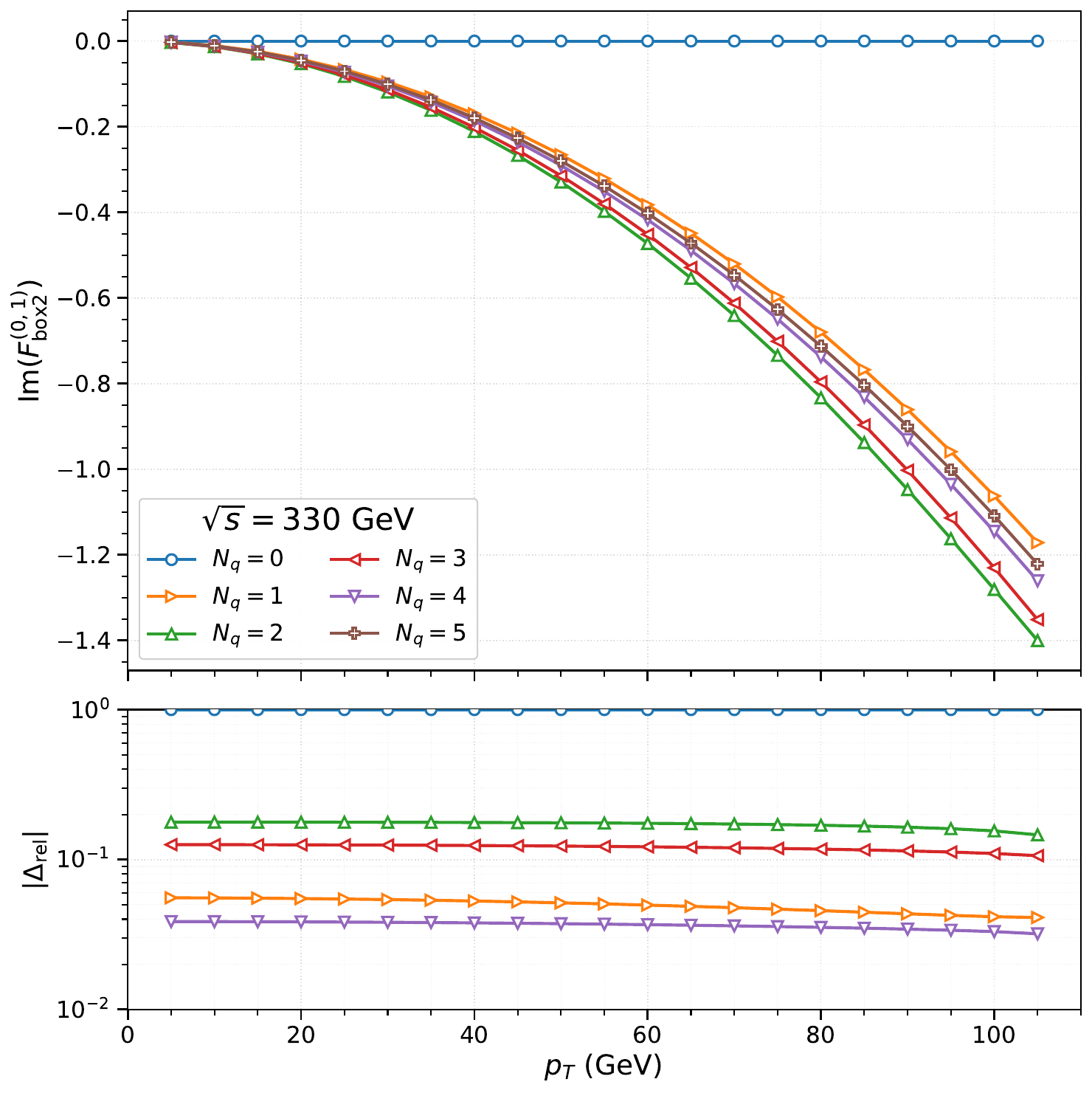} 
    \end{tabular}
    \end{center}
  \caption{\label{fig::f1-t-region-ss330}
  Real and imaginary parts of $F_{\rm box1}$ and $F_{\rm box2}$ for $\sqrt{s}=330$~GeV as a
function of $p_T$. The upper panels show the approximations
incorporating successively higher orders given by $N_q$. The
$y$-axis scaling was chosen to make the curves distinguishable;
consequently, the leading-order approximation (blue curve) is not
visible in all panels. The lower panels show the relative difference
with respect to the $N_q=5$ approximation.
  }
\end{figure}

\begin{figure}[t]
    \begin{center}
    \begin{tabular}{cc}
    \includegraphics[width=.45\textwidth]{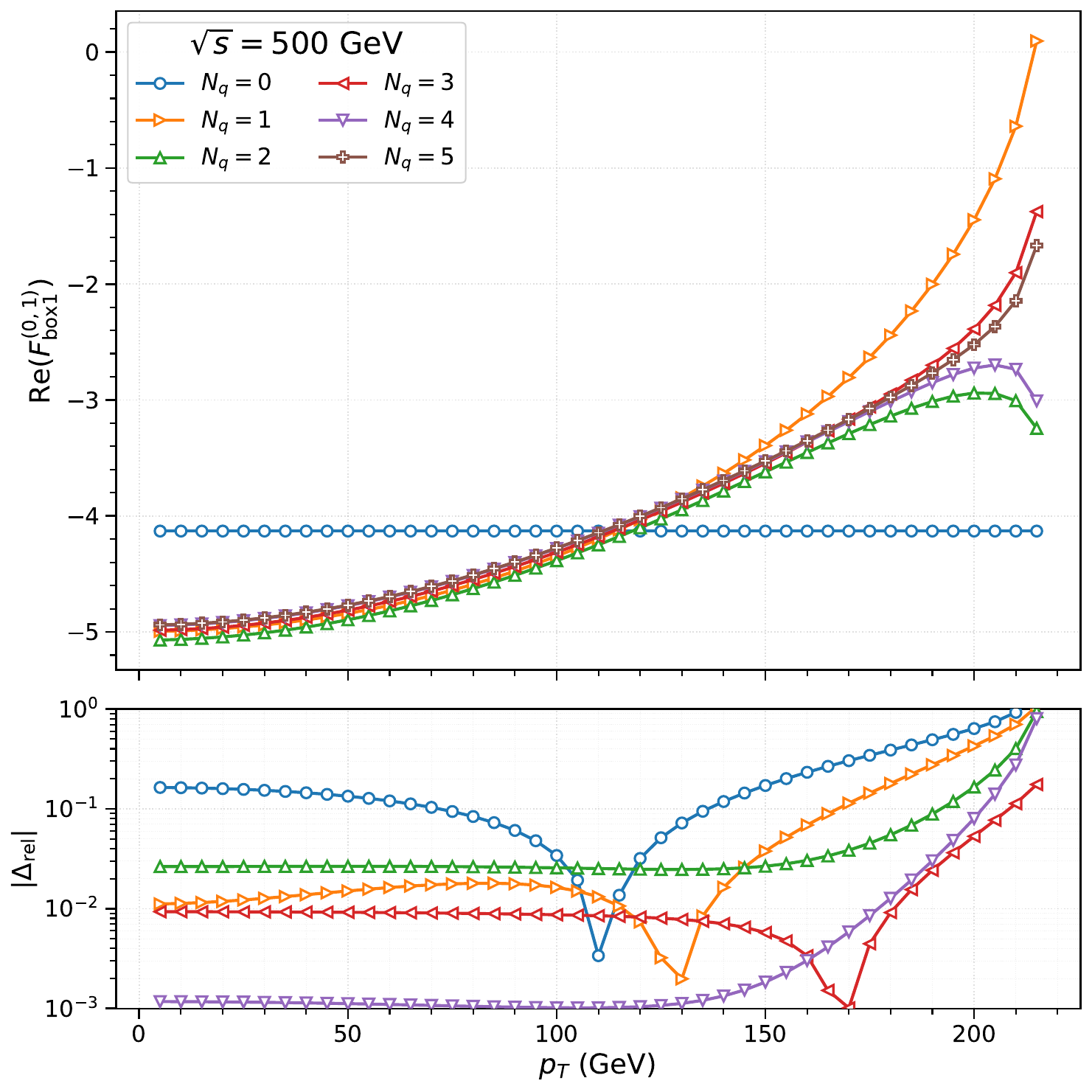} &
    \includegraphics[width=.45\textwidth]{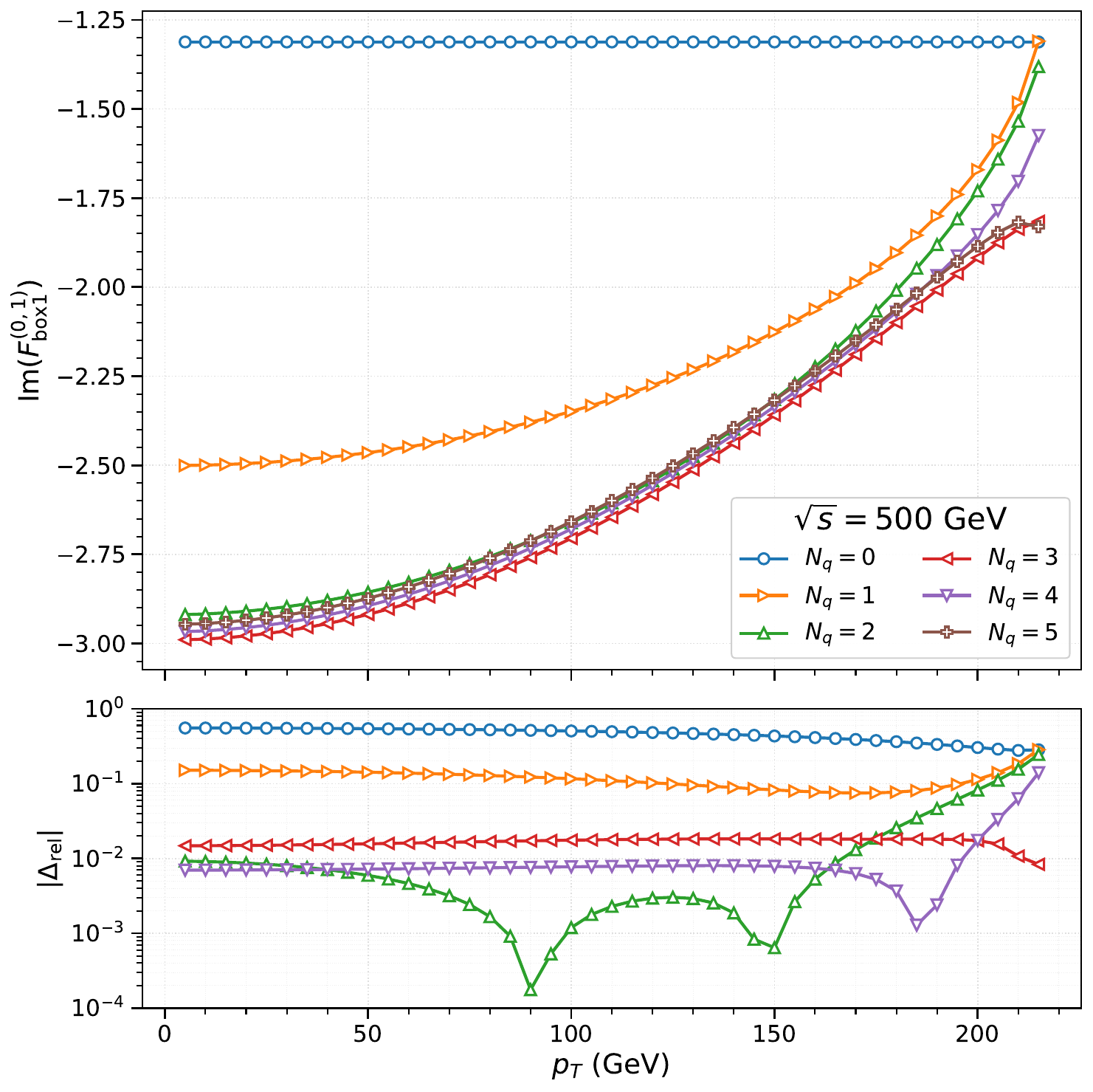} \\
    \includegraphics[width=.45\textwidth]{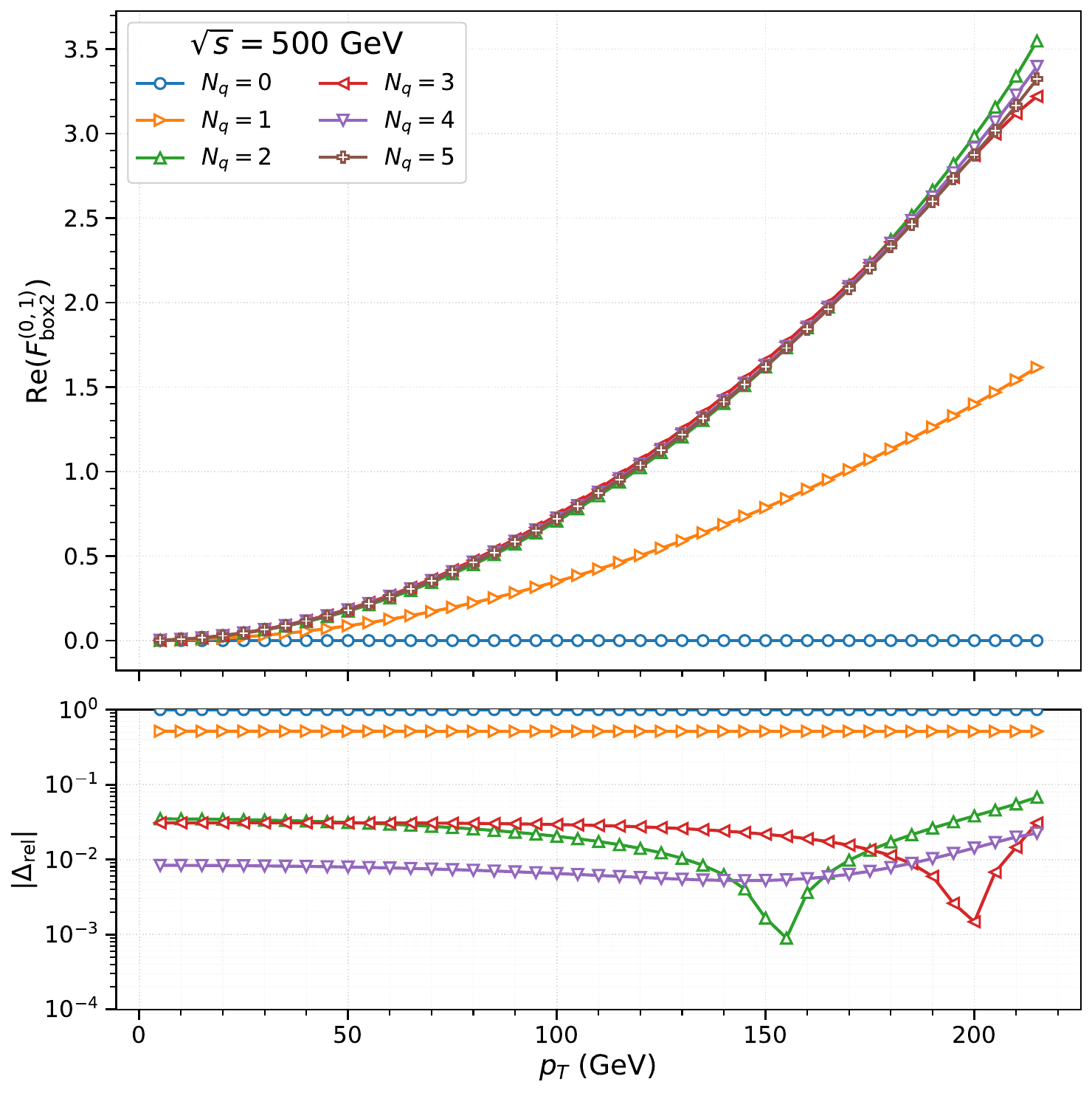} &
    \includegraphics[width=.45\textwidth]{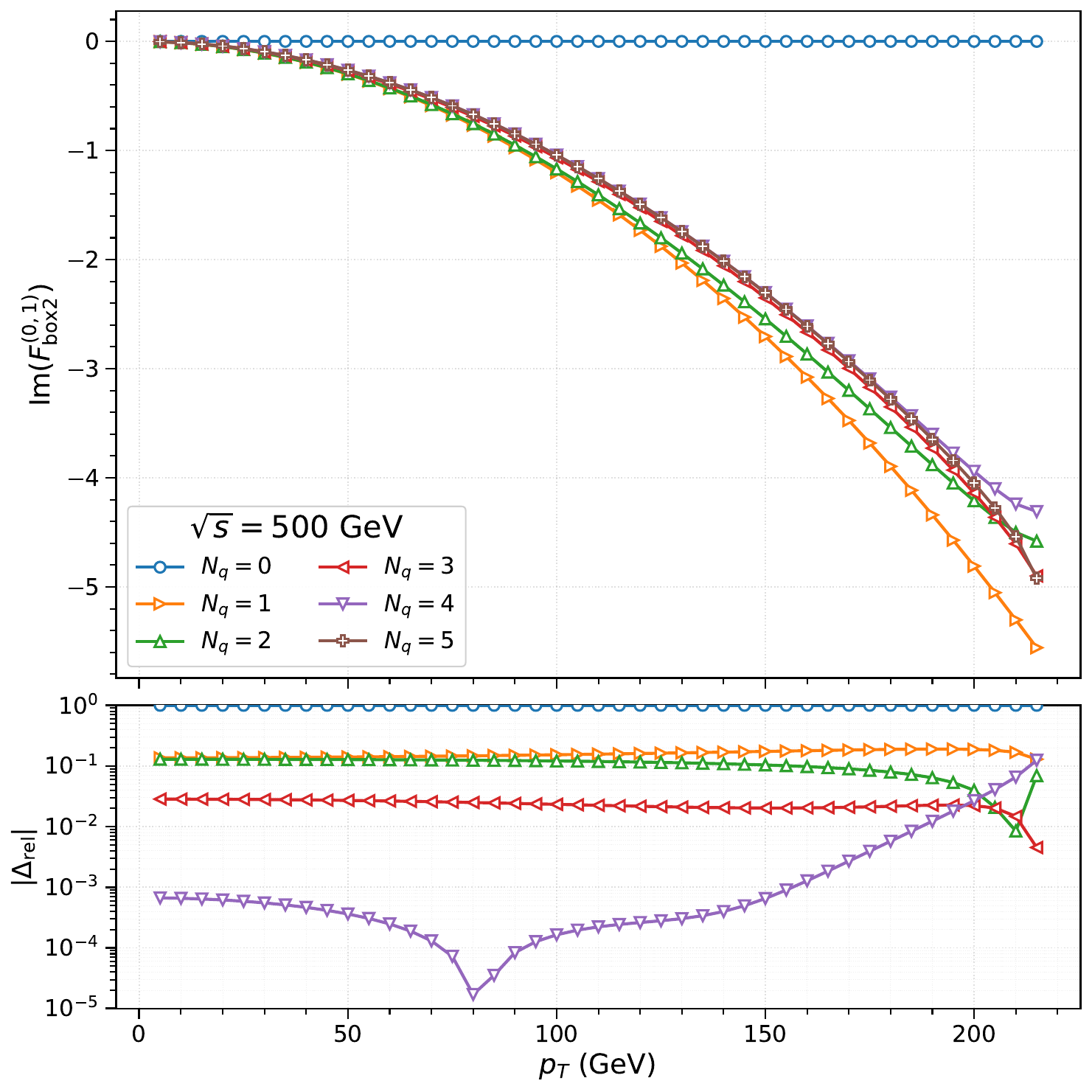} 
    \end{tabular}
    \end{center}
  \caption{\label{fig::f1-t-region-ss500}
  Real and imaginary parts of $F_{\rm box1}$ and $F_{\rm box2}$ for $\sqrt{s}=500$~GeV as a
function of $p_T$. The upper panels show the approximations
incorporating successively higher orders given by $N_q$.
The lower panels show the relative difference
with respect to the $N_q=5$ approximation.
  }
\end{figure}

\begin{figure}[t]
    \begin{center}
    \begin{tabular}{cc}
    \includegraphics[width=.45\textwidth]{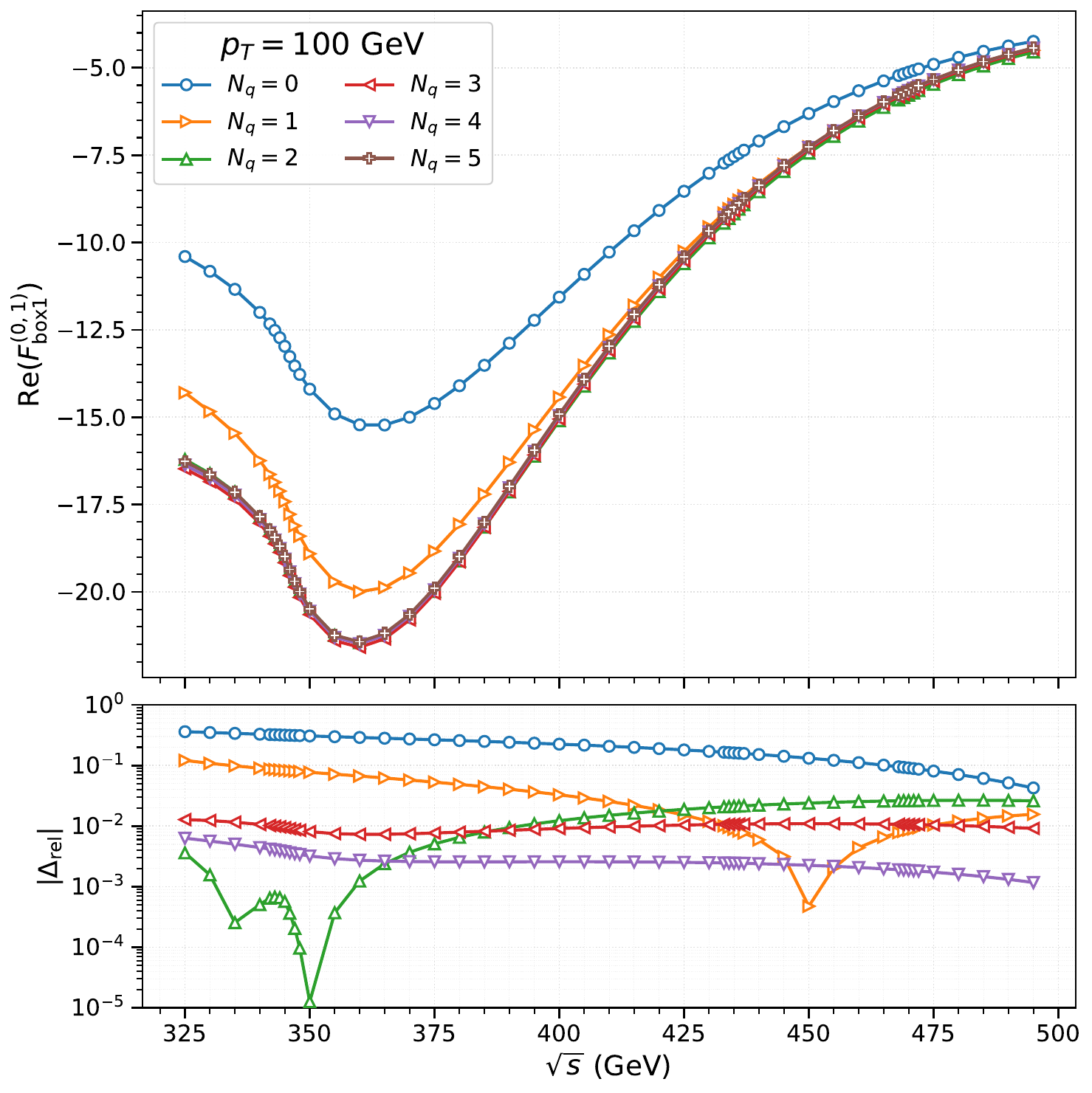} &
    \includegraphics[width=.45\textwidth]{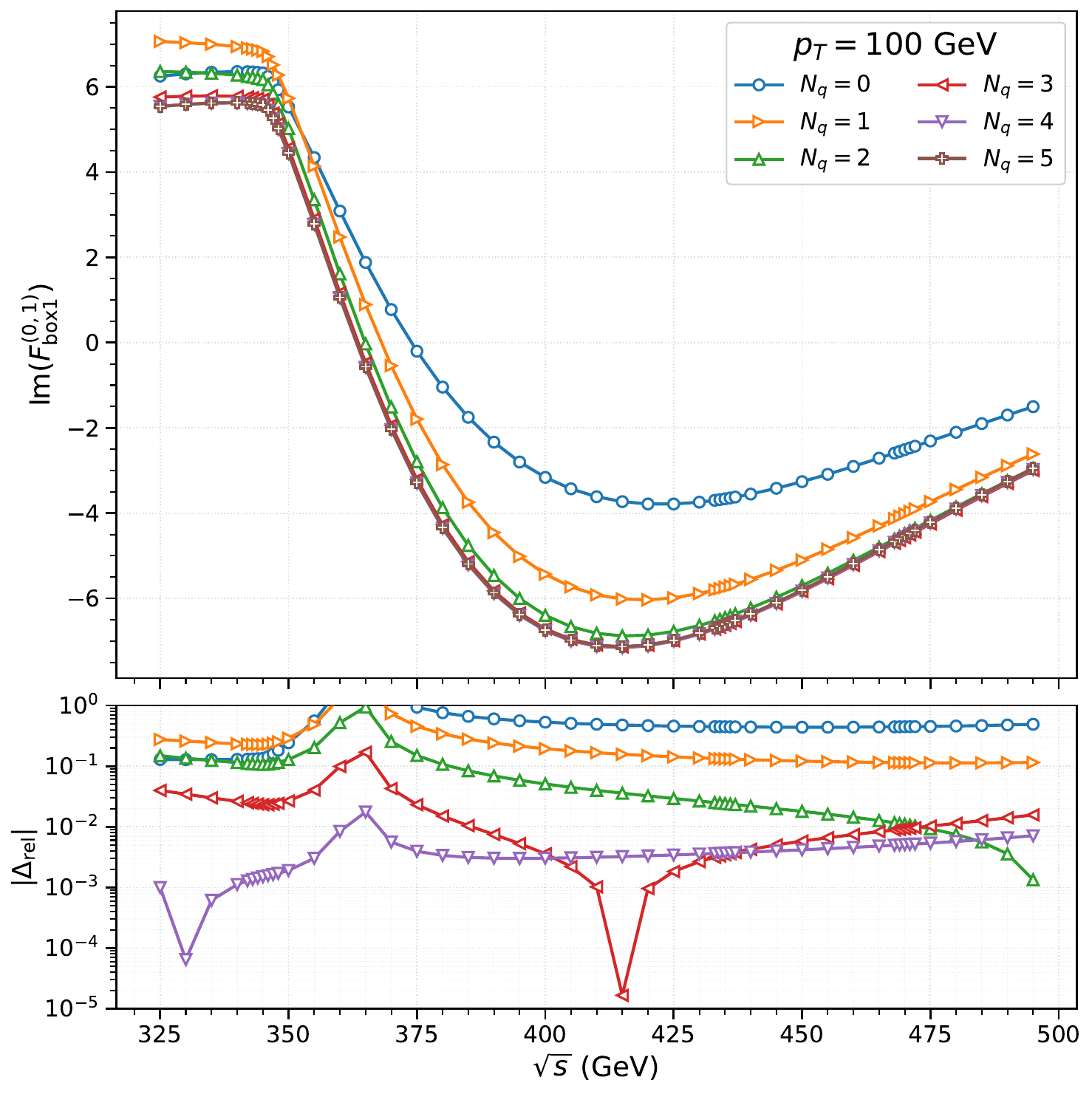} \\
    \includegraphics[width=.45\textwidth]{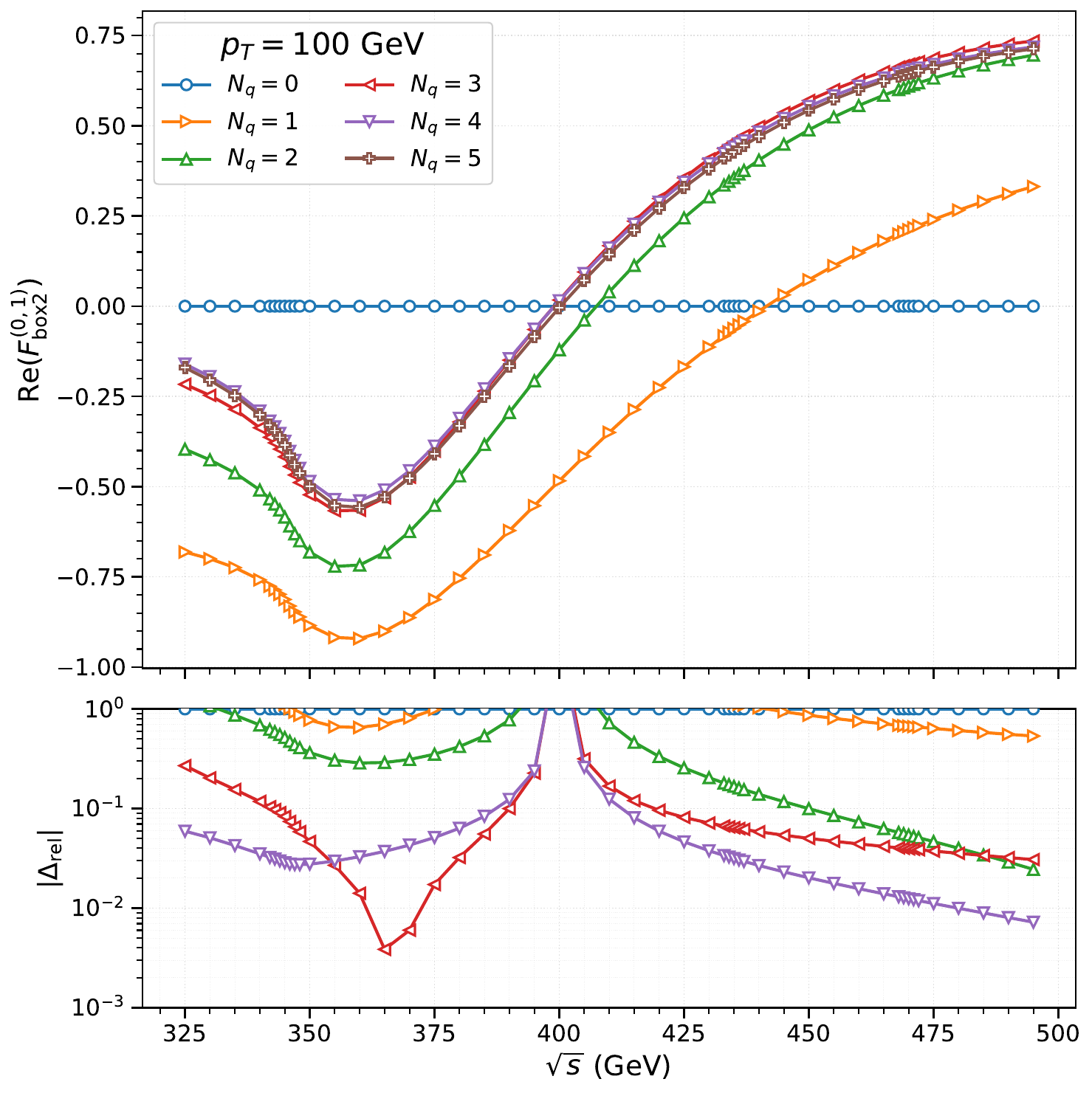} &
    \includegraphics[width=.45\textwidth]{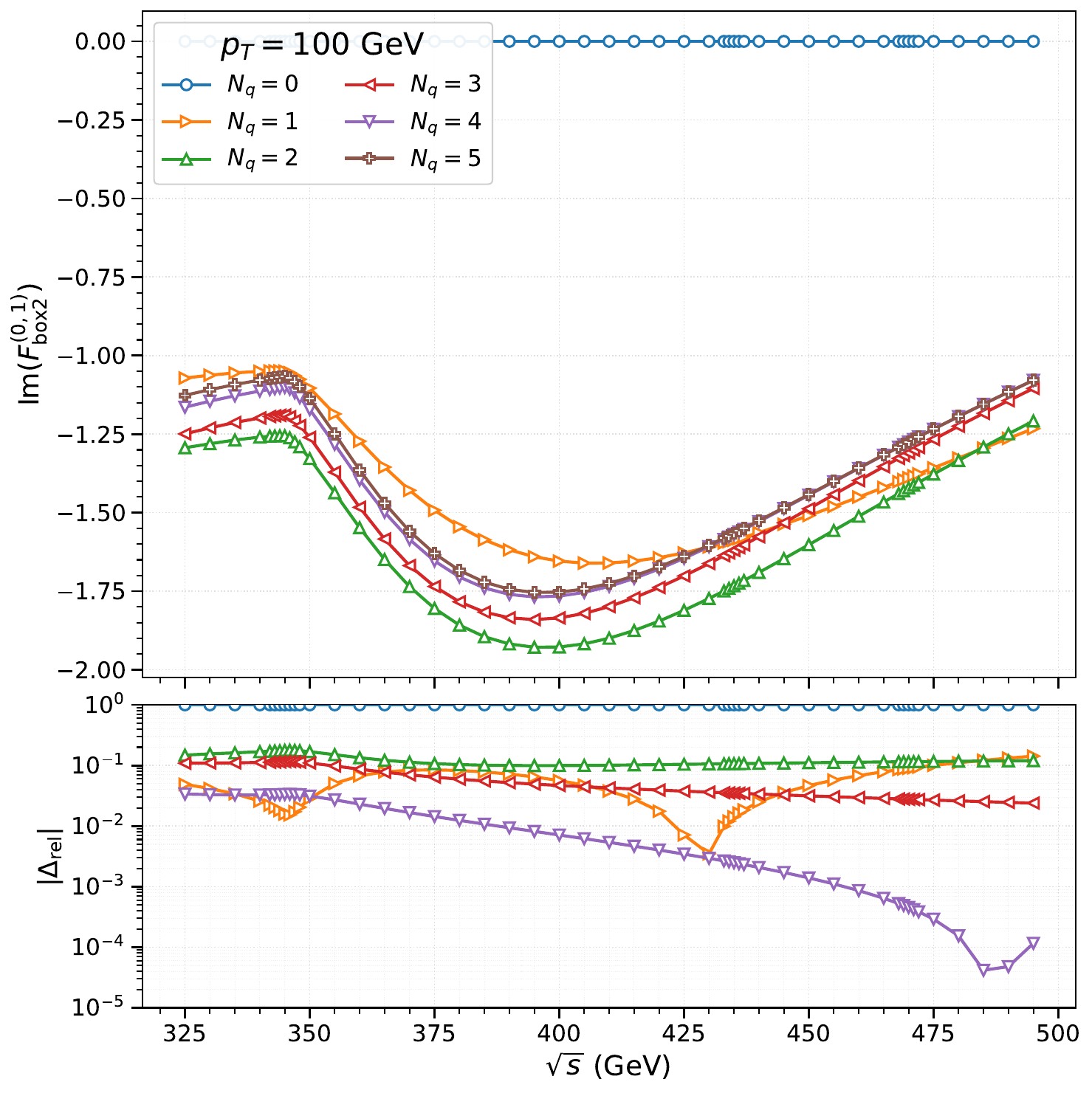} 
    \end{tabular}
    \end{center}
  \caption{\label{fig::f1-t-region-pt100}
  Real and imaginary parts of $F_{\rm box1}$ and $F_{\rm box2}$ for $p_T=100$~GeV as a
function of $\sqrt{s}$. The upper panels show the approximations
incorporating successively higher orders in $t$. The lower panels show the relative difference
with respect to the $N_q=5$ approximation.
  }
\end{figure}

\begin{figure}[t]
    \begin{center}
    \begin{tabular}{cc}
    \includegraphics[width=.45\textwidth]{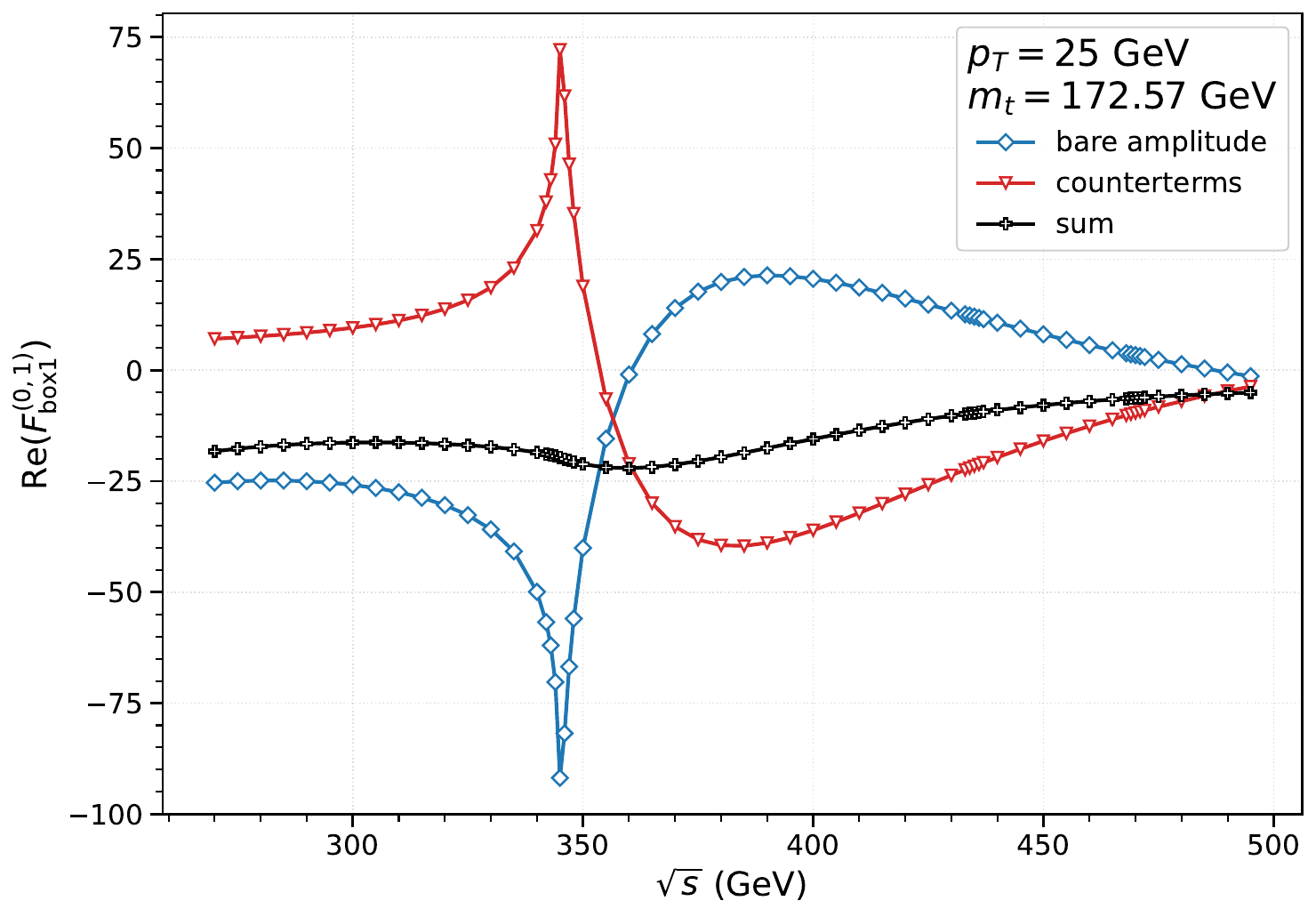} &
    \includegraphics[width=.45\textwidth]{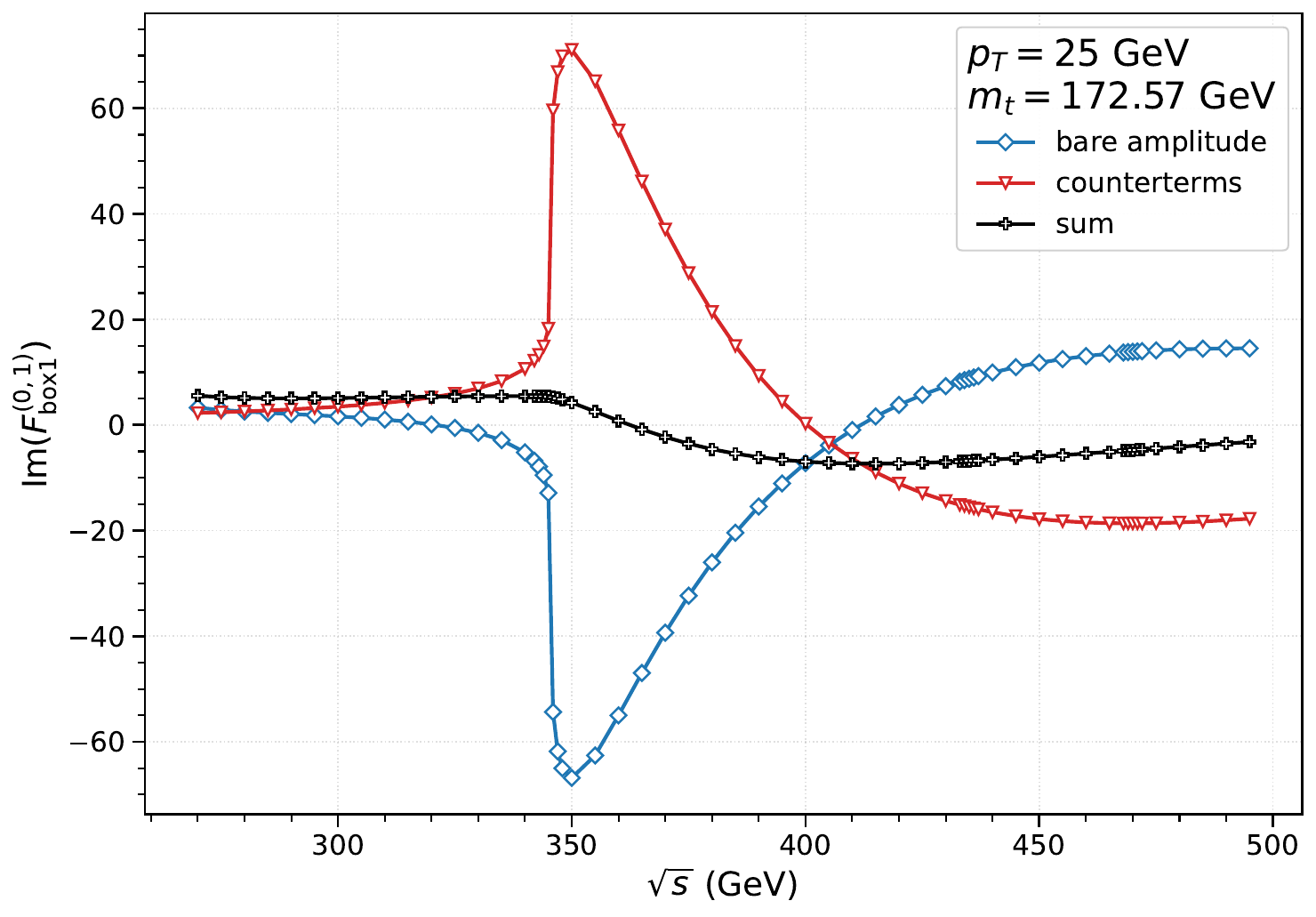} \\
    \includegraphics[width=.45\textwidth]{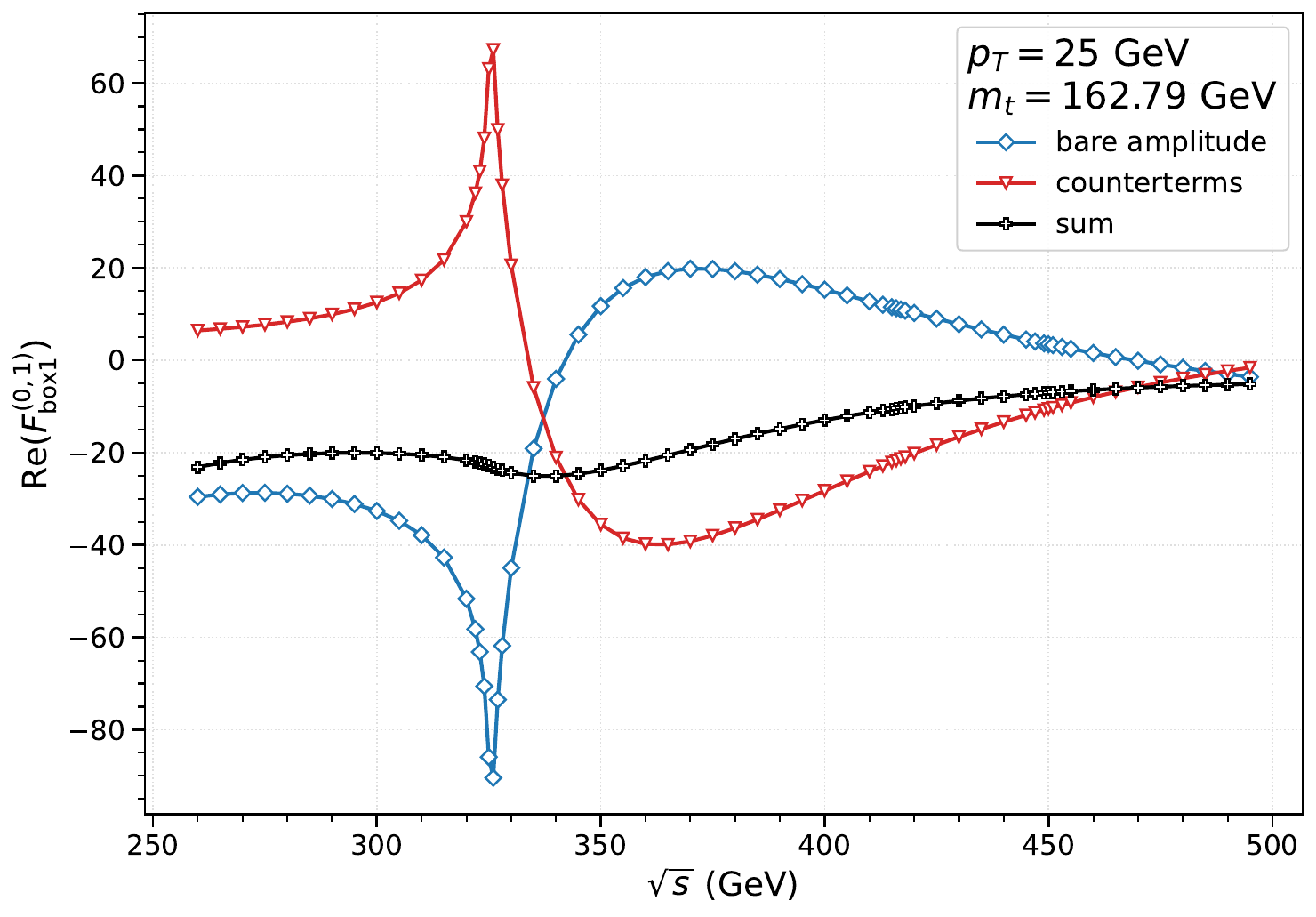} &
    \includegraphics[width=.45\textwidth]{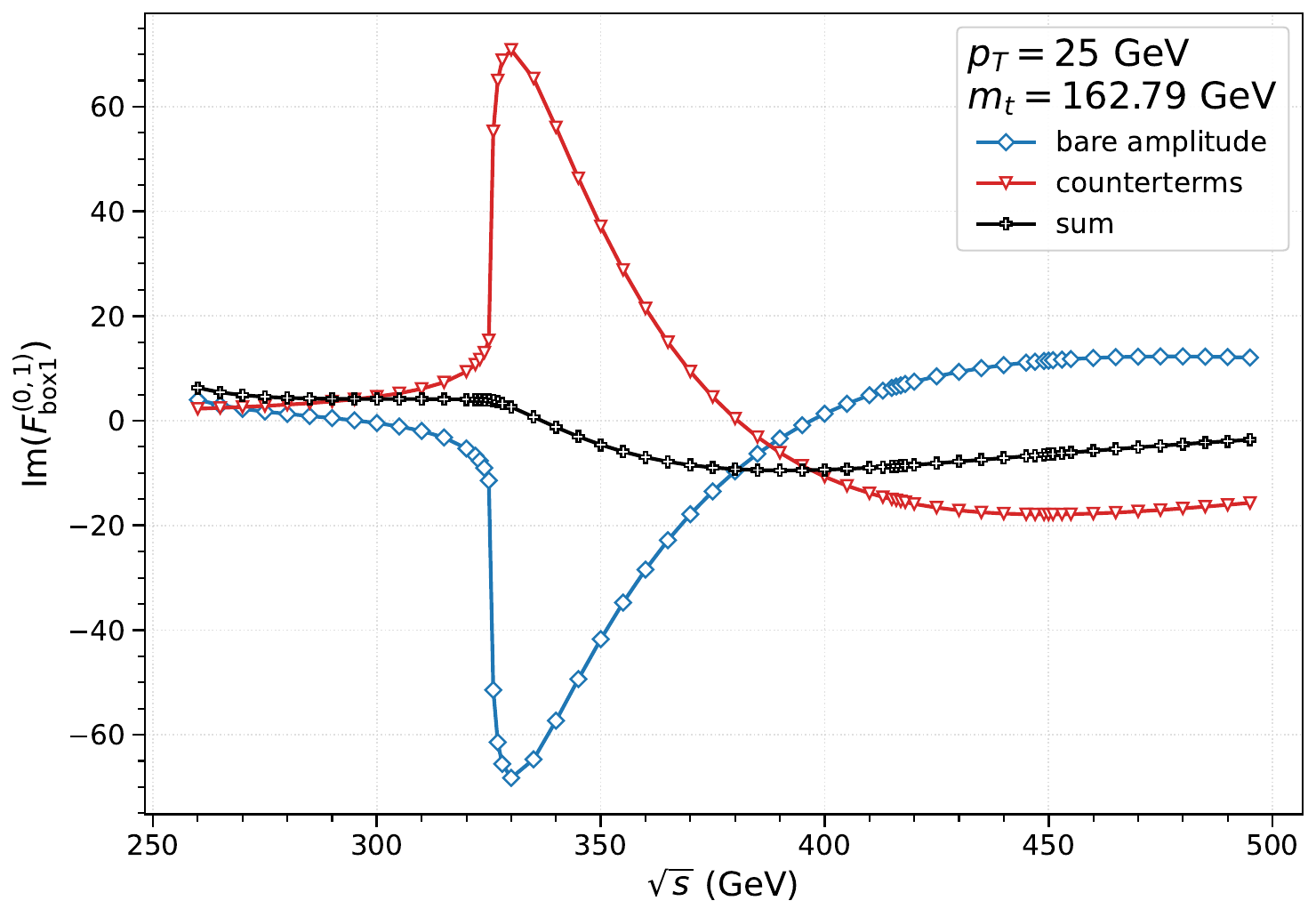}
    \end{tabular} 
    \end{center}
  \caption{\label{fig::f1-t-region-pt25_cancel}
  Real and imaginary parts of $F_{\rm box1}$ for $p_T=25$~GeV as a function of 
  $\sqrt{s}$. We show the bare two loop amplitude, the counterterm 
  contribution and their sum as obtained using the $\delta q$ expansion. In the upper panels the top quark mass is set to $m_t=172.57$~GeV and in the lower panels to $m_t=162.79$~GeV. In both cases, large cancellations can be observed around the two top quark 
  threshold. 
  }
\end{figure}

In Fig.~\ref{fig::f1-t-region-ss330}, we show the real and imaginary parts
of $F_{\rm box1}$ and $F_{\rm box2}$ as functions of $p_T$ for $\sqrt{s}=330$~GeV, i.e.~below the threshold for the production of two top quarks.
The upper panels show successive higher-order terms of the
$\delta q$ expansion
as indicated by $N_q$ in the legend,
while the lower panels display the relative deviation
from the best approximation, $\Delta_{\rm rel}$, which includes 
expansion terms up to and including $N_q=5$.
We omit results refined via Pad\'e approximation, as
they would coincide with the $N_q=5$ curve in this region of the phase
space. In all cases, we observe rapid convergence. For $F_{\rm box1}$, the
relative deviation between the approximations incorporating terms up
to $N_q=4$ and $N_q=5$, respectively, is well below one percent. In the
case of $F_{\rm box2}$, the relative deviation is on the order of a few
percent; however, the absolute contribution of $F_{\rm box2}$ is significantly
smaller than that of $F_{\rm box1}$.
In some cases, lower-order terms in the $\delta q$ expansion appear to
provide better approximations than those that include terms up to
order $N_q=5$. This is the case, for example, for the real part of $F_{\rm box1}$
shown in the top-left panel of Fig.~\ref{fig::f1-t-region-ss330}. This
behaviour should be considered coincidental. Furthermore, the
difference from the best approximation is small.

Fig.~\ref{fig::f1-t-region-ss500} shows the real and imaginary parts
of $F_{\rm box1}$ and $F_{\rm box2}$ for $\sqrt{s}=500$~GeV.  It is once again evident
that the $\delta q$ expansion can reliably cover this region of the phase
space.  For most values of $p_T$, the $\delta q$ expansion provides
approximations with an accuracy of about one percent or better.  The
precision only begins to decrease at $p_T \gtrsim 200$~GeV.  In this
regime, however, the result improved via a Pad\'e approximation yields
good approximations, as was already demonstrated in
Section~\ref{sec::scalar}.
The peak-like structure visible in some of the lower panels
arises from intersections with the curves that contain terms up to $N_q=5$;
that is, the numerator of the quantity $\Delta_{\rm rel}$ becomes small.

Next we discuss results where $p_T$ is kept fixed while $\sqrt{s}$ is
varied. In Fig.~\ref{fig::f1-t-region-pt100} we set $p_T=100$~GeV and
focus on values of $\sqrt{s}$ near the threshold for the production of
two top quarks.  For $F_{\rm box1}$, we again observe rapid convergence of the
$\delta q$ expansion; the discrepancy between the $N_q=4$ and $N_q=5$ terms is
below the one-percent mark.  The situation for $F_{\rm box2}$ is somewhat less
favourable: here, the two highest-order terms agree to within
approximately 3\% at $\sqrt{s}=350$~GeV, with this discrepancy
decreasing to one percent or less at higher energies.  Note however
that $F_{\rm box2}$ as compared to $F_{\rm box1}$ is numerically sub-leading, as already mentioned above.  Naturally,
if a form factor exhibits a zero-crossing, as is the case for $F_{\rm box2}$ at
$\sqrt{s}\approx 400$~GeV, the relative uncertainty becomes large.
In Fig.~\ref{fig::f1-t-region-pt100}, it is not necessary to construct
Pad\'e approximants, since, for $p_T=100$~GeV, we are well within the
convergence region of the $\delta q$ expansion.

In Fig.~\ref{fig::f1-t-region-pt25_cancel}, we show $F_{\rm box1}$ as a function of
$\sqrt{s}$ for a fixed $p_T=25$~GeV. 
In the top and bottom rows, we use different numerical values for the top-quark mass, corresponding to the QCD pole and  $\overline{\rm MS}$ quark mass, respectively.\footnote{As far as electroweak corrections are concerned, we always renormalize the top-quark mass in the on-shell scheme.}
The purpose of this plot is to
demonstrate that a strong cancellation occurs between the
bare two-loop amplitude and the contribution from
the counterterms. As can be seen in Fig.~\ref{fig::f1-t-region-pt25_cancel},
both the real and imaginary parts of the un-renormalized two-loop
amplitude (red curves) exhibit a pronounced, peak-like structure at
$\sqrt{s}$ values near the threshold for top-quark pair
production. The contribution from the counterterms (blue curve) shows
a similar structure, but with the opposite sign. In contrast, the
renormalized form factor (black line) follows a relatively smooth path
as it passes through the threshold. In the threshold region, this
cancellation is about one order of magnitude.

Fig.~\ref{fig::f1-t-region-pt25_cancel} further demonstrates that our approach is capable of
precisely resolving characteristic structures in the
cross section, without any additional computational cost. In this
specific case, the analytical structure of the threshold behaviour is
incorporated into the ``expand-and-match'' procedure.
We want to stress, that the large cancellation already occurs between the expanded bare and counterterm amplitudes and therefore does not affect the precision around the threshold.
This can also be seen in Fig.~\ref{fig::f1-t-region-pt100}, where no deterioration of the expansions around the two top threshold is observed.

\subsection{\label{sub::combine}Combining the forward and high-energy expansions}

In this section, we compare the forward expansion for the form factors
$F_{\rm box1}$ and $F_{\rm box2}$ with the results of the high-energy expansion from
Ref.~\cite{Davies:2026wbx}.

\begin{figure}[t]
    \begin{center}
    \begin{tabular}{cc}
    \includegraphics[width=.45\textwidth]{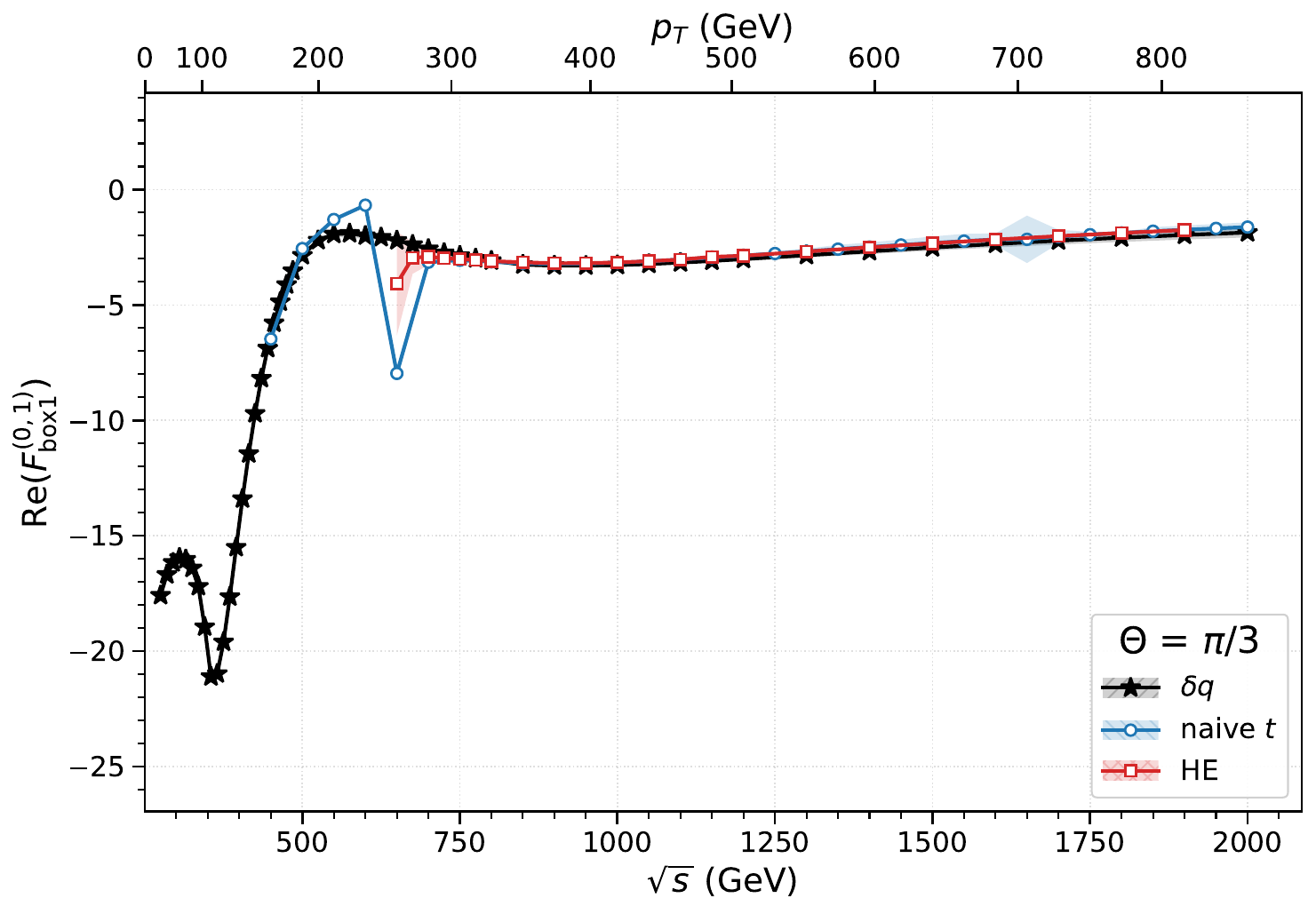} &
    \includegraphics[width=.45\textwidth]{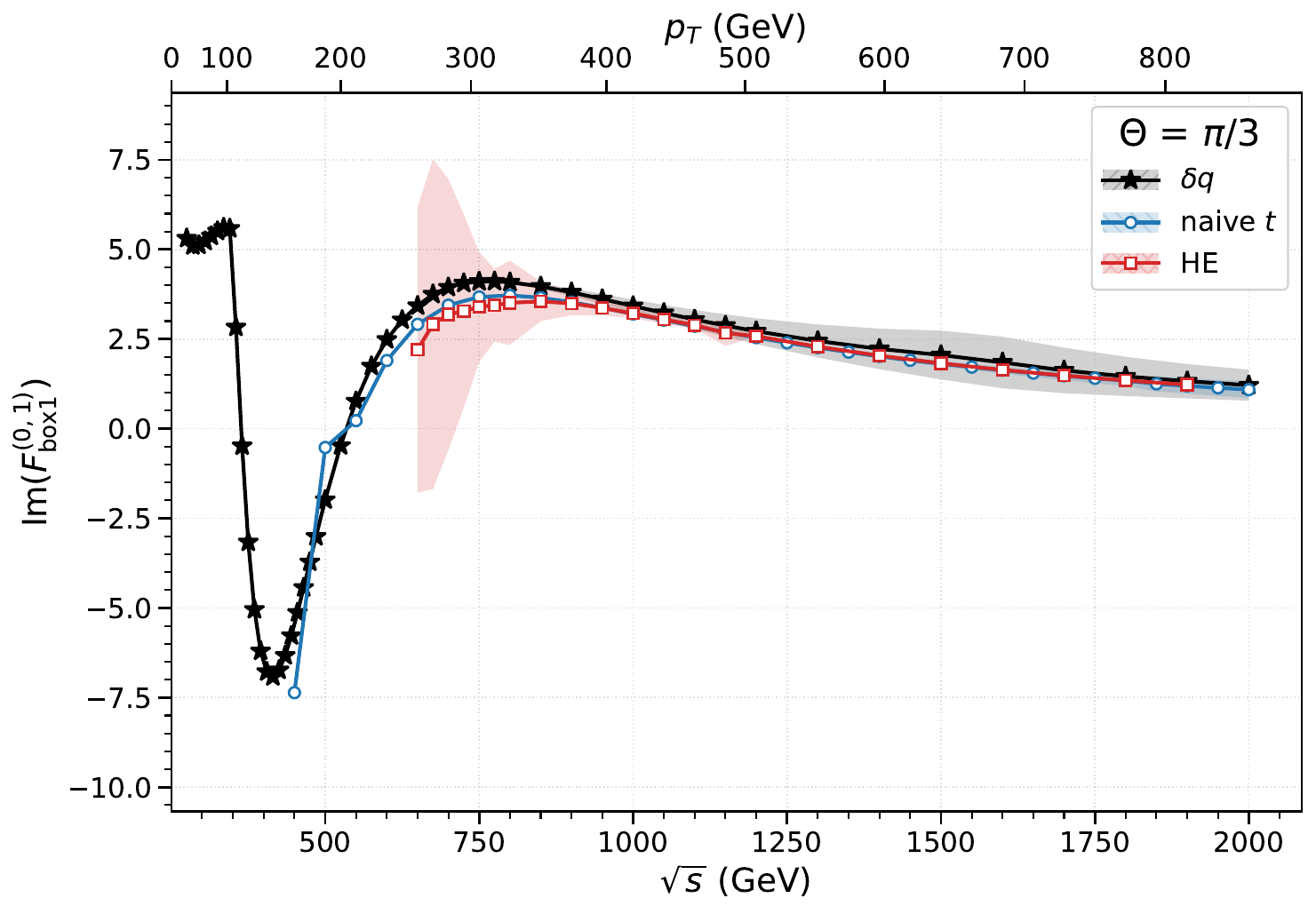} \\
    \includegraphics[width=.45\textwidth]{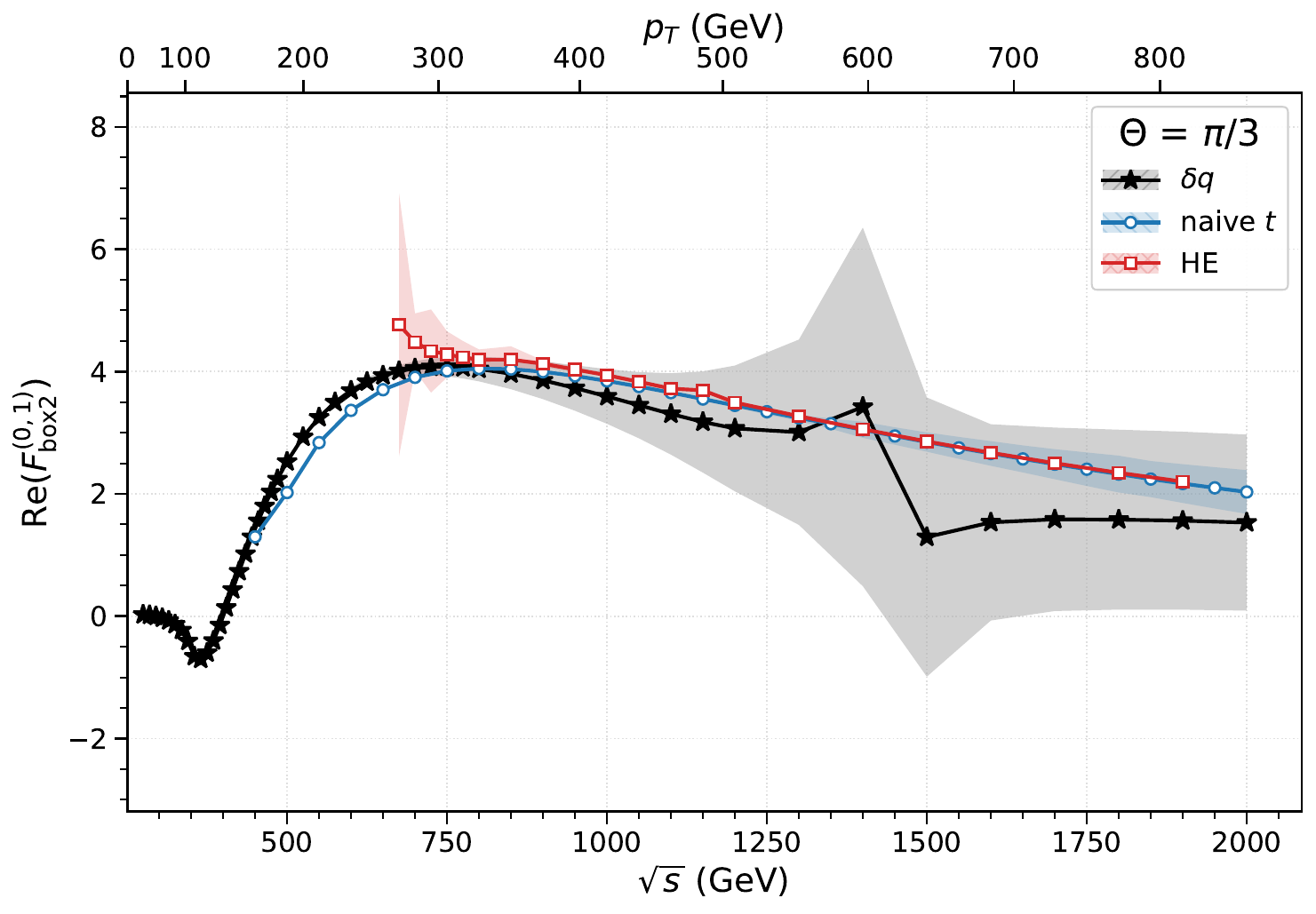} &
    \includegraphics[width=.45\textwidth]{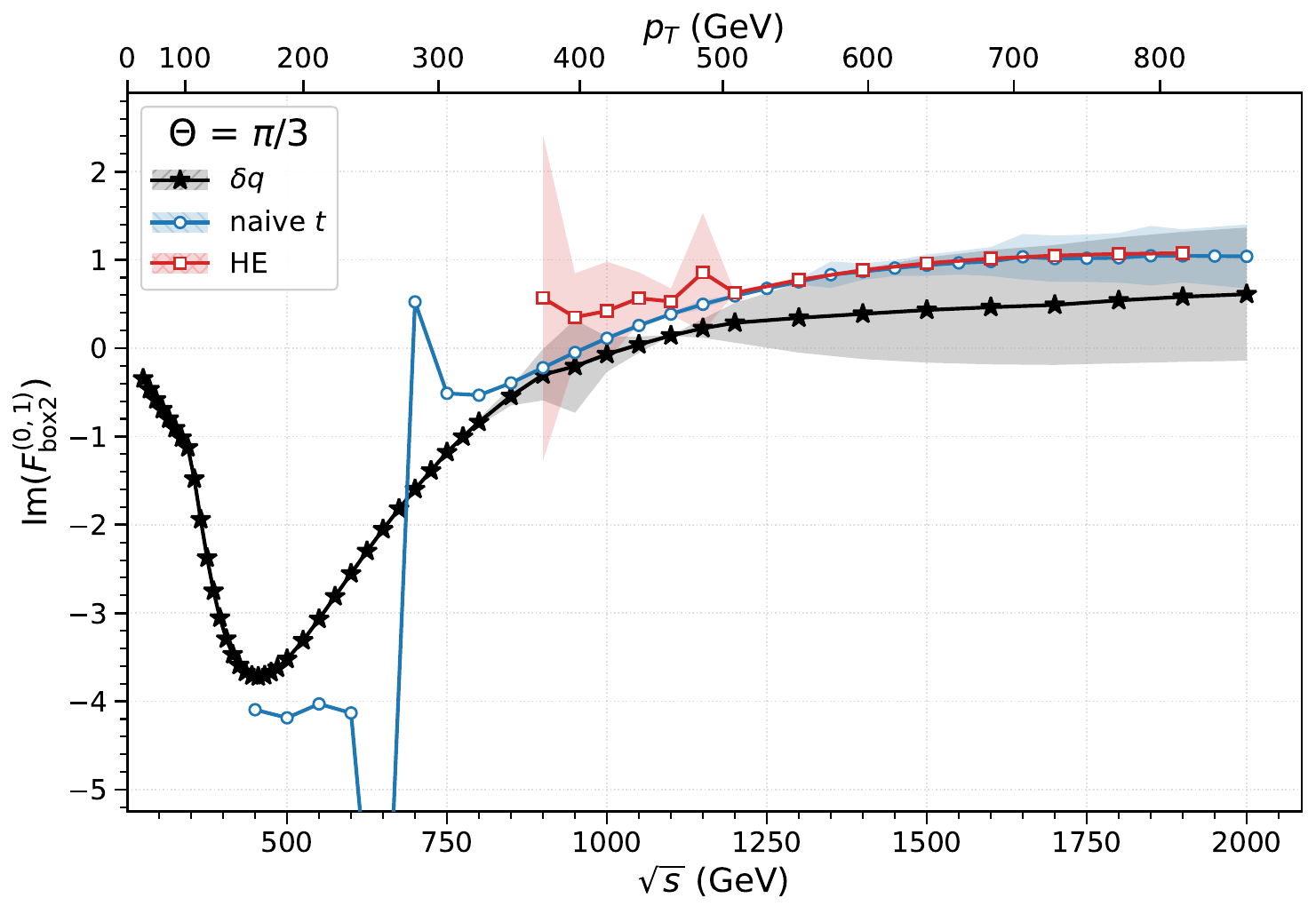}
    \end{tabular}
    \end{center}
  \caption{\label{fig::theta-pio3}
  Form factors $F_{\rm box1}$ and $F_{\rm box2}$ as a function of $\sqrt{s}$ with $\Theta=\pi/3$. 
  In each case we show both versions of the forward limit and high-energy expansion. The uncertainties due to the Pad\'e procedure are shown by coloured bands.
  }
\end{figure}

In Fig.~\ref{fig::theta-pio3}, we present the results for the form
factors at a fixed scattering angle of $\theta=\pi/3$ as a function of $\sqrt{s}$ in the range
between $250$~GeV and $2000$~GeV. For reference, the corresponding
$p_T$ values are indicated on the upper axes. These plots cover the
low-energy regime, where both $\sqrt{s}$ and $p_T$ are small, the two-
and three-particle thresholds, and the high-energy regime,
characterized by larger values of $\sqrt{s}$ and $p_T$. The results
for the three different expansions (the $\delta q$ expansion, the
naive-$t$ expansion, and the high-energy expansion) are shown in
three distinct colours, as indicated in the legend. The $\delta q$
expansion is valid in the left-hand portion of the plots, where
$p_T\lesssim 300$~GeV and $\sqrt{s}\lesssim 800$~GeV. This region
includes, in particular, the two- and three-particle thresholds,
which are well captured by the $\delta q$ expansion by
construction. This is especially true for the threshold for the
production of two top quarks, where the form factors exhibit
significant variations.

In the region around $\sqrt{s} = 800$~GeV, we observe that the
$\delta q$ expansion and the naive-$t$ expansion overlap in all
cases. The visible (minor) discrepancies arise because the naive-$t$
expansion lacks higher-order $\delta_B$ terms.\footnote{It should be
noted that the $\delta q$ expansion accounts for the exact dependence
on $\delta_B$.} This is the regime where we transition from one
expansion to the other. In this part of the phase space, the
high-energy expansion already shows good agreement with the naive-$t$
expansion for the real parts of $F_{\rm box1}$ and $F_{\rm box2}$. For the imaginary
parts, however, this agreement is only reached at larger values of
$p_T$. Nevertheless, a continuous transition from the naive-$t$
expansion to the high-energy expansion can be defined in all cases. A
very similar pattern emerges for other values of the scattering angle
$\theta$.

\begin{figure}[t]
    \begin{center}
    \begin{tabular}{cc}
    \includegraphics[width=.45\textwidth]{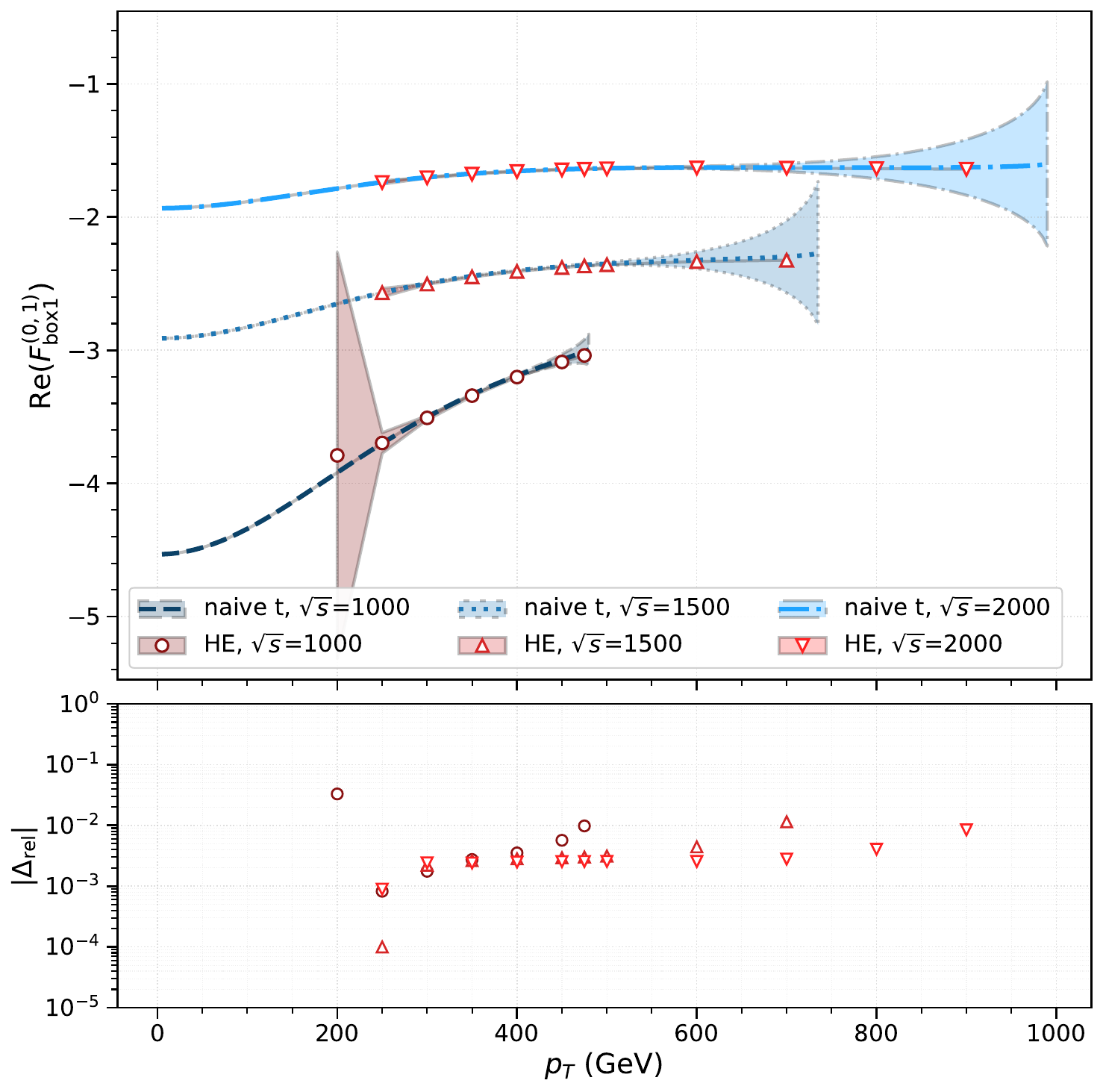} &
    \includegraphics[width=.45\textwidth]{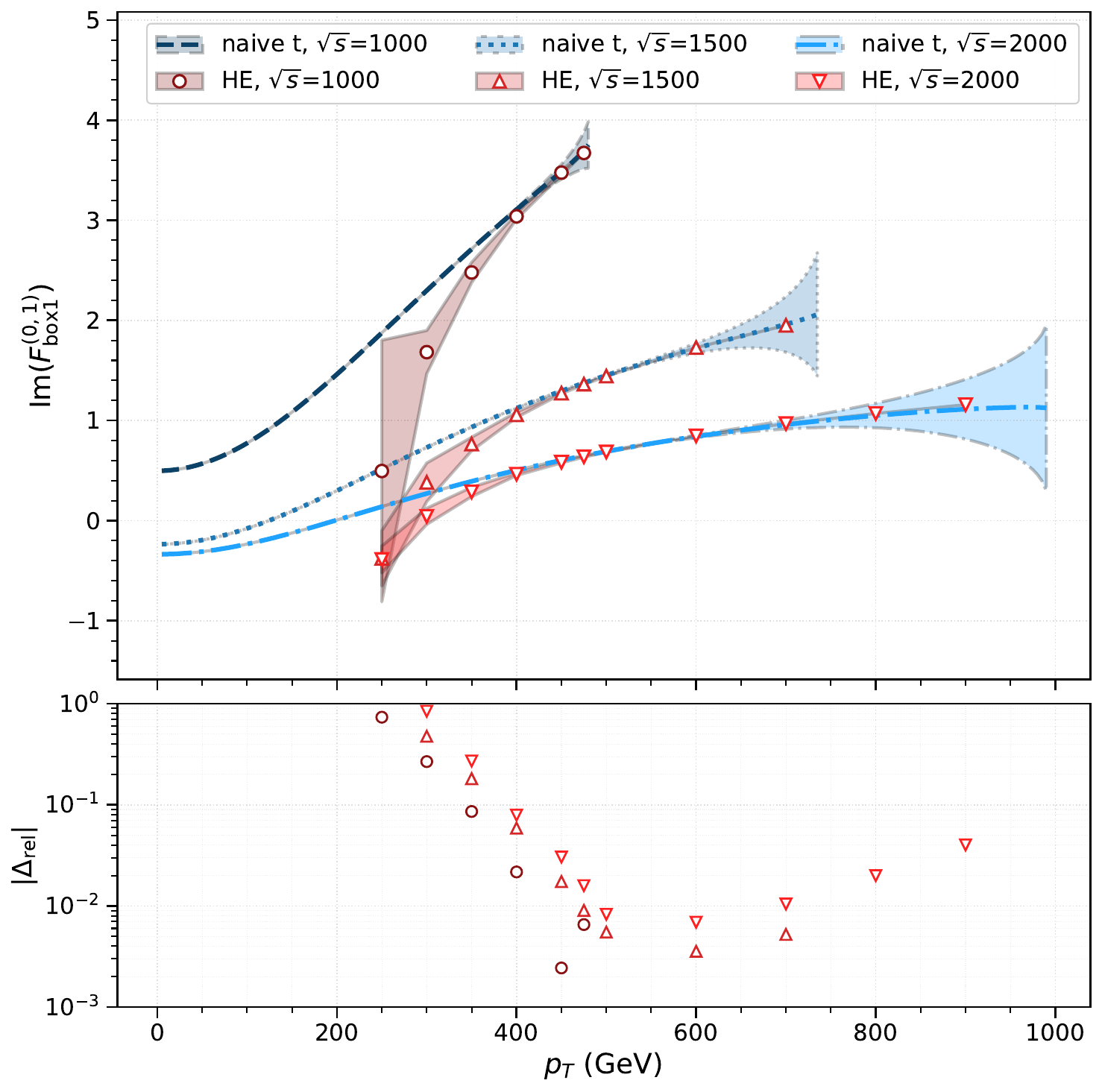} \\
    \includegraphics[width=.45\textwidth]{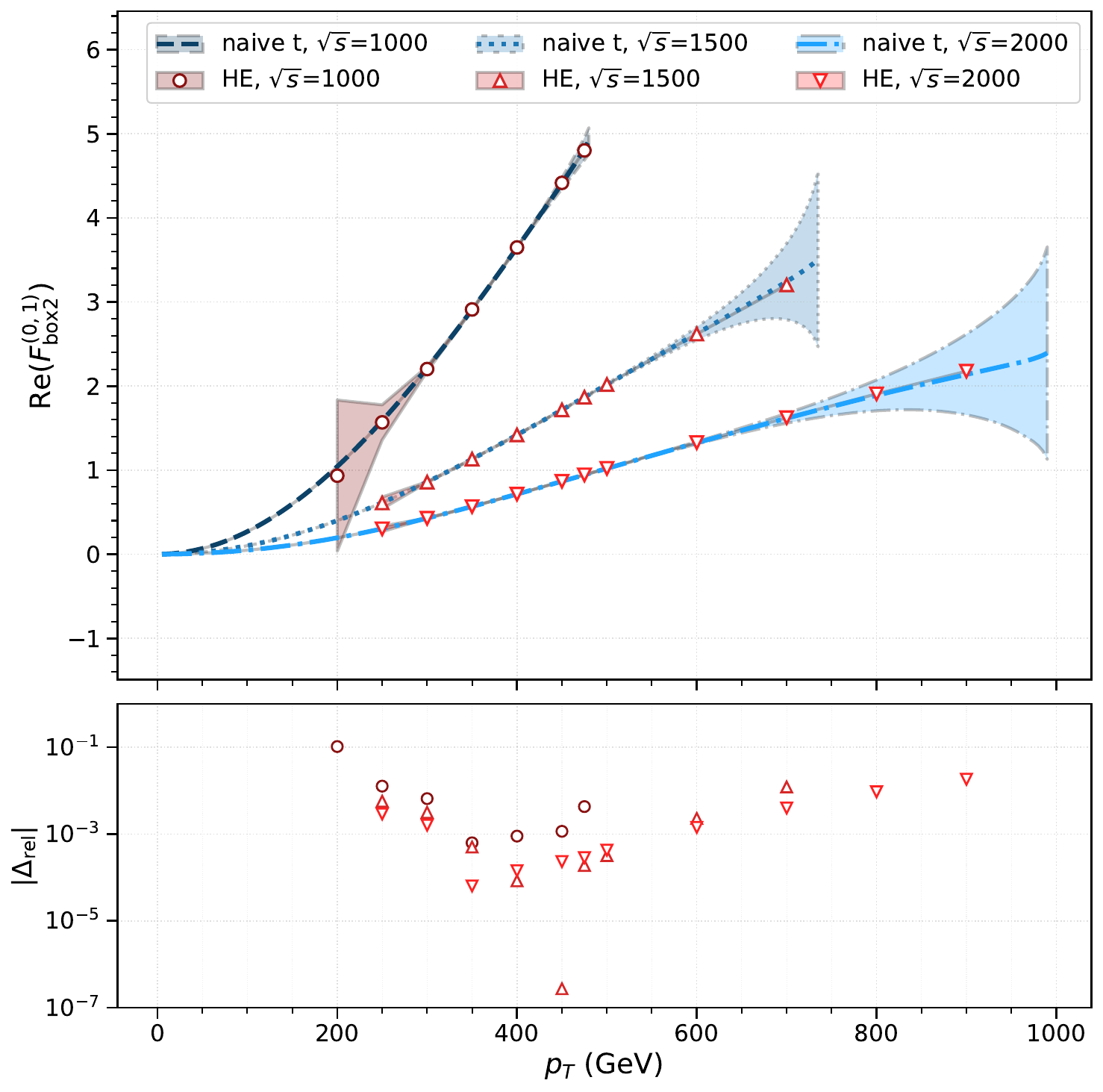} &
    \includegraphics[width=.45\textwidth]{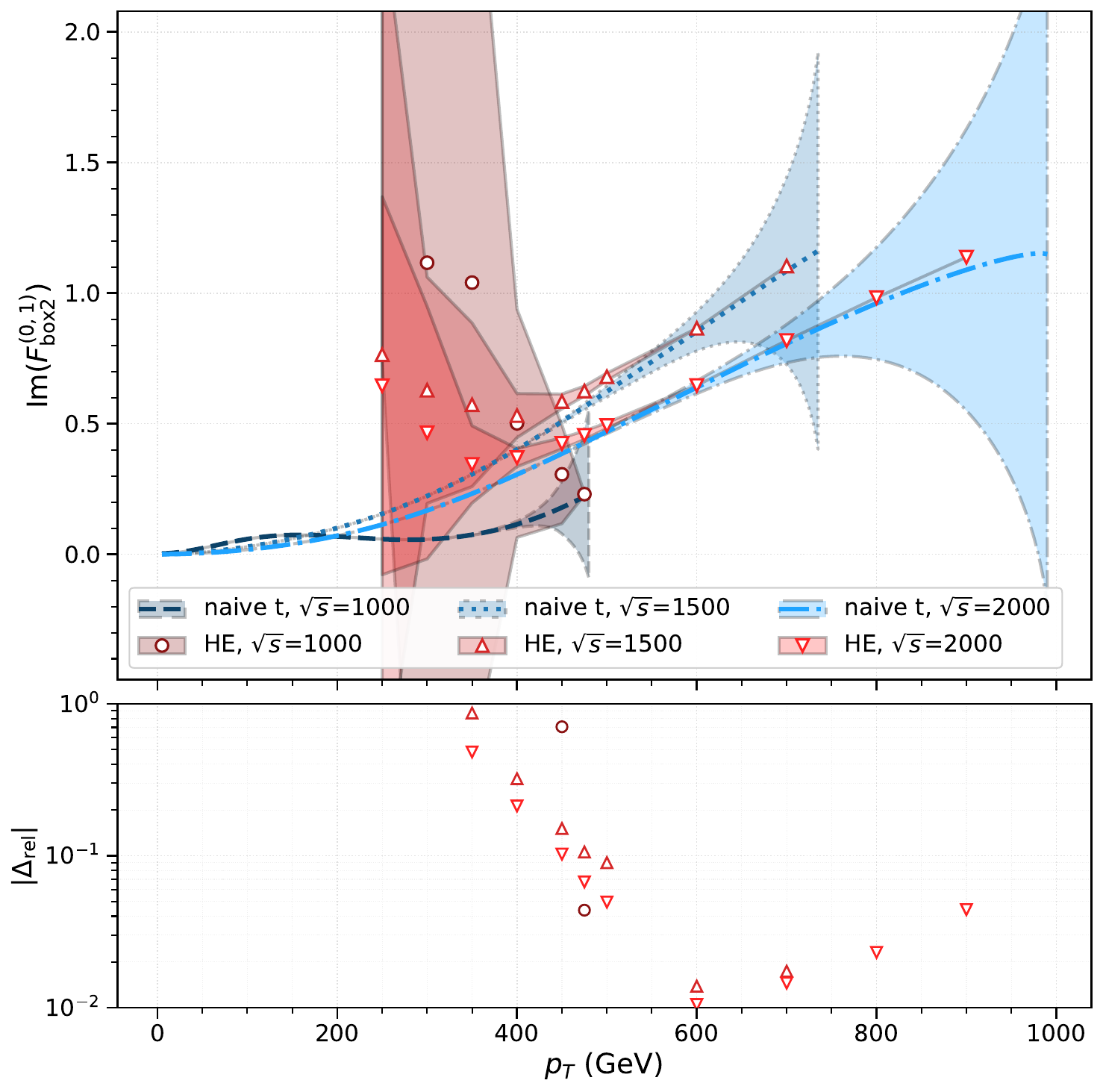}
    \end{tabular}
    \end{center}
  \caption{\label{fig::fixed-sqrts}
  Form factors $F_{\rm box1}$ and $F_{\rm box2}$ as a function of $p_T$ for fixed values of $\sqrt{s}$.
  In each case we show the 
  best approximation of the forward and high-energy expansion. The uncertainties due to the Pad\'e procedure are shown by coloured bands. 
}
\end{figure}

\begin{figure}[t]
    \begin{center}
    \begin{tabular}{cc}
    \includegraphics[width=.45\textwidth]{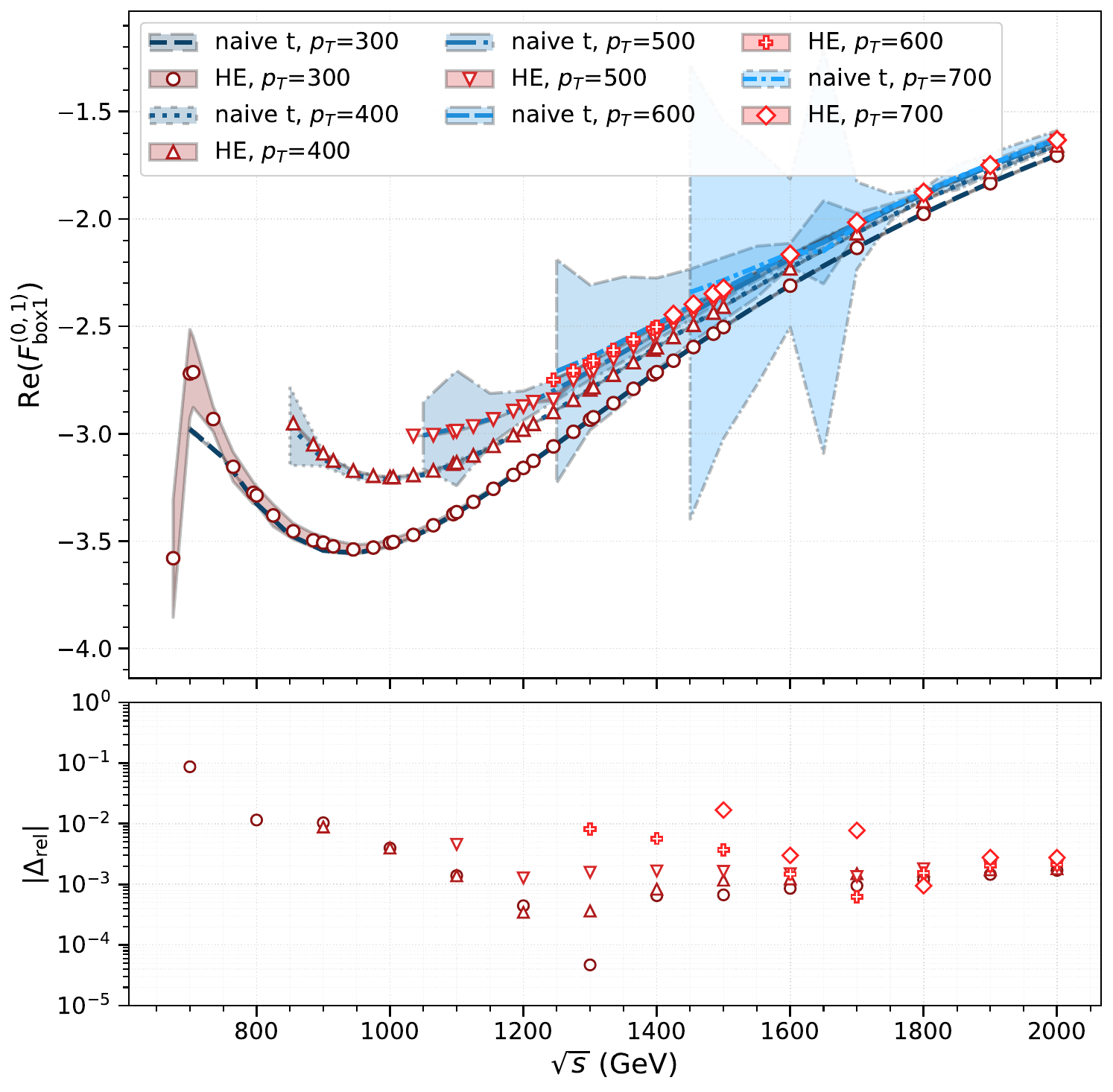} &
    \includegraphics[width=.45\textwidth]{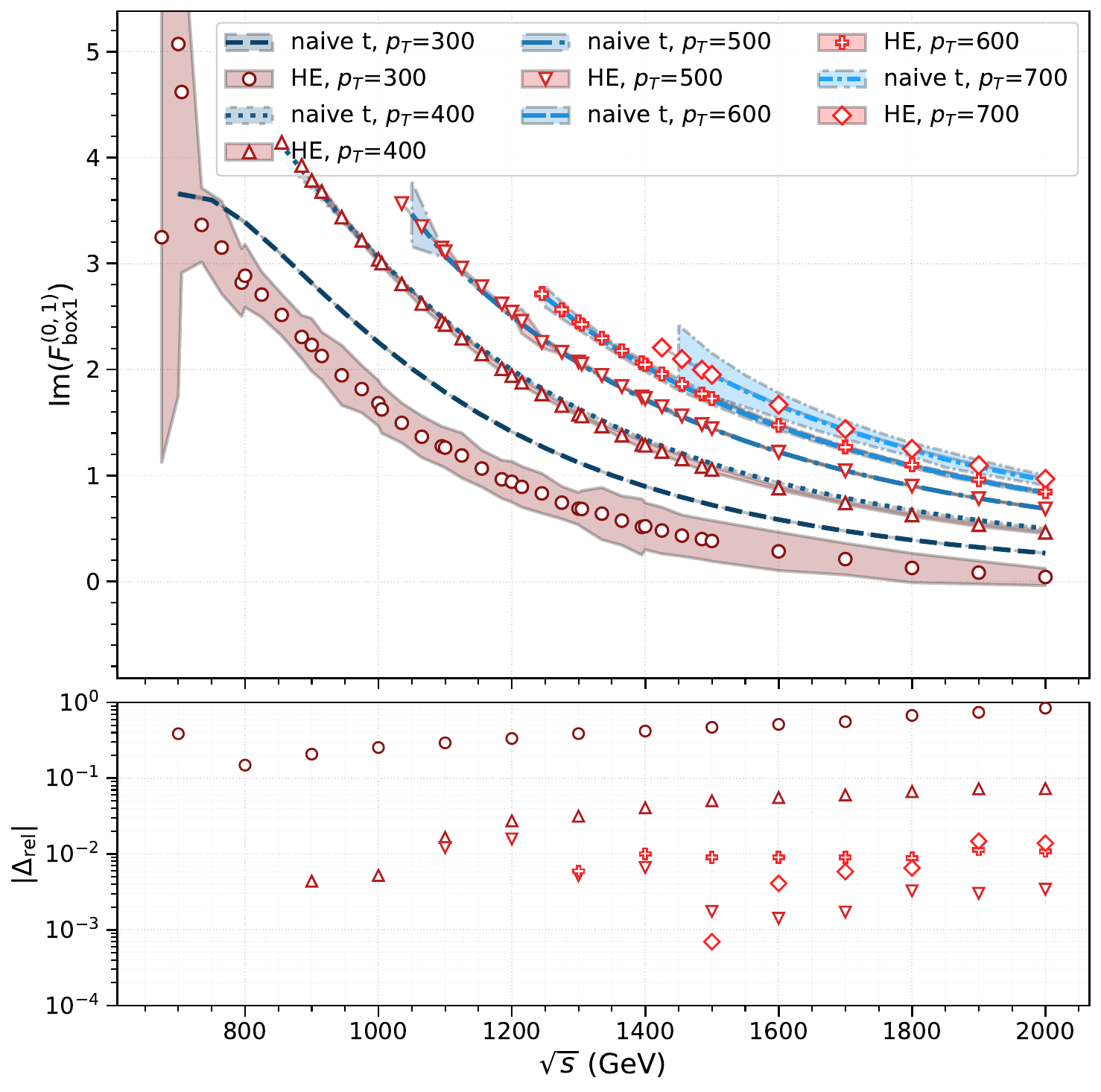} \\
    \includegraphics[width=.45\textwidth]{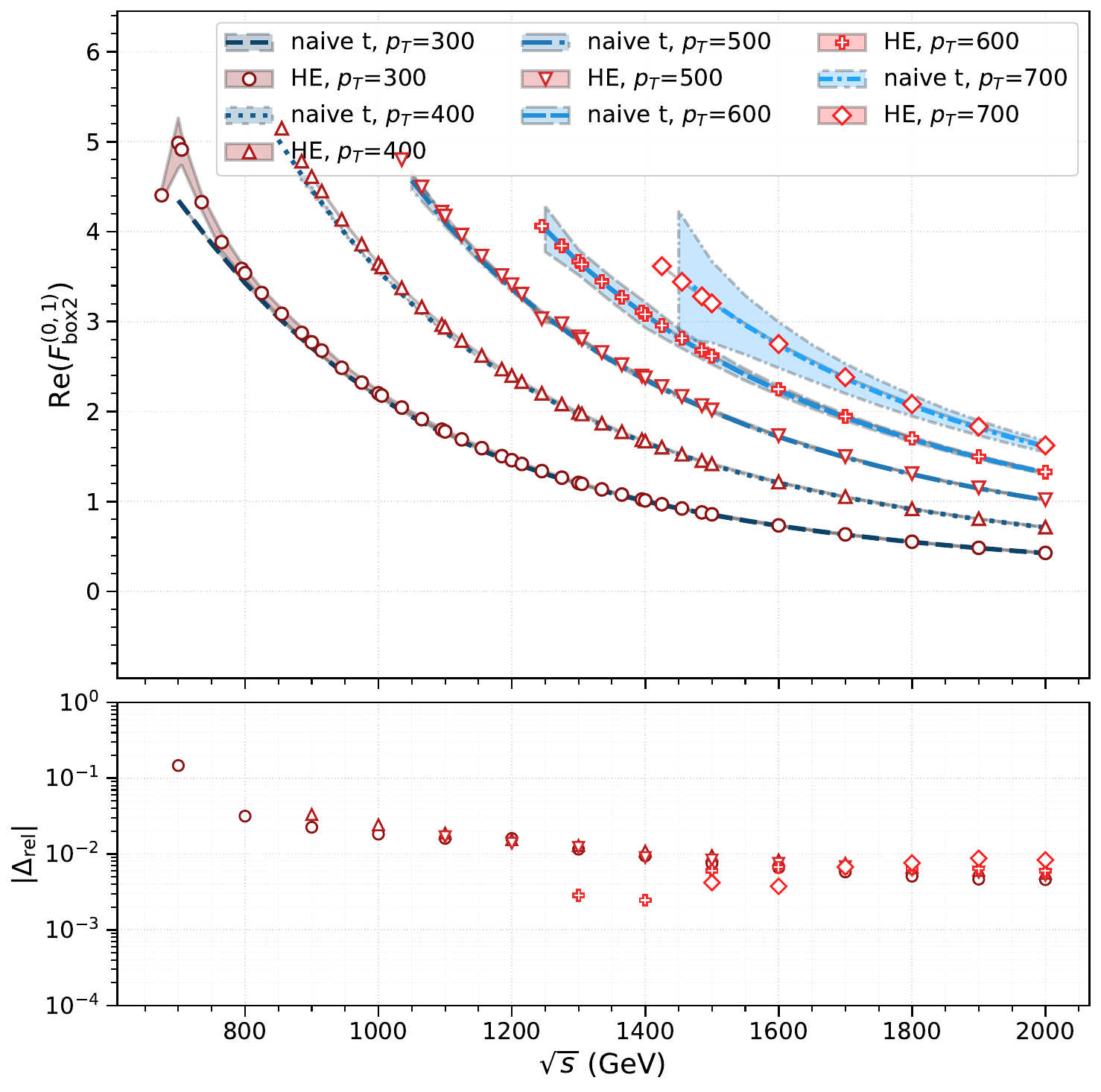} &
    \includegraphics[width=.45\textwidth]{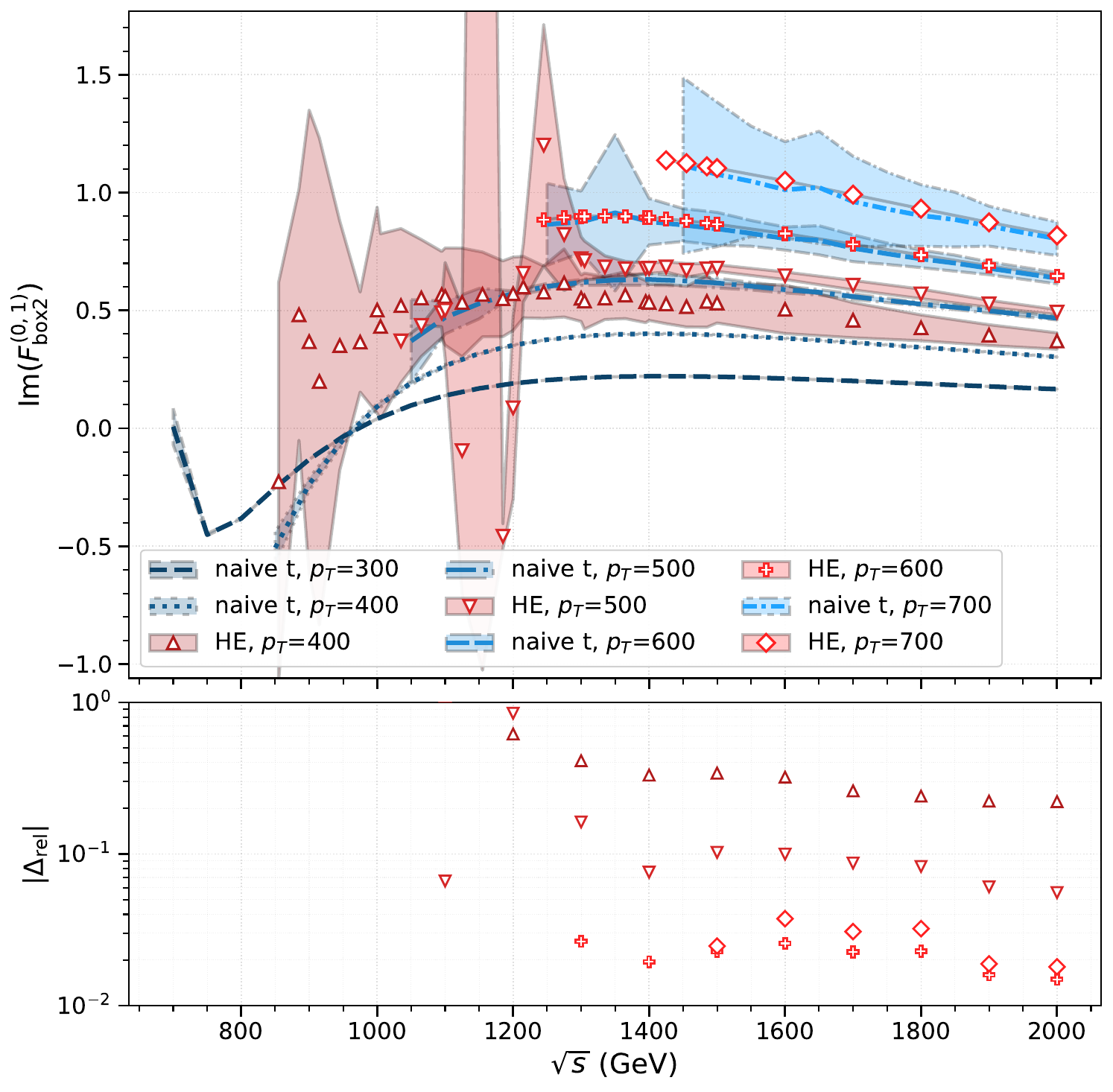} \\   
    \end{tabular}
    \end{center}
  \caption{\label{fig::fixed-pT}
  Form factors $F_{\rm box1}$ and $F_{\rm box2}$ as a function of $\sqrt{s}$ for fixed
  values of $p_T$.
  In each case we show the 
  best approximation of the forward and high-energy expansion. The uncertainties due to the Pad\'e procedure are shown by coloured bands. 
  For $\mbox{Im}(F_{\rm box2})$ we do not show 
  the high-energy results for $p_T=300$~GeV
  as they are subject to large uncertainties, even
  for large values of $\sqrt{s}$.
}
\end{figure}

In Figs.~\ref{fig::fixed-sqrts} and~\ref{fig::fixed-pT} we show real and imaginary parts for the
form factors $F_{\rm box1}$ and $F_{\rm box2}$. We either keep
$\sqrt{s}$ or $p_T$ fixed and show the dependence on the respective other variable.
For the forward\footnote{In Figs.~\ref{fig::fixed-sqrts} 
and~\ref{fig::fixed-pT} we use the naive-$t$ expansion.} and high-energy 
expansions we present
the best approximations which includes, besides the Pad\'e improvement,
also the acceleration in the $\delta_B$ expansion.
Each fixed-$p_T$ curve starts at the lowest $\sqrt{s}$ value
which is kinematically allowed.
Similarly, each fixed-$\sqrt{s}$ curves stops at the largest
kinematically allowed $p_T$ value.
In the lower panels we show the relative deviations
between the forward and high-energy expansions.

Note, that we only show results for intermediate values of $p_T$.
For smaller $p_T$ it is sufficient to use the forward expansion
for all allowed $\sqrt{s}$ values as has been shown in Section~\ref{sub::t_exp}.
Similarly, for larger $p_T$ values the phase space is covered by the
high-energy expansions as has been shown in Ref.~\cite{Davies:2026wbx}.

Next, we use Fig.~\ref{fig::fixed-sqrts} to identify a $p_T$ range in which a transition between the naive-$t$ expansion and the high-energy
expansion is possible, where we consider the following $\sqrt{s} \in \{1000, 1500, 2000\}$~GeV values. It should be
noted that for smaller values of $\sqrt{s}$, the maximum $p_T$ value
is sufficiently small such that the forward expansion yields good
results; see Section~\ref{sub::t_exp} for further details. For $F_{\rm box1}$, we find that in the range $p_T \in
[300, 500]$~GeV, both expansions agree to within less than one
percent. The situation is similar for the real part of $F_{\rm box2}$, where
the agreement is only slightly less precise. For the imaginary part of
$F_{\rm box1}$, we can identify a range of $p_T \in [400, 450]$~GeV in which
the naive-$t$ expansion and the high-energy expansion agree to
within a few percent. 

The situation for the imaginary part of $F_{\rm box2}$ is
different: here, $p_T \gtrsim 600$~GeV is required for the high-energy
expansion to exhibit good convergence. For $\sqrt{s} = 1000$~GeV,
neither the naive-$t$ expansion nor the high-energy expansion yields
to precise results for the largest possible values of $p_T$ close to $500$~GeV. This phase-space region corresponds to the case $t\approx u$, where naturally the convergence of the naive-$t$ expansion is limited.
Consequently, for the imaginary part of $F_{\rm box2}$, there
exists a region in phase space for values of $p_T\approx 500$~GeV and $t\approx u$ where our
approximations carry relatively large uncertainties. It should be
noted, however, that the absolute value of $\mbox{Im}(F_{\rm box2})$ is close
to zero and significantly smaller than in the other three cases. A
larger uncertainty therefore has no significant impact on the
numerical value of the squared matrix element.

In the following, Fig.~\ref{fig::fixed-pT} shows the form factors as a
function of $\sqrt{s}$ for the $p_T$ values identified in
Fig.~\ref{fig::fixed-sqrts}. These plots confirm the conclusions drawn
above. In particular, for the real and imaginary parts of $F_{\rm box1}$ as
well as the real part of $F_{\rm box2}$ for  $p_T$ between $400$~GeV
and $500$~GeV, we observe that the
agreement between the two approximations is within a few percent or
better for all values\footnote{The high-energy expansion
exhibits larger Pad\'e uncertainties at the lower end of the
kinematically allowed $\sqrt{s}$ range. In practice, however, this
poses no issue, as we can always rely on the result from the forward
expansion in this region.} of $\sqrt{s}$. 
For the imaginary part of $F_{\rm box2}$, the transition region must be set at
higher $p_T$ values, since even at $p_T=500$~GeV the high-energy
expansion exhibits larger uncertainties, particularly for
$\sqrt{s}\lesssim 1400$~GeV, whereas the forward expansion is
sufficiently stable.  On the other hand, the high-energy expansion
yields more stable results at $p_T=600$~GeV and higher.

Figures~5 and~6 in Ref.~\cite{Davies:2026wbx} contain the counterparts to the
plots shown in Figs.~\ref{fig::fixed-sqrts} and~\ref{fig::fixed-pT},
displaying exclusively the high-energy expansion.\footnote{It 
should be noted that the uncertainty bands for the curve at
$p_T=400$~GeV are larger in the present work. The reason for this is
the acceleration in $\delta_B$, which leads to large uncertainties when
applied to an unstable series expansion. However, this has no
practical consequences, as the forward expansion yields stable results.}
The significant
advantage of the forward expansion is particularly evident in the
plots at fixed $\sqrt{s}$ for $p_T \lesssim 300$ GeV.

\subsection{Triangle form factor}

In this section, we present exact results for the form factor $F^{(0,1)}_{\text{tri1}}$. All one-particle-irreducible diagrams contributing to this form factor have no dependence on $q_3$ and thus
the exact results for these diagrams are provided (in semi-analytic form) through the leading term of the $\delta q$ expansion. The one-particle-reducible contribution factorizes into one-loop integrals and is known analytically from Ref.~\cite{Zhang:2024rix}.

\begin{figure}[t]
    \centering
    \includegraphics[width=.6\textwidth]{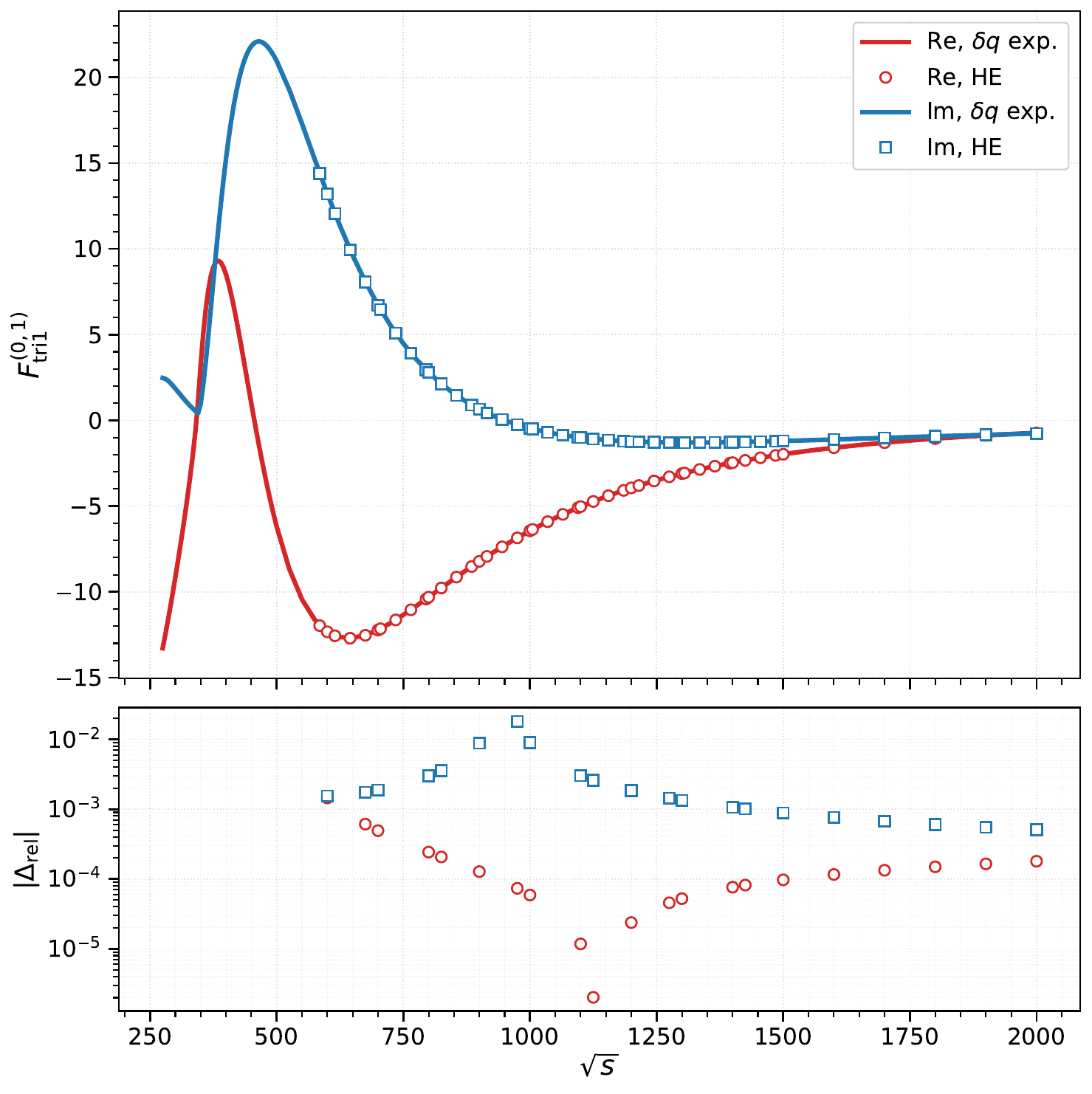} 
  \caption{\label{fig::f1tri}
  Real and imaginary parts of $F_{\text{tri1}}$ as a function of $\sqrt{s}$ computed once with the high-energy expansion and once using the $\delta q$ expansion, i.e., the exact result. The lower panels shows the relative difference
of the high-energy expansion to the exact result, which is obtained as zeroth order of the $\delta q$ expansion.
  }
\end{figure}

In Fig.~\ref{fig::f1tri} the exact results for both the real and imaginary part of $F^{(0,1)}_{\text{tri1}}$ are shown and compared to the high-energy expansion~\cite{Davies:2026wbx}. The high-energy expansion converges well to the exact result with an uncertainty below the percent level for both the real part and the imaginary part. 
The uncertainty of the result based on the $\delta q$ expansion is
limited solely by the ``expand-and-match'' procedure; we estimate it to
be approximately 10 decimal digits.
At the threshold for the production of two top quarks, similar to the contribution of the box-type diagrams, we do not observe a pronounced peak.
It should be emphasized that, by its very design, the
``expand-and-match'' procedure reproduces the analytical behaviour using
high-precision numerical coefficients.

\section{\label{sec::concl}Conclusions and outlook}

In this work, we calculate the NLO electroweak corrections for the
production of Higgs boson pairs in the forward limit. We employ two
types of expansion; the first, referred to as the ``$\delta
q$ expansion'', is a Taylor expansion in the Mandelstam variable $t$ and
the mass of the external Higgs bosons, accounting for terms up to the
order $t^i (m_H^2)^j$ where $i+j=5$. In this approach, we retain the
exact dependence on the partonic centre-of-mass energy $s$, the
top-quark mass $m_t$, and the mass $m_B$ (where $m_B \in \{m_H, m_t,
m_W\}$). This is carried out within the framework of the
``expand-and-match'' approach, which facilitates two-dimensional series
expansions around suitably chosen expansion points. For the second
type of forward expansion, the so-called ``naive-$t$ expansion'', we
expand the integrands in terms of $m_B - m_t$, which allows for a deeper expansion in $t$.

The $\delta q$ expansion yields good results with an accuracy of
about one percent for smaller values of
$p_T\lesssim 300$~GeV, which accounts for 95\% of the integrated
LO cross section. It also reproduces the 
two- and three-particle thresholds present in the
Feynman diagrams. The naive-$t$ expansion, by contrast, extends to
larger values of $\sqrt{s}$ and $p_T$, in some cases even up to $p_T
\approx 600$~GeV, due to its greater expansion depth.

The combination of the two forward-expansion approaches covers a
region in phase space that provides the main contribution to the total
cross section. We estimate that the numerically dominant contribution
to the form factors is approximated with an accuracy on the order of
a few percent. Contributions that are less significant numerically (such
as the imaginary part of $F_{\rm box2}$) exhibit a larger relative uncertainty;
however, their contribution to hadronic cross sections is small.

We also discuss the combination of the forward expansion with the
high-energy expansion, which was calculated in
Ref.~\cite{Davies:2026wbx}. We show that there is good agreement
between the naive-$t$ expansion and the high-energy expansion over a
realistically large region of the phase space. The precise transition
point between the different expansions depends on the specific form
factor.

The major advantage of our approach is that there is an explicit
dependence on all input parameters. Once implemented in the \texttt{C++}
library \texttt{ggxy}~\cite{Davies:2025qjr}, it will thus be 
possible to
calculate arbitrary hadronic distributions in a fast and 
flexible
manner.

We refrain from providing explicit 
(semi-)analytic expressions for the form 
factors. Once implemented in \texttt{ggxy}, 
they will be easily accessible and available 
for use by other groups for various purposes.
Instead to facilitate cross checks and validation, we provide an ancillary file~\cite{progdata} containing results for the form factors at selected points in phase space; see Appendix~\ref{app::data}.



\section*{Acknowledgements}

This work is supported by the Deutsche Forschungsgemeinschaft (DFG, German
Research Foundation) under grant 396021762 --- TRR 257 ``Particle Physics
Phenomenology after the Higgs Discovery''.
The work of J.D.~is supported by STFC Consolidated Grant ST/X000699/1. 
K.S. is supported by the European Union under the Marie Sk{\l}odowska-Curie Actions (MSCA) Grant 101204018.
We thank Hantian Zhang for useful discussions.


\begin{appendix}

\section{\label{app::photon}Photon contribution to $F_{\rm box1}$ and $F_{\rm box2}$}

\begin{figure}[t]
    \begin{center}
    \begin{tabular}{cc}
    \includegraphics[width=.45\textwidth]{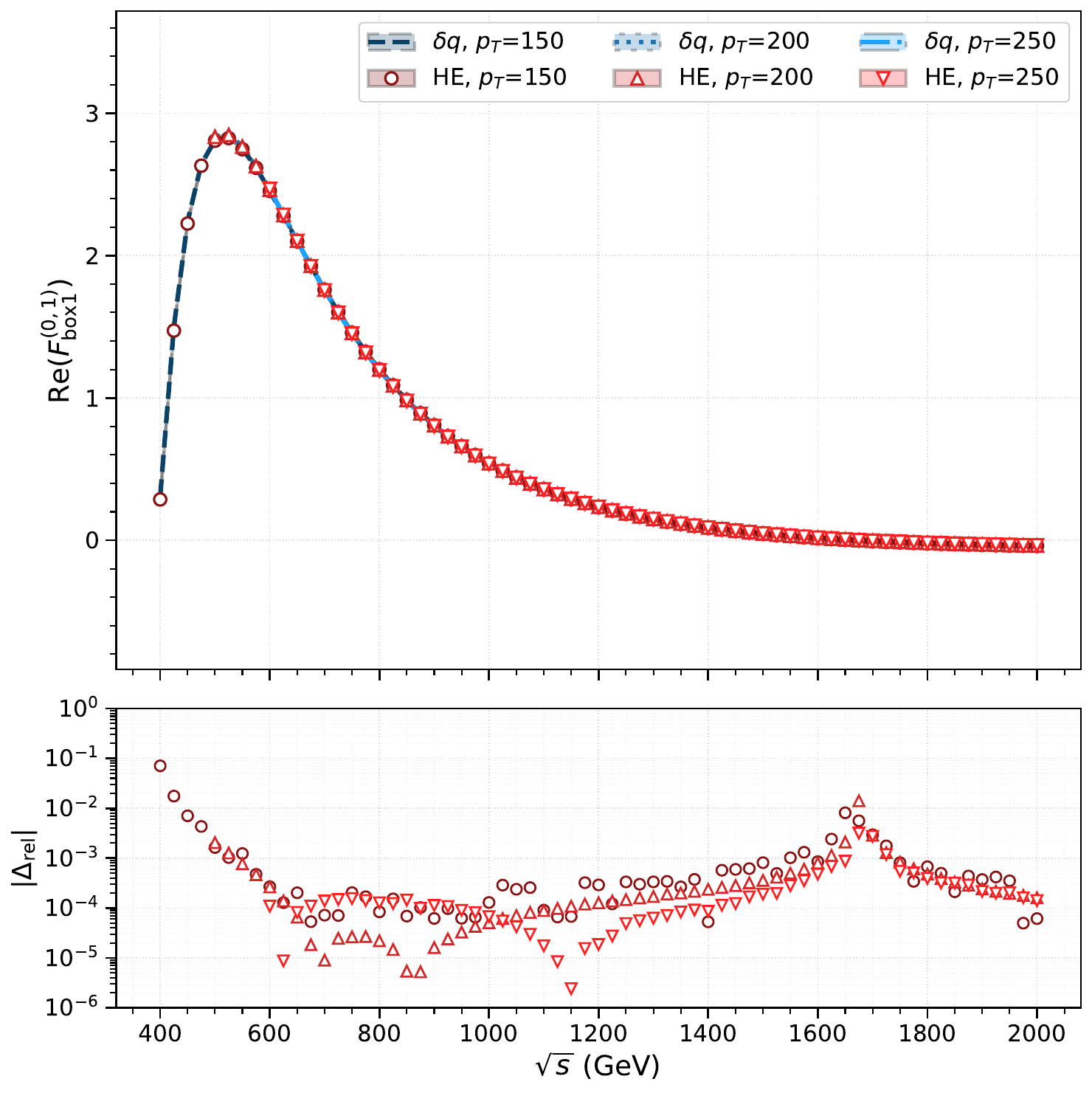} &
    \includegraphics[width=.45\textwidth]{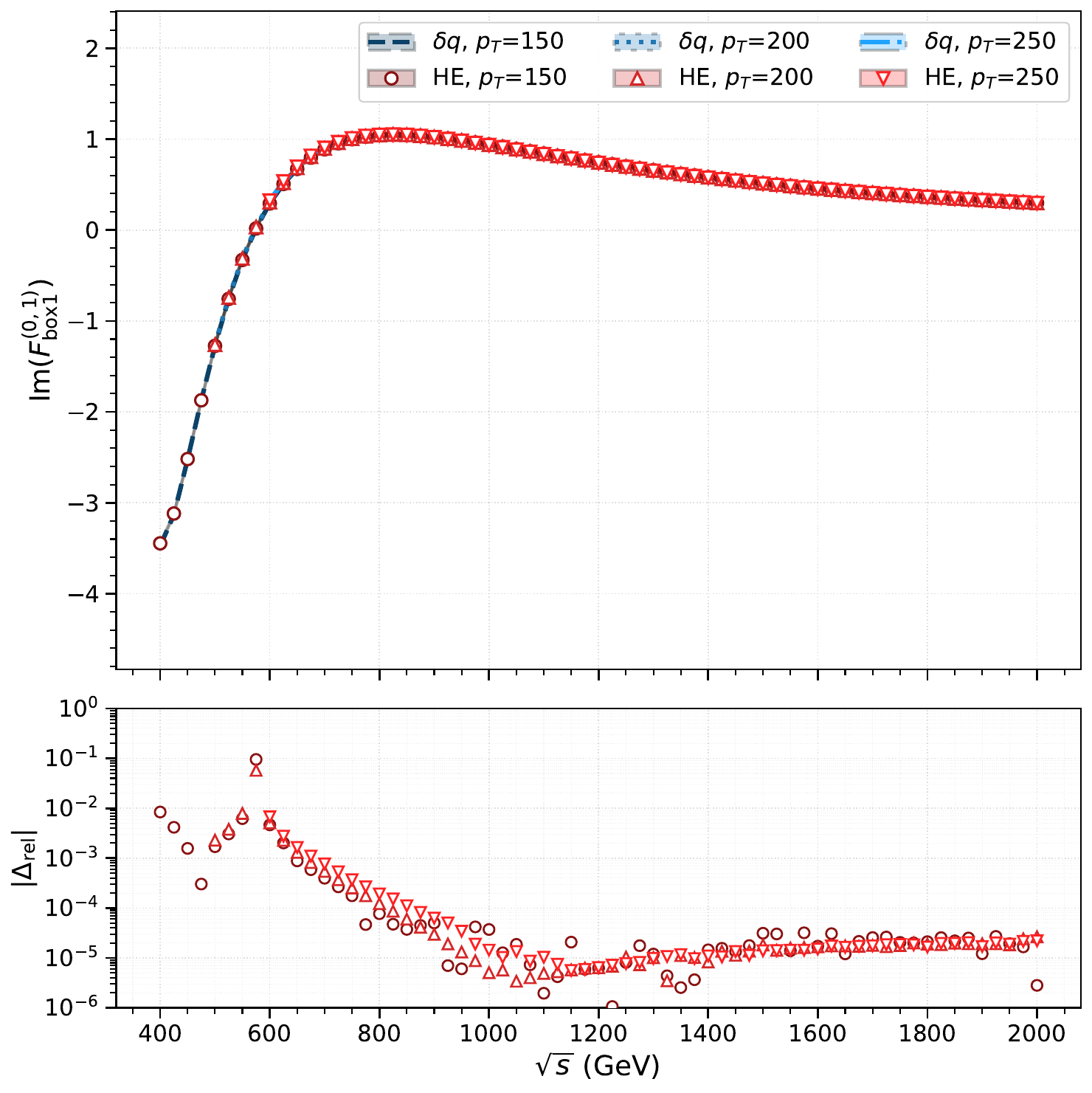} \\
    \includegraphics[width=.45\textwidth]{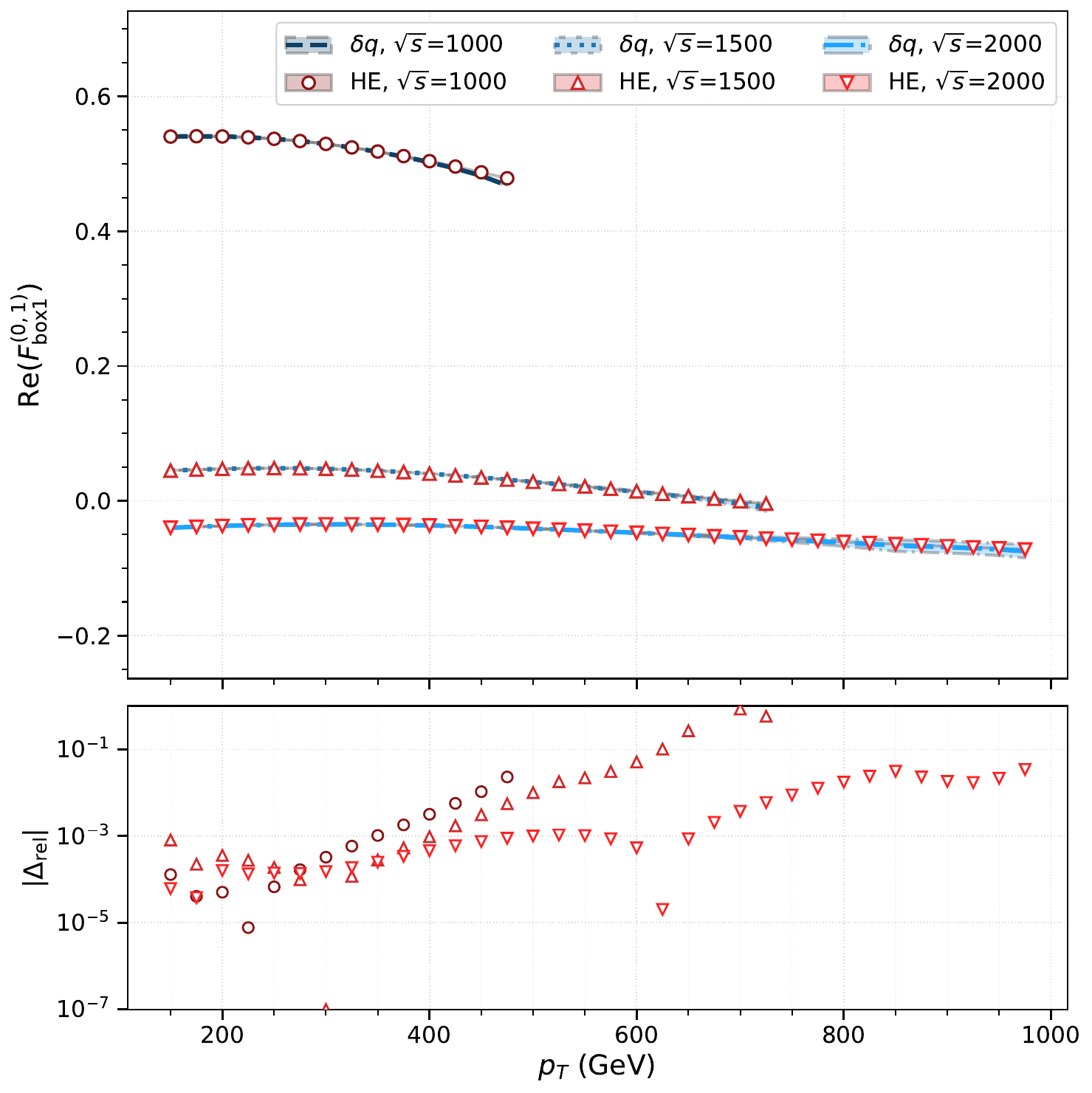} &
    \includegraphics[width=.45\textwidth]{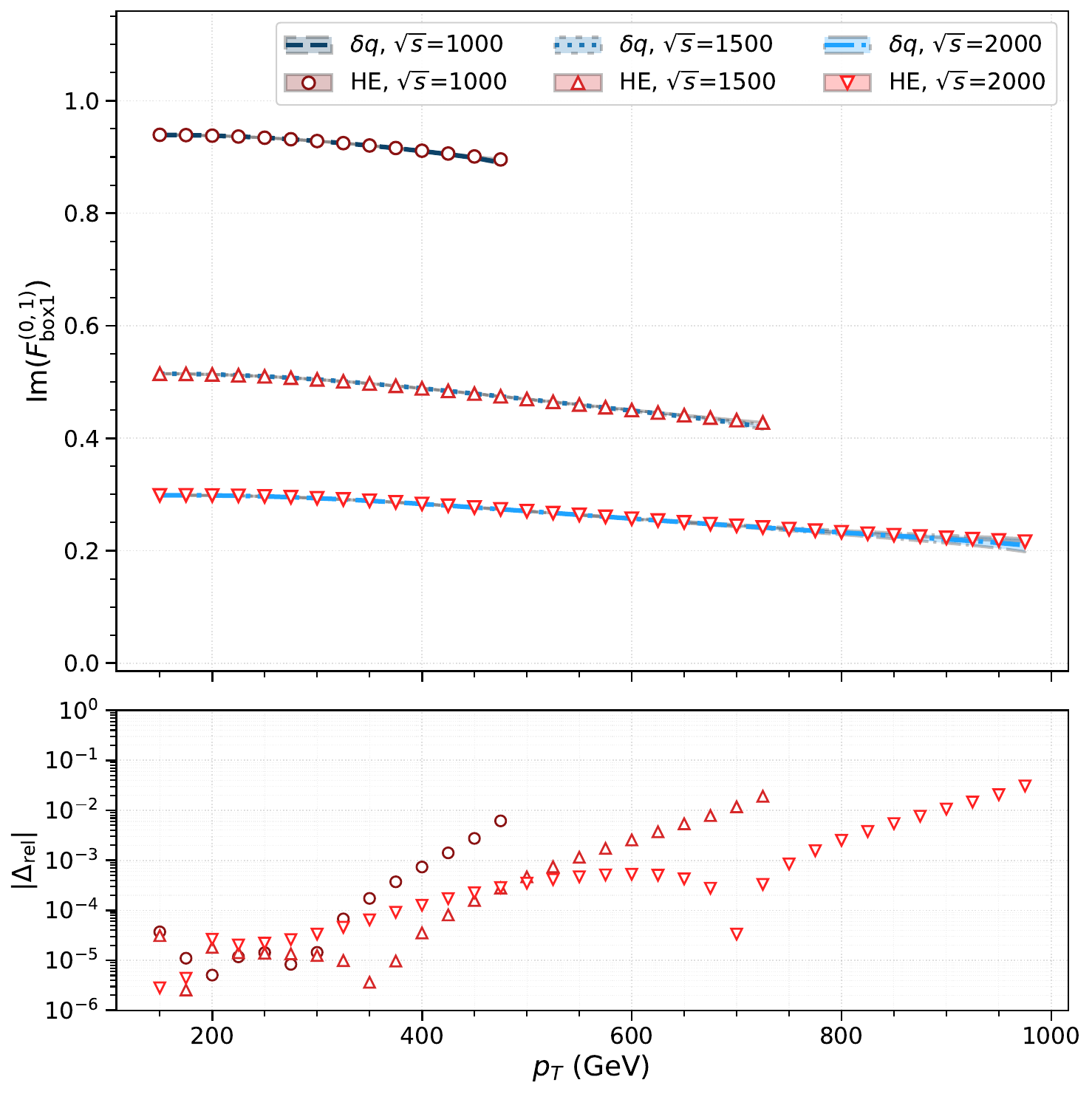} 
    \end{tabular}
    \end{center}
  \caption{\label{fig::f1-photon}
  Photon contribution to form factor $F_{\rm box1}$ as a function of $\sqrt{s}$ (top row) and $p_T$ (bottom row). In each case we show the 
  best approximation of the forward and high-energy expansions. The uncertainties due to the Pad\'e procedure are shown by coloured bands. The lower panels show the relative deviation of the forward and high-energy expansions.}
\end{figure}

\begin{figure}[t]
    \begin{center}
    \begin{tabular}{cc}
    \includegraphics[width=.45\textwidth]{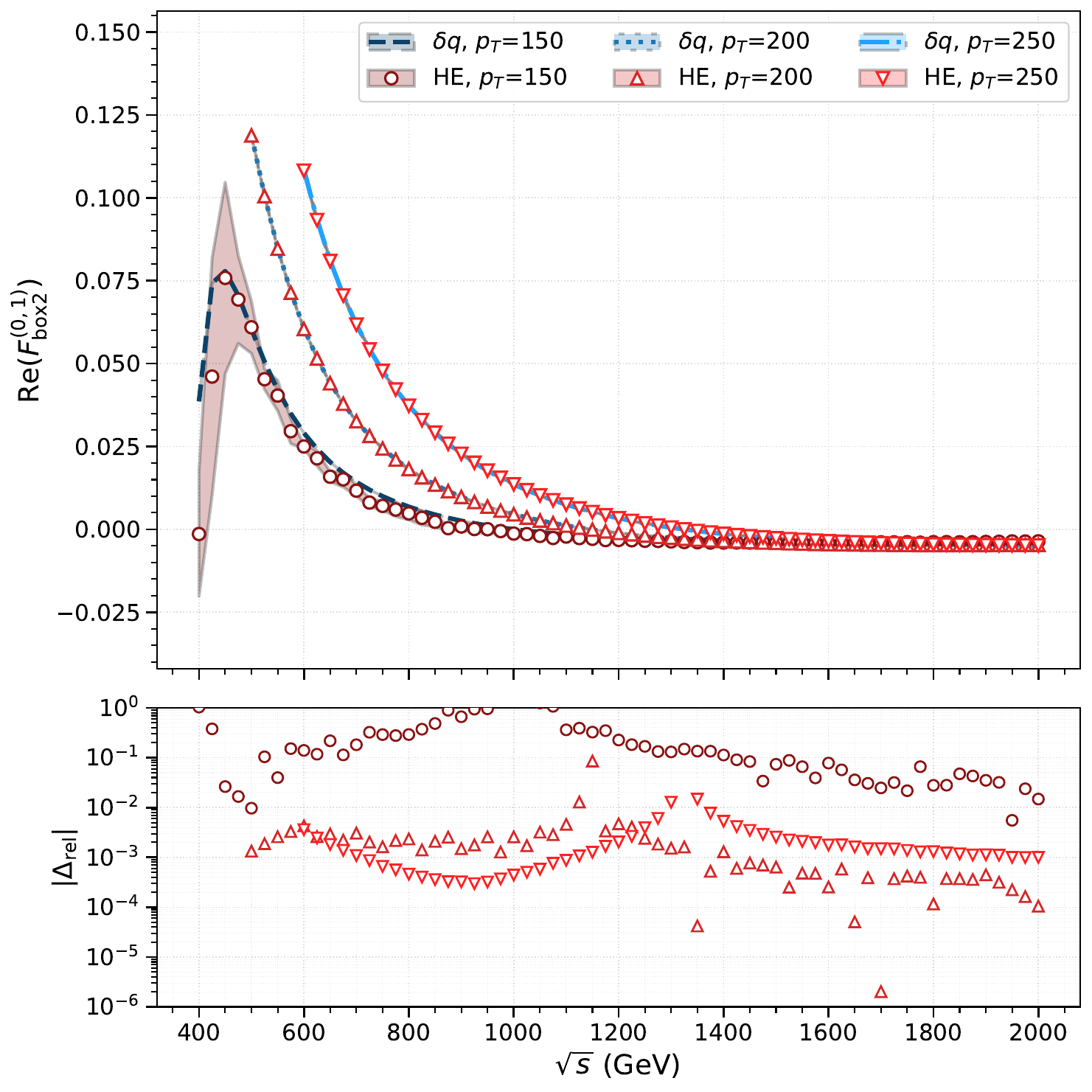} &
    \includegraphics[width=.45\textwidth]{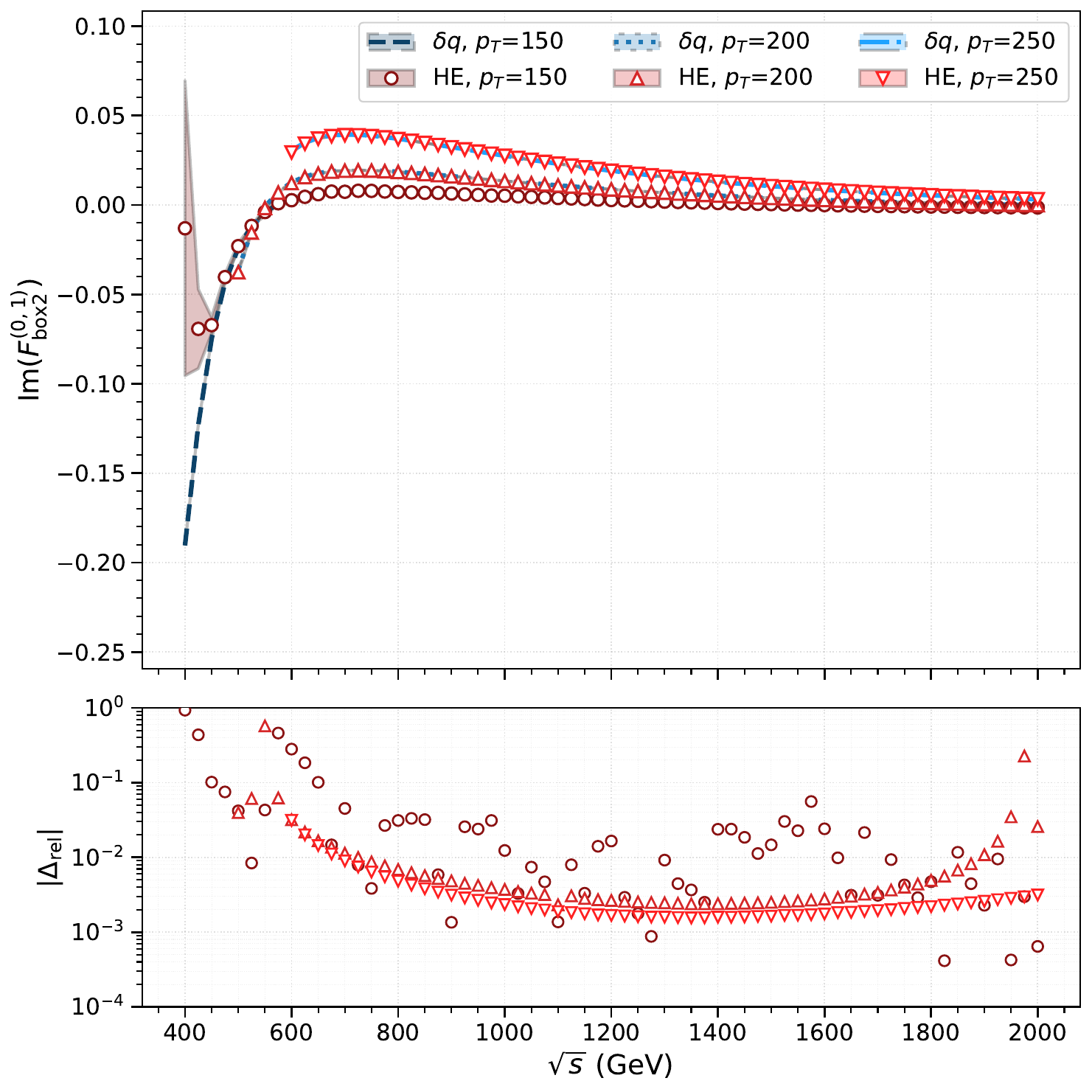} \\
    \includegraphics[width=.45\textwidth]{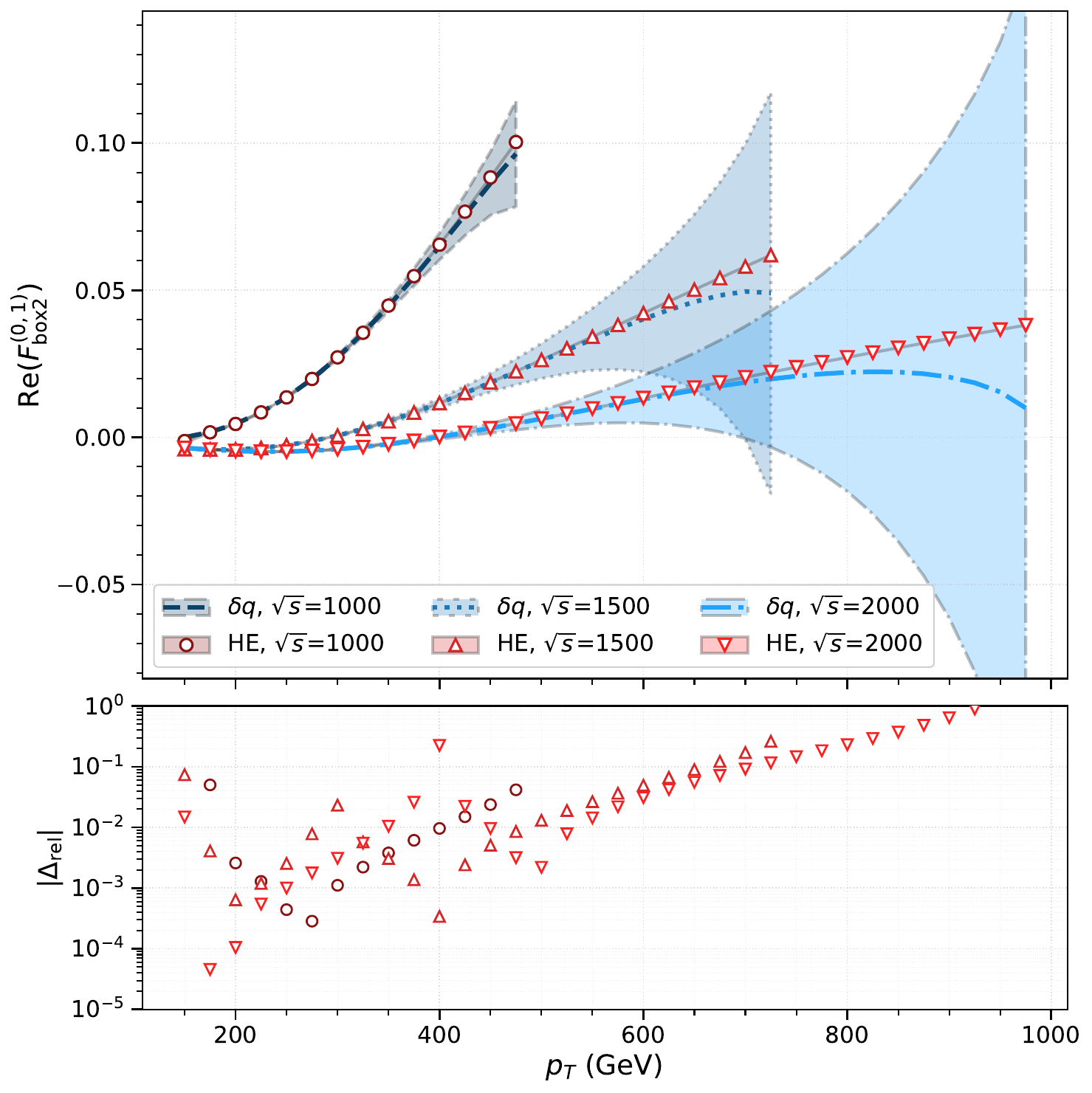} &
    \includegraphics[width=.45\textwidth]{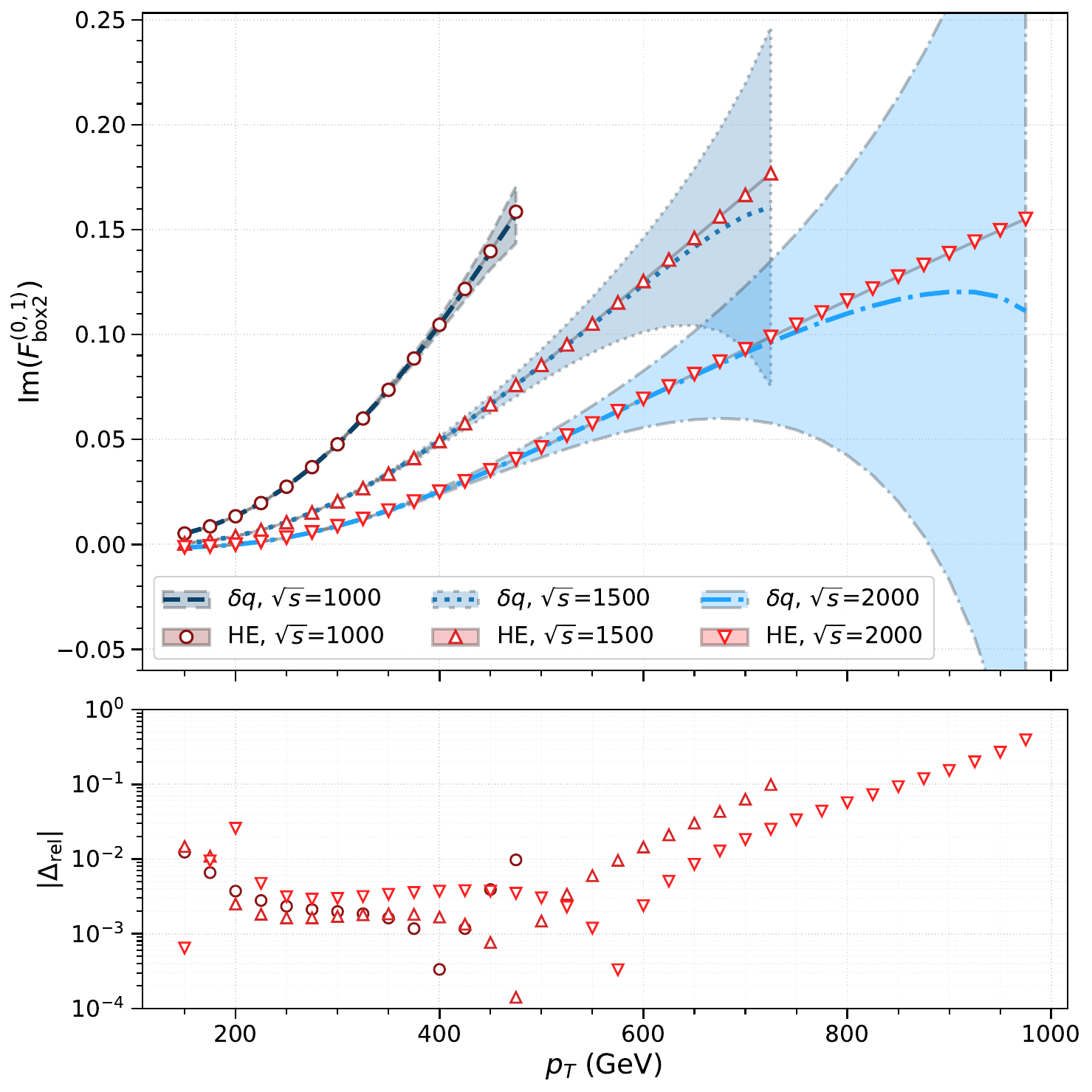} 
    \end{tabular}
    \end{center}
  \caption{\label{fig::f2-photon}
  Same as Fig.~\ref{fig::f1-photon} but for form factor $F_{\rm box2}$.
  }
\end{figure}

In this appendix we present results for the photon contribution to the NLO EW corrections of $gg\rightarrow HH$. The required Feynman diagrams where automatically computed as part of our computational setup. They agree with the $C_F$ colour factor contribution of the NLO QCD corrections of $gg\rightarrow HH$ after adjusting the couplings. We validated this relation with the help of the  \texttt{C++} library \texttt{ggxy}~\cite{Davies:2025qjr}.

In Figs.~\ref{fig::f1-photon} and~\ref{fig::f2-photon} we present the results of the forward and high-energy expansion for the photon contribution which is obtained by using \texttt{ggxy}. We note that for the photon contribution the $\delta q$ expansion and naive-$t$ expansion coincide, such that the implemented results for the forward expansion in \texttt{ggxy} correspond to the combination of the naive-$t$ expansion up to $m_H^4$ and $t^6$ with additional expansion terms in the external mass obtained from the $\delta q$ expansion with $N_q\le 4$. Both the real part and the imaginary part are presented once  as a function of $p_T$ with fixed values of $\sqrt{s}$ and once the other way around. In all panels we observe great agreement between the forward expansion and the high-energy expansion for $200\leq p_T \leq 400$~GeV and all given values of $\sqrt{s}$. As expected the high-energy expansion does not perform well for smaller values and the forward limit expansion for higher values of $p_T$.

In the lower panels, which show the relative deviation between the
forward expansion and the high-energy expansion, it is important to
identify a range of $\sqrt{s}$, respectively $p_T$ values where both
expansions are in good agreement. The upper panels of
Figs.~\ref{fig::f1-photon} and~\ref{fig::f2-photon} demonstrate that
there is indeed a broad range of $\sqrt{s}$ values for which this
holds true, provided that $p_T \ge 200$~GeV. For smaller values, the
$t$ expansion yields good results. This is supported by the lower
panels, which show that for all $p_T$ values close to $200$~GeV,
the agreement is within one percent for all corresponding $\sqrt{s}$
values.

\section{\label{app::data}Results for the form factors for sample phase-space points.}

As supplementary material, we provide numerical results for the form
factors at selected points in phase space. All of these points were
chosen to lie within the radius of convergence for both the $\delta q$
expansion and the naive-$t$ expansion. The results are distributed
across six files, which contain the real and imaginary parts of the
form factors $F_{\text{tri}1}$, $F_{\mathrm{box}1}$, and
$F_{\mathrm{box}2}$. They can be downloaded from~\cite{progdata}.

All files contain the following information: $\sqrt{s}$, $p_T$,
leading-order results, the un-renormalized NLO result, the contribution from counterterms, and the
renormalized NLO contribution. For the numerical result of the form
factor, we also provide an uncertainty, which was determined using the
Pad\'e method. The additional systematic error is not included in the
files; it amounts to approximately 1\% for $p_T \lesssim 300$ GeV and
ranges between 5\% and 10\% in the region where the naive-$t$ expansion
is applied, as explained in the main body of the paper. In all
cases, we employ the same normalization as in
Section~\ref{sec::results} and, consistent with the main body of the
paper, explicitly exclude the photon contribution. For the divergent
results, we use an $\overline{\rm MS}$-like normalization, meaning we
set the Euler constant $\gamma_E$ and $\log(4\pi)$ to
zero. Furthermore, the renormalization scale is set to $\mu =
100$~GeV. All other input parameters are chosen as described in
Section~\ref{sec::results}. For the top-quark mass, we use $m_t^{\rm
OS}$.

\end{appendix}

\bibliographystyle{jhep}

\bibliography{inspire.bib,extra.bib}

\end{document}